\documentclass[prd,floatfix,amsmath,amssymb,10pt,nofootinbib,twocolumn,superscriptaddress,letterpaper]{revtex4}
\usepackage{amsmath,amssymb}
\usepackage{graphicx}
\usepackage{epstopdf}
\usepackage{color}
\usepackage{slashed}
\usepackage{wasysym,stmaryrd}
\usepackage{multirow}
\usepackage{array}
\usepackage{rotating}
\usepackage{url}
\usepackage[bookmarks=false]{hyperref}
\usepackage{hypernat}

\newcommand{\beastie}{{\it Glaucomys volans}}
 
\newcommand{\ourVec}[1]{\left( #1 \right)}
\newcommand{\ourSet}[1]{\left\{ #1 \right\}}
\newcommand{\ourSmallSet}[1]{\{ #1 \}}
\newcommand{\ourMaxMinBracs}[1]{\left[ #1 \right]}

\newcommand{\parent}[1]{\ensuremath{{\mathbb{P}_{#1}}}}

\newcommand{\visible}[1]{\ensuremath{{\mathbb{V}_{#1}}}}
\newcommand{\invisible}[1]{\ensuremath{{\mathbb{I}_{#1}}}}

\newcommand{\genericT}{\ensuremath{T}}
\newcommand{\ourT}{\ensuremath{\top}} 
\newcommand{\ourperp}{\ensuremath{\vee}}
\newcommand{\myT}{\ensuremath{\perp}}
\newcommand{\myL}{\ensuremath{\parallel}}

\newcommand{\massless}{{\ensuremath{\circ}}}

\newcommand{\parentset}{\ensuremath{\mathcal{P}}}
\newcommand{\visset}{\ensuremath{\mathcal{V}}}
\newcommand{\invisset}{\ensuremath{\mathcal{I}}}
\newcommand{\invismassset}{\ensuremath{\tilde\mu}}
\newcommand{\invismassseti}[1]{\ensuremath{\tilde\mu_{#1}}}
\newcommand{\invisvelset}{\ensuremath{\tilde v}}
\newcommand{\invisvelseti}[1]{\ensuremath{\tilde v_{#1}}}
\newcommand{\visassign}[1]{\ensuremath{{\mathcal{V}_{#1}}}}
\newcommand{\invisassign}[1]{\ensuremath{{\mathcal{I}_{#1}}}}
\renewcommand{\th}{\ensuremath{^\textrm{th}}}

\newcommand{\trueparentmass}{{\ensuremath{M}}}
\newcommand{\chiM}{\ensuremath{\mmass}}
\newcommand{\chiV}{\ensuremath{\mv}}
\newcommand{\chiO}{\ensuremath{\mo}}
\newcommand{\chiS}{\ensuremath{\mSS}}
\newcommand{\ourtick}{\checkmark}

\newcommand{\numparents}{\ensuremath{N}}
\newcommand{\numvis}{\ensuremath{N_\mathcal{V}}}
\newcommand{\numinvis}{\ensuremath{N_\mathcal{I}}}
\newcommand{\numvisi}[1]{\ensuremath{\left|{\visassign{#1}}\right|}}
\newcommand{\numinvisi}[1]{\ensuremath{\left|{\invisassign{#1}}\right|}}
\newcommand{\flox}{{\cal F}}

\newcommand{\comp}[1]{\ensuremath{{\mathbf #1}}}
\newcommand{\hide}[1]{}

\newcommand\ourmm[1]{\ensuremath{{  \mathcal{M}   }\ourSmallSet{#1}}}
\newcommand\ourmmsq[1]{\ensuremath{{   \mathcal{M}   }^2\ourSmallSet{#1}}}
\newcommand\pairtitle[2]{The \ourmm{#1,#2} parental mass bound}

\newcommand{\MTTWO}{\ensuremath{M_{\genericT2}}}

\newcommand{\MCT}{\ensuremath{M_{C\genericT}}}

\newcommand{\MEFF}{\ensuremath{m_{\mathrm{eff}}}}

\newcommand{\ROOTSHAT}{\ensuremath{\hat{s}^{1/2}}}

\newcommand{\beq}{\begin{equation}}
\newcommand{\eeq}{\end{equation}}
\newcommand{\beqs}{\begin{equation*}}
\newcommand{\eeqs}{\end{equation*}}
\newcommand{\bea}{\begin{eqnarray}}
\newcommand{\eea}{\end{eqnarray}}
\newcommand{\beas}{\begin{eqnarray*}}
\newcommand{\eeas}{\end{eqnarray*}}

\newcommand{\met}{\slashed{E}_\genericT}
\newcommand{\metvec}{\slashed{\vec{E}}_\genericT}
\newcommand{\mpt}{\slashed{p}_\genericT}
\newcommand{\mptvec}{\slashed{\vec{p}}_\genericT}
\newcommand{\mmass}{{\slashed{\comp M}}}
\newcommand{\mv}{{\slashed{\comp{V}}}}
\newcommand{\mo}{{\slashed{\comp{O}}}}
\newcommand{\mSS}{{\slashed{\comp{S}}}}

\newcommand{\comCL}[1]{{\color{blue}\textsf{\small [CGL: #1]}}}

\newcommand{\comKM}[1]{{\color{red}\textsf{\small [KTM: #1]}}}
\definecolor{purple}{rgb}{0.6,0,0.6}

\newcommand{\particles}{physics objects}
\DeclareMathOperator{\proj}{proj}  
\newcommand{\theExampleFigs}{Figures~\ref{fig:countingParticles}, \ref{fig:countingParticles2parents} and \ref{fig:countingParticles1parent}}

\newcommand{\allevents}{\ensuremath{{\cal E}}}
\newcommand{\ourevents}{\ensuremath{{\cal S}}}
\newcommand{\event}{\ensuremath{{\epsilon}}}
\newcommand{\assumed}{\ensuremath{{\cal A}}}

\begin{document}

\title{A storm in a ``T'' cup:
the connoisseur's guide to 
transverse projections and mass-constraining variables}

\author{A.~J.~Barr}
\affiliation{Department of Physics, Denys Wilkinson Building, Keble Road, Oxford OX1 3RH, UK}
\author{T.~J.~Khoo}
\affiliation{Department of Physics, Cavendish Laboratory, JJ Thomson
Avenue, Cambridge, CB3 0HE, UK}
\author{P.~Konar}
\affiliation{Theoretical Physics Group, Physical Research Laboratory,
Ahmedabad, Gujarat - 380 009, India}
\author{K.~Kong}
\affiliation{Department of Physics and Astronomy, University of Kansas, Lawrence, KA 66045, USA}
\author{C.~G.~Lester}  
\affiliation{Department of Physics, Cavendish Laboratory, JJ Thomson
Avenue, Cambridge, CB3 0HE, UK}
\author{K.~T.~Matchev}
\author{M.~Park}
\affiliation{Department of Physics, University of Florida, Gainesville, FL 32611, USA}

\date{May 13, 2011}

\begin{abstract}
This paper seeks to demonstrate that many of the existing
mass-measurement variables proposed for hadron colliders ($m_T$,
$\MEFF$, $m_{T2}$, missing $\vec{p}_\genericT$, $h_T$, $\sqrt{\hat{s}}_{\rm min}$, etc.) 
are far more closely related to each other than is widely appreciated, and
indeed can all be viewed as a common mass bound specialized
for a variety of purposes.  A consequence of this is that one may
understand better the strengths and weaknesses of each variable, and the
circumstances in which each can be used to best effect.  In order to
achieve this, we find it necessary first to revisit the seemingly
empty and infertile wilderness populated by the subscript ``$T$'' (as
in ``$p_\genericT$'') in order to remind ourselves what this process of
transversification actually means.  We note that, far from
being simple, transversification can mean quite different things to
different people.  Those readers who manage to battle through the
barrage of transverse notation distinguishing ``$\ourT$'' from
``$\ourperp$'' or from ``$\massless$'', and ``early projection'' from ``late projection'', will
find their efforts rewarded towards the end of the paper with (i) a
better understanding of how collider mass variables fit together, (ii) an
appreciation of how these variables could be generalized to search for
things more complicated than supersymmetry, (iii) will depart
with an aversion to thoughtless or na\"ive use of the so-called
``transverse'' methods of any of the popular computer Lorentz-vector
libraries, and (iv) will take care in their subsequent
papers to be explicit about which of the 61 identified variants 
of the ``transverse mass'' they are employing.
\end{abstract}

\maketitle 

\begin{widetext}
\tableofcontents
\end{widetext}

\section{Introduction}

Almost every analysis of data from hadron colliders uses at some 
point a variable which represents a ``projection'' of an energy or 
momentum into the plane transverse to the beams. The typical reason for performing 
these projections is that one does not wish the analysis to be 
sensitive to the unknown momentum -- along the direction of the 
beams -- of the quarks or gluons which collide in the `hard' interaction. 
Given the widespread use of such variables it is perhaps surprising 
that many collider physicists are probably unaware that there exist at least 
two commonly-used ways of projecting of a Lorentz energy--momentum 
vector into the transverse plane, and that these two different 
methods have very different properties when the mass is non-zero
(see Section~\ref{sec:projections} below). 
Furthermore, as explained later in Section~\ref{sec:earlylate},
for each of those transverse projections, there are at least two
inequivalent ways that transverse vectors can be ``added together'', 
each of which has benefits and weaknesses. 
A careful definition of what we mean by a transverse projection 
forms the first part of this paper.

The later part of the paper 
(Sections~\ref{sec:interpreting}--\ref{sec:examples}) 
deals with mass-scale (or energy-scale) 
variables, a variety of which have been proposed in the run-up 
to the LHC data-taking\footnote{For a recent review 
see~\cite{Barr:2010zj}.}. Though some of these variables have 
been constructed from careful consideration of the Lorentz 
symmetries of space-time, others have been created in a 
somewhat {\it ad-hoc} process, after simulations demonstrate 
that they provide good signal-to-background discrimination, 
or that they are highly correlated with the mass of some 
particle or particles. The main aim of this part of the paper is to 
demonstrate that many of these seemingly {\it ad-hoc} 
definitions are in fact not only well-motivated from the 
kinematical perspective, but also that the associated 
variables are more closely related than one might have thought.

\begin{figure}[tbh]
\includegraphics[width=0.9\columnwidth]{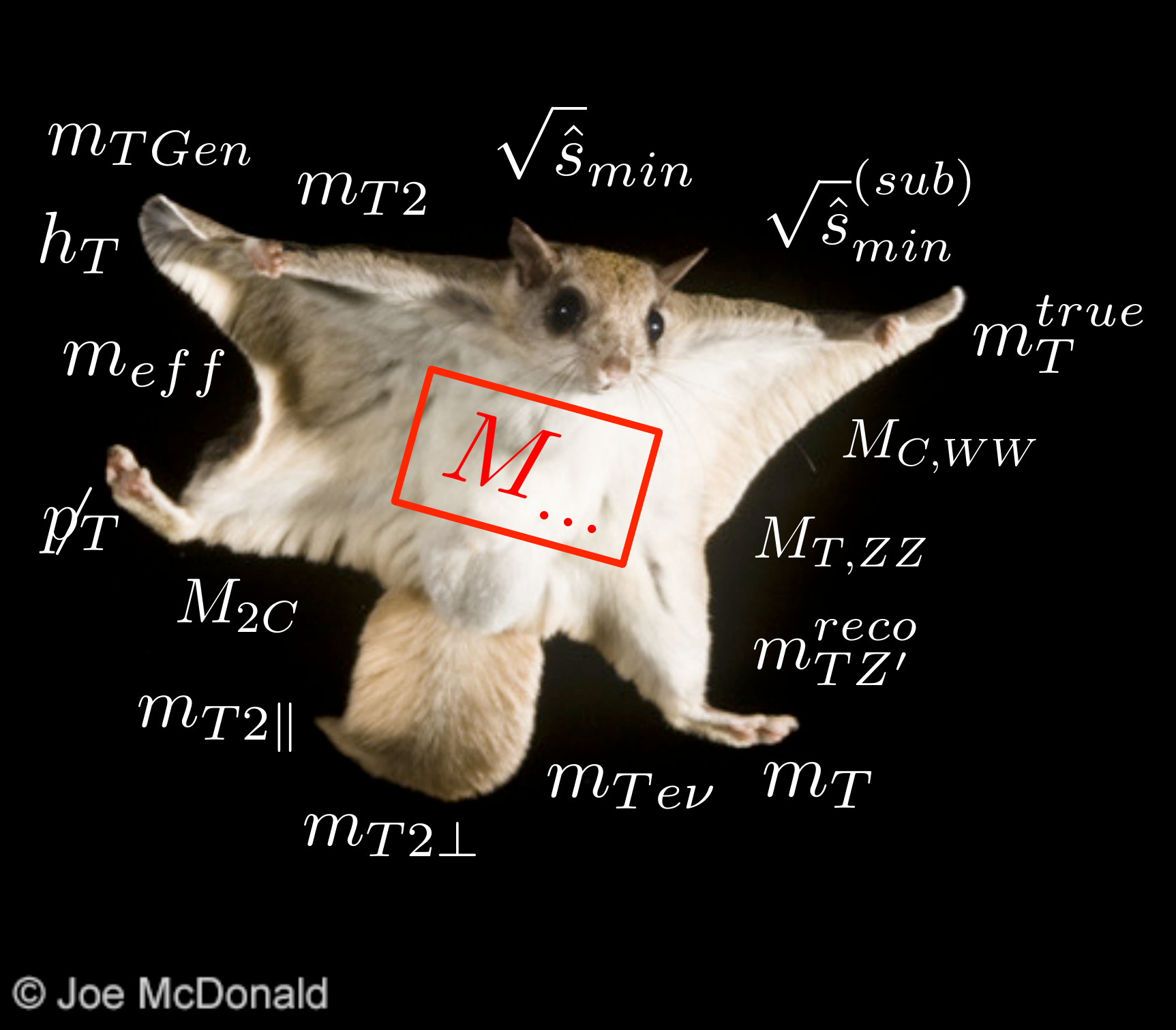}
\caption[The stretched, webbed limbs of the \beastie]{\label{fig@thesquirrel}
The stretched, webbed limbs of the \beastie{}
have been adapted by generations of natural selection to provide an ideal visual illustration
of the various different, yet related, transverse mass variables
(and incidentally provide an appropriate aerodynamic shape for gliding flight).
\newline {\tiny Photograph \copyright{} Joe McDonald.}
}
\end{figure}

Figure~\ref{fig@thesquirrel} illustrates some of the variables
that are found to be connected in ways that are not widely appreciated.
One might argue that we add little to the sum total of human 
knowledge by merely showing the relationships between 
existing variables which are already known to work well in particular roles. 
However, careful study of their similarities and differences 
not only gives insights into why (and under what circumstances) 
these choices are appropriate, it also fits them into a common 
framework -- from which it is straightforward to make 
generalizations to more complex decay topologies.

The paper is organized as follows; first we carefully
define our notation for Lorentz 1+3 vectors and their 
transverse projections in Section~\ref{sec:vecnotation}.
Then in Section~\ref{sec:projections}
we describe the two common but inequivalent transverse projections, 
which we shall denote by subscripts $\ourT$ or $\ourperp$.
We also introduce the special case of a ``massless''
transverse projection, denoted by $\massless$.
In Section~\ref{sec:comparison} we compare the
results from the three different types of projections:
$\ourT$, $\ourperp$ and $\massless$.
In Section~\ref{sec:earlylate} we highlight
the differences between projecting into the transverse 
plane {\em before} or {\em after} forming composite objects.
Section~\ref{sec:interpreting} describes the general 
event topology targeted by new physics searches 
in channels with missing momentum. 
\par
All of those ingredients are put to work in 
Sections~\ref{sec:imass} to \ref{sec:literature},
which contain the main results of this paper.
In Section~\ref{sec:imass} we introduce  
the general class of mass-constraining variables
which can be usefully applied for studying events containing invisible particles.
The set of possible transverse mass variables is extended in 
Section~\ref{sec:doubleT}, where we consider additionally
projected one-dimensional objects.
Some mathematical properties of these mass-constraining variables 
are discussed in Section~\ref{sec:math}.
Some of the variables have previously appeared elsewhere in the literature
and we clarify the corresponding connections in 
Section~\ref{sec:literature}. In Section~\ref{sec:examples}
we illustrate the use of these variables with two simple examples: 
an $s$-channel resonant production process, for which we take 
inclusive Higgs boson production $pp\to h\to W^+W^- \to \ell^+\ell^-+\met$,
and a pair-production process represented by top quark 
production $pp\to t\bar{t}\to b\bar{b}\ell^+\ell^-+\met$.
Section~\ref{sec:conclusions} contains a short summary and conclusions.

Appendix~\ref{sec:libraries} contains a short guide to the 
currently existing computer libraries and codes which can 
be used for computing some of the variables described in 
the main body of the text. 
Appendix~\ref{sec:massbounds} 
provides derivations of extremal mass-bound results
and other general mathematical proofs which are used elsewhere in the paper.

\section{Notation and conventions} 
\label{sec:vecnotation}

\subsection{Labelling momenta and their components}

In general, capital letters ($P$, $Q$, $M$, $E$, etc.) 
will refer to genuine 1+3 dimensional vectors,
while lowercase letters ($p$, $q$, $m$, $e$, etc.) 
will refer to ``less than 1+3'' dimensional constructs.
Lower indices $i,j,\ldots$ label {\em individual} final state particles, 
while lower indices $a,b,\ldots$ are used for parent particles and the corresponding 
{\em collections} of final state particles defined below in Sec.~\ref{sec:interpreting}.
We also use upper indices $\mu, \nu,\ldots$ to label the components of 1+3 vectors,
and upper indices $\alpha, \beta, \ldots$ to label the components
of the projected 1+2 dimensional transverse ``vectors'' of the 
types defined in Section~\ref{sec:projections}.  
The 1+3 metric $g_{\mu\nu}$ is $\textrm{diag}(1,-1,-1,-1)$ and the 
1+2 dimensional metric $g_{\alpha\beta}$ is $\textrm{diag}(1,-1,-1)$.
Thus the 1+3 energy-momentum vector for some particle is written 
$P^\mu = \ourVec{ E, \vec{P} }$ 
and the corresponding mass denoted by a capital $M$:
\begin{equation}
M^2 = P^\mu P_\mu = E^2 - \vec{P}^{\, 2}.
\end{equation}

\par

\begin{figure}
\begin{center}
\includegraphics[width=0.99\columnwidth]{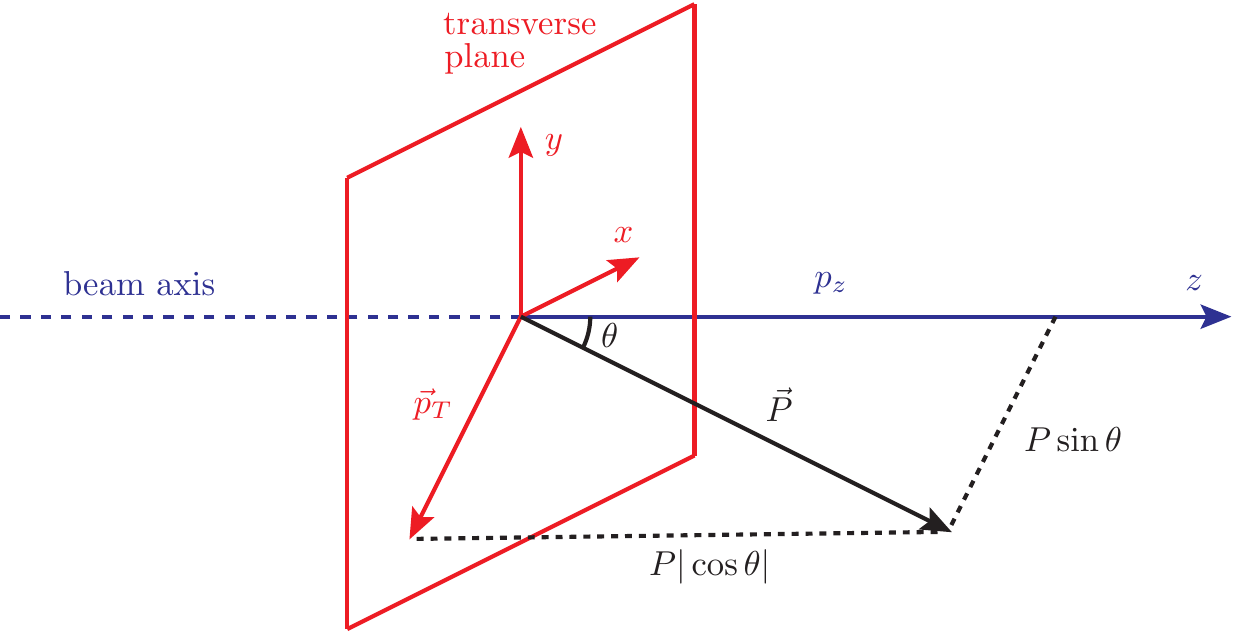}
\caption[Orientation of axes]{\label{fig:Tprojection} 
The standard geometry of a collider experiment.
The $z$ axis (in blue) is oriented along the beam,
while the $x$ and $y$ axes (in red) define the transverse plane.
Any 3-dimensional vector $\vec{P}$ can be uniquely decomposed 
into a longitudinal component $p_z$ and a transverse component
$\vec{p}_T$.}
\end{center}
\end{figure}

As illustrated in Fig.~\ref{fig:Tprojection},
any 3-dimensional vector $\vec{P}$ can be trivially decomposed 
into a transverse and a longitudinal component:
\begin{equation}
\vec{P} \equiv \ourVec{ \vec{p}_\genericT, p_{z} }.
\label{Pdecompose}
\end{equation}
The transverse momentum $\vec{p}_\genericT = \ourVec{p_x,p_y}$
of the particle is, of course, 2-dimensional, so it has a 
lowercase ``p''. Similarly, the longitudinal momentum $p_z$
is 1-dimensional, and is also lowercase. By contrast, 
the energy $E$ measured in the detector is a component of a ``1+3 dimensional thing'',
since it is given in terms of the 1+3 dimensional mass $M$
and the 3-dimensional momentum $\vec{P}$:
\beq
E = \sqrt{M^2 + \vec{P}^{\,2}} = \sqrt{M^2 + \vec{p}_\genericT^{\,2} + p_z^2}.
\eeq

When it comes to projecting geometric 3-vectors like $\vec{P}$,
the decomposition shown in eq.~(\ref{Pdecompose}) and Fig.~\ref{fig:Tprojection}
is unambiguous. One has no other choice -- the very
definition of the transverse plane requires one simply to dispose of
the $z$-component to arrive at $\vec{p}_{\genericT}=\ourVec{p_x,p_y}$.  
All the transverse projections considered in this paper (and any others
that one might invent) must share this property, or else they cannot
justify being so named.

\par

However, ``projecting'' the time-like component $E$ is not, in itself,
a well defined operation. What does it mean?  There is not a single
correct answer, but rather a number of different answers, each with
different properties and motivations.  How one should (and even {\em
whether} one should) project time-like components of 1+3 Lorentz
vectors is dependent on what one is trying to achieve.

\subsection{Labelling transverse projections}
\label{sec:labproj}

In the particle physics literature, one can find evidence of at least
three different types of ``transverse projection'' being applied to
(1+3)-Lorentz vectors -- although this diversity is not obvious at
first glance, as the majority of papers do not explicitly state which
projection they are using.\footnote{This may be because all forms turn out to be equivalent for massless particles.}  Even those papers which
define the projection explicitly, usually neither comment on {\em
why} the particular choice was made, nor comment on what would happen
were another projection to have been used.

One of the main objectives of this paper is to place these three main
types of projection side-by-side so that their differences, the things
that they share, and their respective uses can be directly compared.
Before we describe them in more detail, we make some remarks about
notation.

In the literature, {\bf all} of the types of projection are labelled
by the {\bf same} symbol: the letter ``$\genericT$''.  Since in this document we need to
clearly distinguish the three types of projection, it is necessary for
us to create our own notation for each - and we use the three symbols
``$\ourT$'', ``$\ourperp$'' and ``$\massless$'' for that purpose.

 We will continue to use the letter ``$\genericT$'' to indicate ``generic''
 transverse quantities, {\it i.e.}~quantities which are either common
 to all projections (e.g.~the transverse momentum (2)-vector
 $\vec{p}_\genericT$ already commented upon, the missing tranverse
 momentum vector $\mptvec$, or the transverse upstream visible
 momentum vector $\vec{u}_T$ defined below in Section~\ref{sec:topology}) 
 or for quantities which
 for historical reasons carry a transverse subscript, but which may
 not be tied to one type of projection to the exclusion of others
 (e.g.~$h_T$).

Note that certain quantities, such as the so-called ``transverse
energy'' and ``transverse mass'', 
are different in each of the projections.  For this reason
the symbol ``$e_\genericT$'' is effectively meaningless, and should
appear nowhere in this document (outside this sentence) unlike
$e_\ourT$, $e_\ourperp$ and $e_\massless$ (which are all different and
all well-defined).  Similarly, $m_\genericT$
is also ambiguous, and should be specified as being 
$m_\ourT$, $m_\ourperp$ or $m_\massless$.
In contrast, ``$\vec{p}_\genericT$'' is perfectly
legitimate, and indeed (as we have already noted) is equivalent to
$\vec{p}_\ourT$, $\vec{p}_\ourperp$ and $\vec{p}_\massless$:
\beq
\vec{p}_\genericT \equiv \vec{p}_\ourT \equiv \vec{p}_\ourperp \equiv \vec{p}_\massless\,  .
\eeq

\par

\section{Transverse projections}
\label{sec:projections}

In this Section we describe the three different types of projections
``$\ourT$'', ``$\ourperp$'' and ``$\massless$''.
While reading this and the following sections, the reader may find it helpful
to refer to Table~\ref{tab:comparisonoftransversemethods}
for notational reference, and also to see how the results for each
projection compare to those of the others.

\begin{table*}\begin{center}
\renewcommand\arraystretch{1.5}
\begin{tabular}{| c | c | c | c |}
\hline
               & \multicolumn{3}{c|}{{\bf Transverse projection method} }   \\
\cline{2-4}
{\bf Quantity} & {\bf Mass-preserving `\ourT'} & {\bf Speed-preserving `$\ourperp$'} & {\bf Massless `$\massless$'} \\
\hline
Original (4)-momentum & \multicolumn{3}{c|}{$ P^\mu = \ourVec{ E, \vec{p}_\genericT, p_{z}}$ }  \\ 
(1+3)-mass invariant  & \multicolumn{3}{c|}{$ M = \sqrt{E^2 - \vec{p}_\genericT^{\,2} - p_z^2}$ } \\ 
Transverse momentum & \multicolumn{3}{c|}{ $ \vec{p}_\genericT \equiv \ourVec{p_x,p_y} $ } \\
\hline
(1+2)-vectors & 
$ p_\ourT^\alpha \equiv \ourVec{ e_\ourT, \vec{p}_\ourT }$ & 
$ p_\ourperp^\alpha \equiv \ourVec{ e_\ourperp, \vec{p}_\ourperp }$ &
$ p_\massless^\alpha \equiv \ourVec{ e_\massless, \vec{p}_\massless }$ \\
\hline
\parbox{4cm}{{\vskip 2mm}
Transverse momentum \\ under the projection {\vskip 2mm}} 
& $\vec{p}_\ourT \equiv \vec{p}_\genericT$ &  $\vec{p}_\ourperp \equiv \vec{p}_\genericT$ &  $\vec{p}_\massless \equiv \vec{p}_\genericT$ \\
\hline
\parbox{4cm}{{\vskip 2mm} 
Transverse energy \\ under the projection {\vskip 2mm}} & 
$e_\ourT \equiv \sqrt{M^2 + \vec{p}_\genericT^{\,2}}$  & 
$ e_\ourperp \equiv E  \left|{\sin \theta}\right| = |\vec{p}_T|/V $ &
$ e_\massless \equiv |\vec{p}_\genericT|$ \\
\hline
\parbox{4cm}{{\vskip 2mm}
Transverse mass \\ under the projection {\vskip 2mm}} & $m_\ourT^2 = e_\ourT^2 - \vec{p}_\ourT^{\,2}$ &  
$m_\ourperp ^2 \equiv  e_\ourperp^2 - \vec{p}_\ourperp^{\,2}$ & 
$m_\massless^2 \equiv  e_\massless^2 - \vec{p}_\massless^{\,2} = 0$\\
\hline
\multirow{2}{*}{\parbox{4cm}{{\vskip 2mm} Relationship between transverse quantity and its (1+3) analogue {\vskip 2mm}}}
& $m_\ourT = M$ 
&  $m_\ourperp = M \left|\sin\theta\right| $ & $m_\massless = 0$  \\ 
\cline{2-4}
   & \parbox{4cm}{{\vskip 2mm} $\frac{1}{v_{\ourT}}=\frac{1}{V}\sqrt{1+(1-V^2)\frac{p_z^2}{p_T^2}}$ {\vskip 1mm}}    & $v_{\ourperp}=V$  &  $v_\massless = 1$  \\ [2mm]
\hline
\parbox{4cm}{Equivalence classes under $(1+3) \overset{\mathrm{\proj}}{\longmapsto} (1+2)$ } &
\parbox{4cm}{{\vskip 2mm} 
All $P^\mu$ with the same \\ $p_x$, $p_y$ and  $M$ {\vskip 2mm}} &
\parbox{4cm}{{\vskip 2mm} 
All $P^\mu$ with the same \\ $p_x$, $p_y$ and  $V$
{\vskip 2mm}} & 
\parbox{4cm}{{\vskip 2mm} 
All $P^\mu$ with the same \\ $p_x$ and $p_y$
{\vskip 2mm}}\\ [2mm]
\hline
\end{tabular}
\caption[A comparison of transversification methods]{\label{tab:comparisonoftransversemethods} A comparison of the three transversification methods
introduced in Section \ref{sec:projections}.}
\end{center}
\end{table*}

\subsection{The mass-preserving ``\ourT'' projection}
\label{sec:TBGformalism}

The first approach we will describe, which will be denoted by a ``$\ourT$'' subscript, 
is the most common in the mass measurement literature.  
For example it is found in the early literature on the transverse 
mass when it was used to measure the $W$ mass 
\cite{vanNeerven:1982mz,Arnison:1983rp,Banner:1983jy,Smith:1983aa,Barger:1983wf} 
and in the generalization of the transverse mass to pair production, 
namely \MTTWO\ (the stransverse mass) \cite{Lester:1999tx,Barr:2003rg,Lester:2007fq,Gripaios:2007is,%
Cho:2007qv,Barr:2007hy,Cho:2007dh,Ross:2007rm,Nojiri:2008hy,Cho:2008cu,Barr:2008ba,%
Nojiri:2008vq,Cho:2008tj,Cheng:2008hk,Burns:2008va,Barr:2008hv,%
Kim:2009nq,Barr:2009jv,Konar:2009wn,Konar:2009qr,%
Barr:2009wu,Alwall:2009zu,Choi:2010dk} as well as in literature relating 
to \MCT\ \cite{Tovey:2008ui,Serna:2008zk,Polesello:2009rn,Cho:2009ve,Matchev:2009ad,Barr:2010ii} 
and in reviews of the field \cite{Barr:2010zj}.  

\par

In the $\ourT$ projection one
defines the 1+2 dimensional transverse energy\footnote{Note that 
in equation~(\ref{eq:eTdef}), it is the middle expression 
$\sqrt{M^2 + \vec{p}_{\genericT}^{\,2}}$ that we use to justify 
our ``calling'' the LHS a (transverse) ``energy'' -- since 
it is square root of a ``mass squared plus a transverse momentum squared''.  
Someone who saw the right hand expression first, $\sqrt{E^2 - p_{z}^2}$, 
could argue differently, and might reasonably expect us to call the whole 
quantity a "longitudinal mass" -- since it is a square root of 
an ``energy squared minus a longitudinal momentum squared''.  
All this really goes to show is that the ``name'' of the quantity 
is to some extent a matter of convention rather than physics.
} $e_{\ourT}$ and transverse momentum $\vec{p}_\ourT$ 
in terms of the 1+3 dimensional mass $M$ and 1+3 dimensional components according to
\bea
e_{\ourT} &\equiv& \sqrt{M^2 + \vec{p}_{\genericT}^{\,2}} \equiv \sqrt{E^2 - p_{z}^2},
\label{eq:eTdef}\\
\vec{p}_\ourT &\equiv& \vec{p}_\genericT,\\
m_\ourT &\equiv& M.
\label{eq:MourTdef}
\eea
In this case, the components of the 1+2 dimensional  quantity
\beq
p_\ourT^\alpha\equiv \ourVec{ e_{\ourT},\vec{p}_{\ourT}}
\label{defpT}
\eeq
satisfy the mass shell condition
\beq \label{eq:massshellT}
e_\ourT^2 - \vec{p}_\ourT^{\, 2} = m_\ourT ^ 2 = M^2
\eeq
with the 1+3 dimensional mass $M$.

\par

The equivalence class for this projection function 
-- the set of 1+3 vectors which map to the {\em same} 1+2 projected vector
under $\ourT$
-- consists of the set of 1+3 vectors with the same $\vec{p}_\genericT$ and $M$:
\beq
\ourVec{\sqrt{M^2+p_T^2+p_z^2},\vec{p}_T,p_z } 
\overset{\ourT}{\longmapsto} \ourVec{\sqrt{M^2+p_T^2},\vec{p}_T }.
\label{Tproj}
\eeq
The fact that all members of the equivalence class share the same mass is 
what motivates us to call this the ``mass preserving'' $\ourT$ projection.
 
Given its dominant use in 
the literature, it is something of a surprise that the nomenclature of 
the $\ourT$ projection is {\em not} adopted in the commonly used high-energy 
physics computer libraries such as {\tt CLHEP} \cite{Lonnblad:1994kt} 
or {\tt ROOT}~\cite{Antcheva:2009zz} which instead implement 
the alternative $\ourperp$ projection introduced below in Section~\ref{sec:perpmalism}.
The $\ourT$ projection is, however, used in the ``Oxbridge stransverse 
mass library'' \cite{oxbridgeStransverseMassLibrary} 
and the U.C.~Davis \MTTWO\ library \cite{zenuhanStransverseMassLibrary}. 
See Appendix~\ref{sec:libraries} and Table~\ref{tab:libraries} in it
for a summary of library conventions. 

\subsection{The speed-preserving ``$\ourperp$'' projection}
\label{sec:perpmalism}

Alternatively one can follow the method of the {\tt CLHEP} \cite{Lonnblad:1994kt} 
and {\tt ROOT}~\cite{Antcheva:2009zz} libraries
and ``project'' the energy on the transverse plane, 
using the same angle $\theta$ as for the momentum vector.
As alteady seen in Fig.~\ref{fig:Tprojection}, the
magnitude $p_T$ of the transverse momentum $\vec{p}_T$
is related to the magnitude $P$ of the 3-dimensional 
momentum $\vec{P}$ by
\beq
p_T = P\, \sin\theta,
\label{eq:ptsinthetadef}
\eeq
with
\beq
\tan\theta \equiv \frac{p_{T}}{p_{z}}.
\eeq
Thus by analogy with (\ref{eq:ptsinthetadef}) one can define the 
transverse energy in terms of its 1+3 dimensional counterpart $E$ as
\beq
e_{\ourperp} \equiv E \sin\theta.
\label{eq:eperpdef}
\eeq

Then for any {\em individual} 1+3 momentum vector we have the $\ourperp$ version of 
the ``transverse'' components
\bea
e_{\ourperp} &\equiv& E \sin\theta = \frac{p_{T}}{\sqrt{p_{T}^2+p_{z}^2}}\, E,  \label{eq:eperpdef1}\\
\vec{p}_{\ourperp} &\equiv& \vec{p}_{T}, \\
m_{\ourperp} &\equiv& M \sin\theta = \frac{p_{T}}{\sqrt{p_{T}^2+p_{z}^2}}\, M.  \label{eq:mperpdef}
\eea
We can take the angle $\theta$ to be defined in $(0,\pi)$,
so that $e_{\ourperp}$ and $m_{\ourperp}$ are always nonnegative.

\par

In this $\ourperp$ method of projection we can also introduce 1+2 ``vectors'' which  
now have components
\beq
p_{\ourperp}^\alpha\equiv \ourVec{ e_{\ourperp},\vec{p}_{\ourperp}}.
\label{defpperp}
\eeq
The $\ourperp$ projected components obey a different mass shell relation
than the $\ourT$ projected components in \eqref{eq:massshellT}:
\beq
e^2_{\ourperp} - p_{\ourperp}^2 = m_{\ourperp}^2 \le M^2,
\label{imperp}
\eeq
with the 1+2 dimensional $\ourperp$ projected mass $m_{\ourperp}$.

Just as an aside, one could also define the ``longitudinal'' components
in complete analogy to (\ref{eq:eperpdef1})-(\ref{eq:mperpdef})
\bea
e_{z} &\equiv& E\, |\cos\theta| = \frac{|p_{z}|}{\sqrt{p_{T}^2+p_{z}^2}}\, E, \\
p_{z} &\equiv& p_{z}, \\
m_{z} &\equiv& M\, |\cos\theta| = \frac{|p_{z}|}{\sqrt{p_{T}^2+p_{z}^2}}\, M, 
\eea
although in what follows we shall not be making any use of those.
The connection between the 1+3 dimensional quantities and the 
$\ourperp$ 1+2 dimensional components is
\bea
E^2 &=& e^2_{\ourperp} + e^2_{z}, \label{eq:perpEsq} \\ [2mm]
M^2 &=& m^2_{\ourperp} + m^2_{z}\ . \label{eq:perpMsq}
\eea

\par
For massive vectors\footnote{See section~\ref{sec:masslesscomments} 
for comments concerning the massless case.}
the equivalence classes of the $\ourperp$ projection 
are different from those of the $\ourT$ projection.
The mass-shell relation \eqref{imperp} implies that all the 1+3 vectors 
which map to the same 1+2 vector under the $\ourperp$ projection
share the same value of $m_\ourperp=M\sin\theta$ and thus generally do {\em not}
preserve the usual invariant mass $M$, since $m_\ourperp\ne M$ for
any $\theta\ne\frac{\pi}{2}$.

A more physical picture of the equivalence class
of vectors for the $\ourperp$ projection can be 
found by considering the 3-speed of the particle
\beq
V \equiv \frac{P}{E}.
\label{3speed}
\eeq
After the $\ourperp$ projection, the corresponding
2-speed is given by
\beq
v_{\ourperp} \equiv \frac{p_{\ourperp}}{e_{\ourperp}}
= \frac{p_{T}}{e_{\ourperp}}
= \frac{P\sin\theta}{E\sin\theta}
= \frac{P}{E}.
\label{2speed}
\eeq
Eqs.~(\ref{3speed}) and (\ref{2speed})
reveal that the $\ourperp$ projection
is ``speed preserving'', i.e.
\beq
v_{\ourperp} = V,
\eeq
which justifies our choice of subscript notation for this kind 
of transverse projection.
The equivalence class for the $\ourperp$ projection therefore 
consists of all 1+3 vectors with the 
same $\vec{p}_T$ and speed $V$: 
\beq
\ourVec{\frac{\sqrt{p_T^2+p_z^2}}{V},\vec{p}_T,p_z } 
\overset{\ourperp}{\longmapsto} \ourVec{\frac{p_T}{V},\vec{p}_T }.
\label{Perpproj}
\eeq
Note that members belonging to the same equivalence class under 
the $\ourperp$ projection (\ref{Perpproj}) have the same speed, but
different masses, while members of the same equivalence class 
under the $\ourT$ projection (\ref{Tproj}) have the same mass, but different speeds.

\subsection{The massless ``\massless'' projection}
\label{sec:masslessism}
The massless ``\massless'' projection defines components
\bea
e_\massless &\equiv &\left| \vec{p}_\genericT\right|, 
\label{emassless}
\\
\vec{p}_\massless &\equiv &\vec{p}_\genericT
\eea
and thereby defines a massless 1+2 vector of the form
\beq\label{eq:masslessdef}
p_\massless^\alpha = \ourVec{|\vec{p}_\genericT|, \vec{p}_\genericT} .
\eeq
The main feature of this projection is that the 1+2 vector $p_\massless^\alpha$ 
{\em always} has a null invariant
\beq\label{eq:masslessinv}
g_{\alpha\beta}\, p_\massless^\alpha\, p_\massless^\beta \equiv m_\massless^2 = 0.
\eeq

It should be noted that $p_\ourT^\alpha$ and $p_\ourperp^\alpha$ 
have three degrees of freedom ($\ourSet{e_\ourT, p_x, p_y}$ and 
$\ourSet{e_\ourperp, p_x, p_y}$, correspondingly). 
Therefore their equivalence classes
are one-dimensional, and can be parameterized by the coordinate
$p_z$, as indicated in (\ref{Tproj}) and (\ref{Perpproj}). In contrast,
our `\massless' projected vector 
$p_\massless^\alpha$ has only {\em two} degrees of freedom, $p_x$ and 
$p_y$ ---  the time-like component being fully specified from $p_x$ and 
$p_y$ through $e_\massless=|\vec{p}_T|$.
The equivalence class of any $p_\circ^\alpha$ vector
is therefore also a $4-2$ = 2-dimensional object, parameterized by, say, $p_z$ and $E$:
\beq
\ourVec{E,\vec{p}_T,p_z } 
\overset{\massless}{\longmapsto} \ourVec{\left| \vec{p}_\genericT\right|,\vec{p}_T }.
\eeq

\section{Comparison of the different transverse projections}
\label{sec:comparison}

\begin{figure}
\begin{center}
\includegraphics[width=0.99\columnwidth]{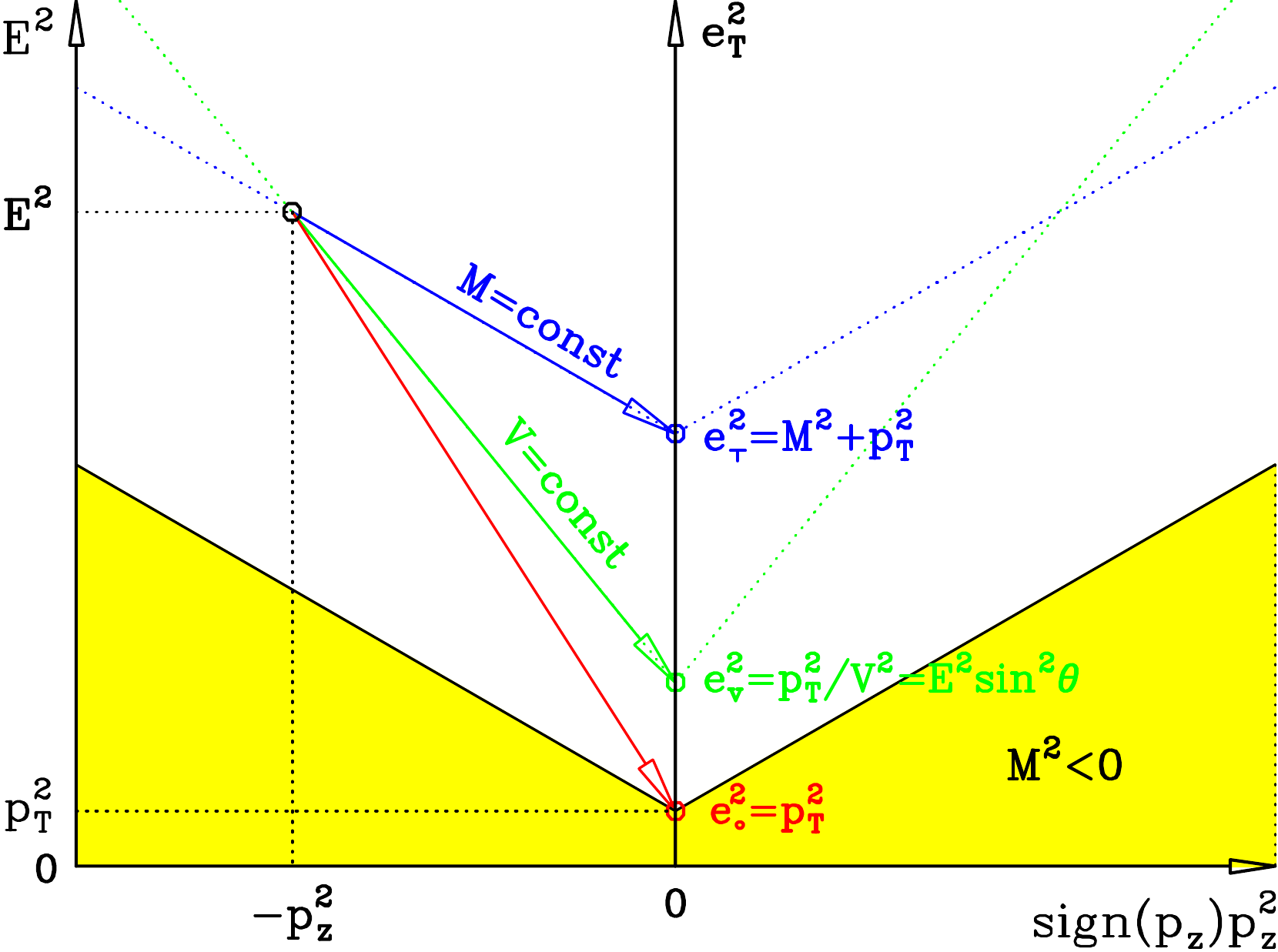}
\caption[E versus pz]{\label{fig:2Dproj} 
A pictorial representation of the three transverse projections
discussed in Section~\ref{sec:projections}. 
The colored arrows represent the mappings under the $\ourT$, 
$\ourperp$ and $\massless$ projections.
The blue and green dotted lines represent the equivalence classes 
of the projected points under the $\ourT$ and $\ourperp$ projections,
respectively.
}
\end{center}
\end{figure}

The three different projections discussed in Section~\ref{sec:projections}
are pictorially represented in Fig.~\ref{fig:2Dproj}.
For a given fixed value of $p_T$, the white region in the figure depicts
all possible allowed values of the energy $E$ 
and the longitudinal momentum $p_z$.
(The yellow-shaded region $E^2<p_T^2+p_z^2$ is forbidden because it corresponds to
a tachyonic particle with $M^2<0$, travelling with superluminal speed.)
In this figure, we consider the plane of energy {\em squared}
versus momentum {\em squared}, and in order to retain the information 
about the sign of the longitudinal momentum component, we plot
${\rm sign}(p_z)p_z^2$, so that the mapping from the $(E, p_z)$-plane to the $(E^2, {\rm sign}(p_z)\,p_z^2)$-plane is one-to-one.

Each of the three transverse projections maps a point with some 
given\footnote{For definiteness, in Fig.~\ref{fig:2Dproj} we 
have chosen an illustration point with $p_z<0$.}  
values of $E$ and $p_z$ onto the 
$p_z=0$ axis as shown. 
In the case of $\ourT$, the projection is
along a line of constant mass $M$ and results in 
transverse energy squared $e_\ourT^2 = M^2 + p_T^2$.
In the $(E^2, {\rm sign}(p_z)\,p_z^2)$-plane, lines of constant 
$M$ are straight lines, which explains our choice of
quadratic power scale on the axes. Fig.~\ref{fig:2Dproj} 
illustrates that the equivalence class of vectors under the 
$\ourT$ projection is one-dimensional: it is represented 
by the two blue dotted straight lines, which can be simply
parameterized by the value of $p_z$.

The $\ourperp$ projection, on the other hand, projects along 
a line of constant speed $V$, as indicated in Fig.~\ref{fig:2Dproj}.
In the $(E^2, {\rm sign}(p_z)\,p_z^2)$-plane, lines of constant 
$V$ are also straight lines, albeit with a different slope.
The resulting value of the transverse energy is 
$e_\ourperp=p_T/V$. The corresponding equivalence class 
of vectors is given by the two green dotted lines, and can
also be parameterized in terms of a single parameter, say $p_z$.

Finally, the massless ``$\massless$'' projection maps {\em any} allowed point 
in the $(E, p_z)$-plane to the massless 1+2 vector with transverse energy
$e_\massless = p_T$. The equivalence class of vectors in this case 
is two-dimensional, and is represented by the whole white 
shaded region in Fig.~\ref{fig:2Dproj}.

\begin{figure}
\begin{center}
\includegraphics[width=0.99\columnwidth]{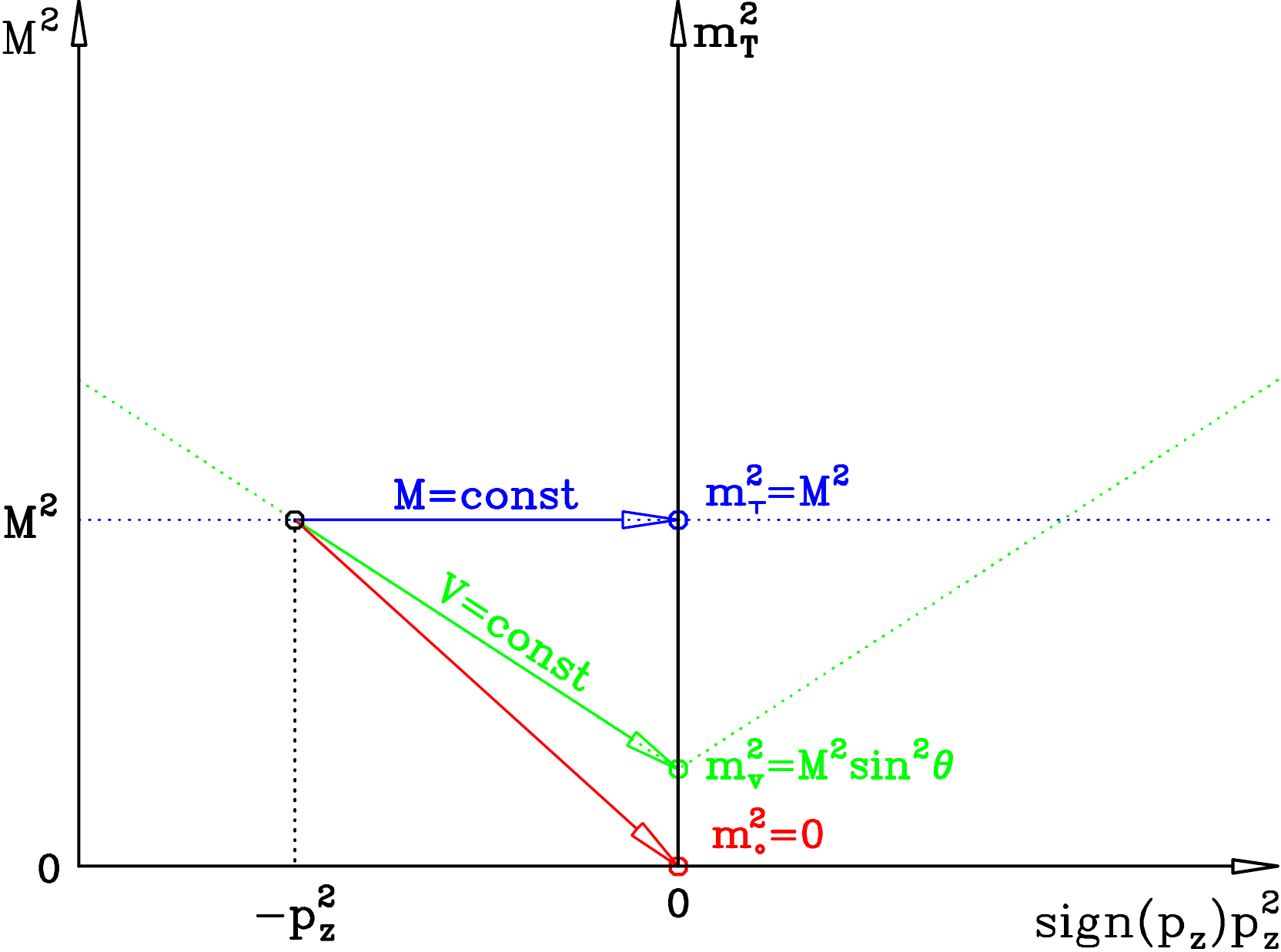}
\caption[M versus pz]{\label{fig:m_vs_pz} 
The same as Fig.~\ref{fig:2Dproj}, but 
plotted in the $(M^2,{\rm sign}(p_z)\,p_z^2)$-plane.}
\end{center}
\end{figure}

All of the previous discussion can be recast in the language of the 
$(M^2,{\rm sign}(p_z)\,p_z^2)$-plane, as shown in Fig.~\ref{fig:m_vs_pz}.
In this case, the whole $M^2\ge 0$ half-plane is allowed, 
and the $\ourT$ projection projects horizontally onto the
$p_z=0$ axis, following the blue arrow. The $\ourperp$ 
projection is also done along a straight line, following 
the green arrow. As before, the equivalence classes for 
the $\ourT$ and $\ourperp$ operations are straight lines, 
while the equivalence class for the ``$\massless$'' case 
is given by the whole $M^2\ge 0$ half-plane.

\subsection{A hierarchy among projections}

As illustrated in Fig.~\ref{fig:m_vs_pz}, 
the definition of each projection 
imposes a hierarchy among the projected masses of the form: 
\beq 
M = m_\ourT \ge m_\ourperp \ge m_\massless = 0.
\label{masshierarchy}
\eeq
We draw attention to this hierarchy here as it will have very 
close analogues in the more complicated experimental mass 
bounds derived from each type of projection in the later sections.

Given the mass hierarchy (\ref{masshierarchy}), 
eqs.~(\ref{eq:massshellT}), (\ref{imperp}) and (\ref{emassless})
imply that a similar hierarchy exists for the projected energies:
\beq
E \ge e_{\ourT} \ge e_\ourperp \ge e_\massless = p_T,
\label{energyhierarchy}
\eeq
which is illustrated in Fig.~\ref{fig:2Dproj}.

\subsection{Equivalence in the massless limit}
\label{sec:masslesscomments}

We note that in the special (but common) case in which the 
original four-vector is massless ($M=0$) all projections 
are equivalent since
\beq
\lim_{M \to 0}{e_\ourT} =  
\lim_{M \to 0}{e_\ourperp} = 
e_\massless = \left|\vec{p}_\genericT\right|
\label{eq:allETsameinmasslesscase}
\eeq
and thus
\beq
\lim_{M \to 0} p_\ourT^\alpha =
\lim_{M \to 0} p_\ourperp^\alpha =
p_\massless^\alpha = \ourVec{|\vec{p}_\genericT|, \vec{p}_\genericT}.
\label{eq:allprojsameinmasslesscase}
\eeq
Clearly the projections are not equivalent for massive particles, 
nor for collections of massless particles (unless they be collinear) 
since collections of massless particles can have large total 
invariant mass -- the equivalence extends only to application 
to {\em individual} massless particles.

\par

In practice, the statement above may also be taken as saying that
all the projections are equivalent in the {\em high-energy limit} --
i.e.~the limit in which the momentum of a particle is much greater
than its mass -- again only at the level of {\em individual}
high-energy particles.

\par

Since all the projections are equivalent in the above limits, and
since most individual reconstructed particles in high-energy physics
experiments satisfy one of those limits due to the small masses of the
leptons and light quarks, one might wonder what all the fuss is about.
However, the importance of the distinctions will be seen to arise and
become very large when we consider {\em composite} particles, i.e.
collections of massless ``daughter'' particles\footnote{The need for considering 
composite particles arises when dealing with short-lived heavy resonances,
which decay promptly to a certain collection of daughter particles,
which in turn are seen in the detector. The energy and momentum
of the parent resonance are correspondingly obtained by 
summing the measured energies and momenta of the daughter particles.}. 
Composite particles are expected to have non-negligible masses, 
even when they consist of sums of (approximately) massless particles.
As we already learned from the simple example considered 
in Figs.~\ref{fig:2Dproj} and \ref{fig:m_vs_pz}, not only do these
composite particles generate very different projected 1+2 vectors,
but the classes of equivalent four-vectors associated with those 
projections are very different as well.

\section{Summing and projecting: early versus late projections}
\label{sec:earlylate}

In forming transverse kinematic variables for composite particles,
one needs to perform two separate operations:
summation of the momentum vectors of the daughter particles,
and projecting into the transverse plane.
The {\em order} of these operations 
does not matter for the two {\em space-like} vector components:
\bea
\sum_i \vec{p}_{i\ourT    }  
& = & \left( \sum_i \vec{P}_i \right)_{\ourT}    , \label{commuteourT} \\ 
\sum_i \vec{p}_{i\ourperp }  
& = & \left( \sum_i \vec{P}_i \right)_{\ourperp} , \label{commuteourperp}  \\
\sum_i \vec{p}_{i\massless}  
& = & \left( \sum_i \vec{P}_i \right)_{\massless}, \label{commutemassless} 
\eea
where we use an index $i$ to label the momenta of the individual daughter 
particles and the sums run over all such daughter particles\footnote{Recall our convention that 
lowercase letters refer to 1+2 dimensional quantities and
capital letters refer to 1+3 dimensional quantities.
Thus in the left-hand-sides of eqs.~(\ref{commuteourT})-(\ref{commutemassless})
we are adding 2-dimensional transverse vectors, while in the 
right-hand-sides we are first adding the corresponding 3-vectors, 
then projecting their sum onto the transverse plane.}.

However, projecting before or after the sum can make a very significant 
difference to the value of the time-like ($e_\ourT$, $e_\ourperp$ or $e_\massless$) 
component of the final 1+2 vector -- and therefore the operations of 
projecting and summing do {\em not} generally commute:
\bea
\sum_i e_{i\ourT    } & \ne & \left( \sum_i E_i \right)_{\ourT    }, \\
\sum_i e_{i\ourperp } & \ne & \left( \sum_i E_i \right)_{\ourperp }, \\
\sum_i e_{i\massless} & \ne & \left( \sum_i E_i \right)_{\massless}.
\eea

One can see clearly how the order makes a difference if one considers 
an extreme case consisting of a pair of massless daughter particles 
travelling in opposite directions along the beam pipe, i.e.~with 1+3 momenta
\bea
P_{1}^\mu &=& \ourVec{E, 0,0,+E},  \\ [1mm]
P_{2}^\mu &=& \ourVec{E, 0,0,-E}.
\eea
If one were to project these 1+3 momenta into the transverse plane before 
summing (a combined operation hereafter called {\em early projection}), 
one would find that the resulting 1+2 dimensional vector
\begin{subequations}
\bea
\sum_i p_{i\ourT}^\alpha 
&=&  p_{1\ourT}^\alpha + p_{2\ourT}^\alpha  
\\
&=& \ourVec{E, 0,0,E}_\ourT + \ourVec{E, 0,0,-E}_\ourT 
\\
&=& \ourVec{0,0,0}+\ourVec{0,0,0} 
\\
&=& \ourVec{0,0,0} \label{eq:nullexample}
\eea
\end{subequations}
is null. A null sum would also be obtained if we had used 
the $\ourperp$ or $\massless$ projections.\footnote{In fact 
for this example we have chosen massless vectors for which 
the `$\ourT$', `$\ourperp$', and `$\massless$' projections are identical.}
However if one were first to sum the Lorentz 1+3 vectors $P_i^\mu$ 
and then later project into the transverse plane 
(hereafter denoted {\em late projection}) one would find that
\begin{subequations}
\bea
\left(\sum_i P_{i}^\mu \right) _\ourT
&=& \left( P_{1}^\mu + P_{2}^\mu \right)_\ourT  \\
&=& \left( \ourVec{E, 0,0,E} + \ourVec{E, 0,0,-E} \right)_\ourT \\
&=&  \ourVec{2E,0,0,0}_\ourT \\
&=& \ourVec{2E,0,0},
\eea
\end{subequations}
which is clearly not the same as was found in \eqref{eq:nullexample}.
This extreme case shows that while projecting early 
has the effect of reducing dependence on 
longitudinal momenta, projecting late means that the 
resultant projected composite retains much more sensitivity to 
the original relative momenta along the beam directions.

This concludes this section, whose main purpose was simply to
highlight the difference between the ``early'' and the ``late''
transverse projection. It also underscores the need to develop
the proper notation to distinguish between these two types of
transverse projections, which we shall do below in 
Section~\ref{sec:earlylatenotation}. 
The differences between the two projections will be further 
illustrated with the physics examples considered in the 
later sections. One may reasonably wonder which one of the 
two projections is more appropriate and should be used. 
In principle, the answer to this question will depend on the analysis 
being performed. If one is initially building a composite particle from 
two leptons, e.g. from a $Z$-boson decay $Z \to e^+e^-$, then 
the relative longitudinal momentum of the positron 
and the electron is probably a safe quantity to retain 
full sensitivity to in one's calculations.
However, in cases where jets at large rapidity $|\eta|$ are 
concerned, the probability of QCD radiation grows rapidly 
as one gets closer and closer to the beam direction.
One will often prefer not to have the high-energy end 
of the composite-particle spectrum dominated by combinations 
of low $|p_\genericT|$, high-energy forward-going jets with 
other low $|p_\genericT|$, high-energy backward-going jets, 
so in this latter case, early projection would probably be 
appropriate. Nevertheless, giving a universal prescription 
for selecting the ``correct'' transverse projection
for {\em collections} of particles is beyond the scope of this paper.
The best method will depend on non-kinematic factors, 
such as the size of any backgrounds, the detector resolution, 
and other factors that will vary from case to case.

\section{Interpreting events}
\label{sec:interpreting}


\subsection{Characterizing an event}
\label{sec:topology}

\begin{figure}[t]
\begin{center}
\includegraphics[width=0.99\columnwidth]{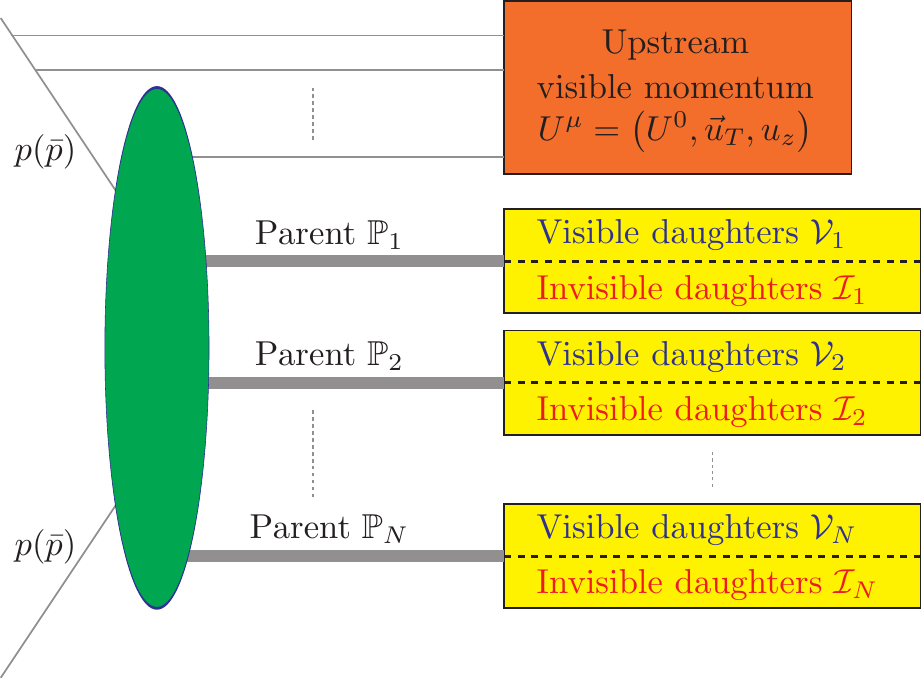}
\caption[The event topology considered]{\label{fig:event} 
The event topology for new physics searches and measurements used in this paper. 
}
\end{center}
\end{figure}

Analysis of an event is a game. The aim of the game is to {\em interpret} the 
available information within a particular framework or hypothesis. 
In this paper we wish to employ a very general framework 
that will be useful for searches and mass measurements 
at hadron colliders ($pp$, $p\bar{p}$ or even $\bar{p}\bar{p}$ for that matter).
Specializations of this framework will then be useful in a wide variety of 
different contexts. The general layout of an event is represented in Fig.~\ref{fig:event}. 
The figure comprises: two incoming objects, denoted by the proton 
lines on the left hand side; an interaction, represented by an 
oval `blob'; and some final state objects, contained within the 
rectangles on the right hand side. Since it is the final state 
objects that provide the kinematic information about the event, 
we now take some time to explain rather carefully what we mean by them.

We define final state objects of two types. A {\em visible} final 
state object is one that leaves a signal in the detector that 
betrays its presence. Those signals may then be reconstructed 
and interpreted as an individual particle -- for example as photon, 
electron or muon -- or
the signals may be indicative of a composite object, such as a 
QCD or tau jet. The ``visible object'' category is deliberately 
allowed to be sufficiently broad as to permit the inclusion of 
very heavy, visibly decaying, composite objects such as $Z$, $W$ or top quarks. 
The classification of a final state object as `visible' here 
implies not only that a signal consistent with the presence of 
some particle has been observed, but also that the full Lorentz 
energy-momentum vector of that particle can be reconstructed from 
the observed signal (to within some experimental precision).  For most heavy 
visible objects (jets, $W$, $Z$, $H$ bosons, \ldots), the four-momentum of the
visible object must be calculated from the vector sum of its constituents.

By contrast, an {\em invisible} final state object is one that 
leaves no direct signal, but the existence of which is demanded 
by the {\em interpretation} of the event being imposed by the analyst.
The numbers, types and masses of any invisible final state 
particles form part of the interpretation of the event.  
The 3-momentum vectors of all invisible particles are {\it a priori} 
unknown, and are constrained only by conservation of the total 
momentum of the event in the plane transverse to the beam.
The general framework can accommodate a final state hypothesis 
in which invisible particles have particular known (or rather 
assumed) masses, but it can also be applied when some or indeed 
all of those invisible particles have unknown masses.

As illustrated in Fig.~\ref{fig:event}, the next step in interpreting the event is to
partition the combined set of all final state objects (visible and invisible)
into subsets, which are represented by rectangles in the figure. 
Each final state object must be found in one and only one such subset.
There is one subset per {\em parent} plus one further subset, the latter being labelled
``upstream visible momentum'' in the figure.

In our interpretation a {\em parent} is any short-lived object 
that is believed to have decayed to produce the visible and 
invisible final state objects in its associated set. The 
term `parent' is usually associated with a short-lived heavy 
state, most often a reasonably narrow resonance (whether 
produced directly in the ``hard scatter'' or from decays 
of even heavier objects).

The general framework presented permits a variety of 
different interpretations for any given event.
For any particular interpretation there is a corresponding 
partitioning of the final state 
into subsets. For example an event which contains evidence 
of an electron and a positron, and which is hypothesized 
to also contain a neutrino and an anti-neutrino, could be 
partitioned into parent/daughter combinations: 
$W^+ \rightarrow \ourSet{e^+,\,\nu_e}$ and 
$W^-\rightarrow \ourSet{e^-,\,\bar{\nu}_e}$ for one analysis; 
however another analysis might find it more appropriate 
to partition those objects according to the interpretation 
$Z^0 \rightarrow \ourSet{e^+,\,e^-}$ and $Z^0\rightarrow \ourSet{\nu_e,\,\bar{\nu}_e}$.

The subset corresponding to any parent may contain any number (including zero) of visible
particles and any number (including zero) of invisible particles --- though 
it is not meaningful to have a totally empty set of daughters. 
The framework is very general, in that
the number of parents can be arbitrary, and
the nature, mass and decay mode of any parent 
need not be related to those of any other.
There is therefore a great deal of freedom in performing the partition into subsets.
We shall later be constraining the masses of the parents 
so the subsets should be chosen to correspond to the descendants 
of the parents whose invariant masses we are interested in.

Figure~\ref{fig:event} also shows the one special (non-`parent') 
subset into which visible final state particles may be allocated. 
That set is labelled ``{\em upstream visible momentum} (UVM)'', 
and is designed to be a `catch-all' that will accommodate any 
visible particle not allocated to any of the parent sets.
This is a special set in the following senses: firstly it is 
permitted to contain (by assertion) only visible objects; 
and secondly, and crucially, final state objects allocated 
to this set are {\em not} used directly to constrain the 
mass of any parent. Objects in this set are only used to 
keep track of overall energy-momentum conservation. 
We do not specify the elements found in this UVM set, but 
in practical applications it almost always contains some 
contribution from ``soft'' particles that are unallocated 
to any parent. Such soft components usually include 
calorimeter energy found outside of jets, and low energy 
jets from multiple parton interactions, and perhaps from initial state radiation (ISR). 
As well as these soft components, one must include any 
other visible objects not associated with any parent. 
The UVM set will often contain more than just `soft' 
activity --- since any type of visible particle can 
end up therein --- possibly including decay products 
(of heavy progenitor particles) that the analyst chose 
not to allocate to any parent.
In practice, every hadron collider event has {\em some} amount of UVM. 
Furthermore, as discussed in 
\cite{Gripaios:2007is,Barr:2007hy,Burns:2008va,Barr:2009jv,Konar:2009qr,Konar:2009wn,Matchev:2009ad}, 
the presence of a {\em significant} amount of UVM can 
in fact be beneficial in mass reconstruction studies.

Apart from the reconstructed physics objects,
another important experimental quantity is the 
{\em missing transverse momentum} in the event.  
This quantity is the experimental
collaboration's best estimate of the amount (and direction) of momentum in any
particular event that has been carried away in the plane transverse to the
beam by invisible particles.
It is an important quantity insofar as we will wish to 
apply the constraint that the missing momentum 
in an event is entirely due to the invisible final state objects.

\subsection{Notation used to characterize events}
\label{sec:eventnotation}

\begin{table*}\renewcommand\arraystretch{1.6}
\begin{tabular}{|c|c|l|c|}
\hline
& Symbol & Meaning  & See also \\
\hline
\multirow{11}{*}{\begin{sideways}objects and sets\end{sideways}}
 & $\left|\cal A\right|$ & Cardinal number (number of elements) of any fininte set $\cal A$. & \multirow{11}{*}{\begin{sideways}Figures~\ref{fig:countingParticles}, \ref{fig:countingParticles2parents} and \ref{fig:countingParticles1parent}\end{sideways}} \\
& $\parent a$ & $a\th$ parent $(a \in\ourSet{ 1,2,\ldots,\numparents})$& \\
& $\parentset$ & Set of all parents $\parentset \equiv \ourSet{\parent 1, \parent 2, \ldots , \parent \numparents}$& \\
& $\visassign{a}$ & Set of visible final state objects associated with the $a\th$ parent& \\
& $\invisassign{a}$ & Set of invisible final state objects associated with the $a\th$ parent& \\
& $\visset \equiv \bigcup_a   \visassign{a}$ &  Set of all visible final state objects ($\visset \equiv\ourSet{\visible 1, \visible 2, \ldots,\visible{\numvis}}$)& \\
& $\invisset \equiv \bigcup_a \invisassign{a}.$ & Set of all invisible final state objects  ($\invisset \equiv\ourSet{\invisible 1, \invisible 2, \ldots,\invisible{\numinvis}}$)& \\
& $N \equiv \left| \parentset \right|$ & Number of parents assumed for the interpretation being applied & \\
& $N_{\cal V} \equiv \left|\visset\right|$ & Total number of visible final state objects & \\
& $N_{\cal I} \equiv \left|\invisset\right|$ & Total number of invisible final state objects & \\
& indices & $\left\{\parbox{10cm}{\raggedright For notational purposes, indices are used interchangably with the names of the particles they identify.  For example: ``$\visassign{a}$'' and ``$\visassign{\parent a}$'' are equivalent; ``$i \in \visset$'' and ``$i\in\ourSet{1,2,\ldots,\numvis}$'' are equivalent; ``$a \in \parentset$'' and ``$a\in\ourSet{1,2,\ldots,\numparents}$'' are equivalent, etc.}\right\}$ & \\  
\hline
\multirow{5}{*}{\,\begin{sideways}1+3 momenta\end{sideways}\,} 
& $P^\mu_i = \ourVec{E_i, \vec{p}_{i\genericT}, p_{iz}}^\mu$ & 1+3 momentum components of the $i\th$ final state visible object ($i \in \visset$)& \\
& $Q^\mu_i = (\tilde{E}_i, \vec{q}_{i\genericT},q_{iz})^\mu$ & Hypothesized 1+3 momentum components of the $i\th$ final state invisible  ($i \in \invisset$)& \\
& $\comp{P}_a^\mu \equiv \sum_{i\in \visassign{a}} P_i^\mu$ & Sum of 1+3 momentum components of visible objects belonging to parent $\parent a$& \eqref{eq:defPa}\\
& $\comp{Q}_a^\mu \equiv \sum_{i\in \invisassign{a}} Q_i^\mu$ & Sum of 1+3 momentum components of invisible objects belonging to parent $\parent a$& \eqref{eq:defQa} \\
& $U^\mu \equiv \ourVec{ U^0, \vec{u}_T, u_z}^\mu$ & Total 1+3 momentum components of the `UVM' set& \eqref{eq:theetmissassumption} \\
\hline
\multirow{12}{*}{\begin{sideways}derived quantities\end{sideways}}
& $\mptvec$ & Missing transverse momentum two vector (magnitude $|\mptvec| = \mpt$) & \eqref{eq:theetmissassumption}\\
& $M_a\equiv M_{\parent a}$  & Mass of the $a\th$ parent ($a \in \parentset$)& \\
& $M_i\equiv M_{\visible i}$  & Mass of the $i\th$ visible final state object ($i \in \visset$)& \\
& $\tilde M_i\equiv M_{\invisible i}$  & Hypothesized mass of the $i\th$ invisible ($i \in \invisset$)& \\
& $\invismassseti a\equiv\ourSet{{\tilde M_i} \mid {i\in\invisassign{a}}}$ & Set of hypothesised masses of the invisibles associated with parent $\parent a$& \\
& $\invismassset \equiv \bigcup_a \invismassseti a$ & Set of the hypothesised masses of all invisibles & \\
& ${\cal M}_a$ & Hypothesized 1+3 dim. invariant mass of the composite parent particle $\parent a$  & \eqref{eq:parentmass}\\
& $\comp{M}_a$ & 1+3 dim. invariant mass of the visibles in $\visassign{a}$ & \eqref{Macompvis}\\
& $\tilde{\comp{M}}_a$ & 1+3 dim. invariant mass of the invisibles in $\invisassign{a}$ & \eqref{Macompinv}\\
& $V_i$ & 3-speed of the $i\th$ visible ($i \in \visset$) & \\
& $\tilde{V}_i$ & Hypothesized 3-speed of the $i\th$ invisible ($i \in \invisset$)& \\
& $\invisvelseti{a} \equiv \ourSet{\tilde V_i \mid i \in \invisassign a} $ & Hypothesised 3-speeds of the invisibles associated with parent $\parent a$ & \eqref{eq:deftildevela}\\
& $\invisvelset \equiv  \bigcup_a \invisvelseti{a}$ & Set of hypothesised 3-speeds of all the invisibles & \eqref{eq:deftildeveltot}\\ 
\hline
& $\mmass_a \equiv \sum_{i\in \invisassign a}\ourMaxMinBracs{\tilde M_i}$ & Sum of the masses of those invisibles associated with parent $\parent a$ & \eqref{eq:chimdefined}\\
& $\chiM \equiv \ourSet{\mmass_a \mid a\in \parentset}$ & Set of all `invisible particle mass sum parameters' & \eqref{Mslashadef}\\
& $\mv_a \equiv \max_{i\in \invisassign{a}}  \ourMaxMinBracs{ \tilde V_i }$ &  Largest hypothesised 3-speed of any invisible associated with parent $\parent a$ & \eqref{eq:chivadefined}\\
& $\chiV\equiv\ourSet{{\mv_a}\mid{a\in\parentset}}$ & Set of all `maximum invisible 3-speed parameters' & \eqref{eq:chivdefined}\\
\hline
\multirow{2}{*}{\begin{sideways}1+2 d\end{sideways}}
& $p_{i\genericT}^\alpha = \ourVec{e_{i\genericT},\,\vec{p}_{i\genericT}}^\alpha$ & 1+2 dim. projected energy-momentum vector for the $i\th$ visible & 
\multirow{2}{*}{Sec.~\ref{sec:projections}}\\
& $q_{i\genericT}^\alpha = \ourVec{\tilde{e}_{i\genericT},\,\vec{a}_{i\genericT}}^\alpha$ & Hypothesized 1+2 dim. projected energy-momentum vector for the $i\th$ invisible & \\
\hline
\end{tabular}
\caption[Notation]{\label{tab:eventnotation}
Notation used in the description of events.
}
\end{table*}

We require considerable amount of notation to describe events and the 
hypotheses and interpretations that we layer on top of them.    
We have summarized the notation we have adopted in 
Table~\ref{tab:eventnotation} --- and we recommend that readers 
immediately compare the first section of that table with any 
of the three small concrete examples provided in \theExampleFigs\ in 
order to follow later sections.   For the simplest pieces of 
notation, Table~\ref{tab:eventnotation} serves as the primary 
definition.  Notation that requires more explanation will be 
described in more detail either below or at first point of use.

The $N$ parents are labelled $\parent a$, $(a=1,2,\ldots,N)$. 
The set of observed visible (hypothesized invisible) daughters 
associated with $\parent a$ is labelled $\visassign{a}$ ($\invisassign{a}$).  
Since no visible or invisible particle has more than one parent, 
we have $\visassign a \bigcap \visassign b = 0$ and $\invisassign a \bigcap \invisassign b=0$ 
when $a\ne b$, and so the number of visible (invisible) particles may either 
be written as the sum of the number of visible (invisible) daughters 
of each parent 
$\numvis = \sum_{a=1}^{\numparents} \numvisi{a}$,\ \  
($\numinvis = \sum_{a=1}^{\numparents} \numinvisi{a}$) 
or as the number of elements from the set of all visible 
(invisible) daughters $\numvis = \left|\visset\right|$ 
($\numinvis = \left|\invisset\right|$) where 
$\visset=\bigcup_{a=1}^\numparents \visassign a$ ($\invisset=\bigcup_{a=1}^\numparents \invisassign a$).

As seen in Table~\ref{tab:eventnotation}, in our conventions
the letter ``P'' (``p'') will be used to denote measured 
momenta, and the letter ``Q'' (``q'') will be used for the momenta 
of any invisible or hypothesized particles.  Correspondingly,
the individual 4-momenta $P^\mu_i$, ($i\in {\cal V}$),
of the visible daughters are measured and known, 
while the individual 4-momenta $Q^\mu_i$, ($i\in {\cal I}$), 
of the invisible daughters are not measured and remain unknown. 
We denote the masses of the visible final state particles by $M_i$ 
and those of the hypothesized invisible final state particles by 
$\tilde M_i$.   Similarly, we will find it convenient to denote 
the 3-speeds of the visible final state particles as $V_i$ and 
the 3-speeds of the hypothesized invisible final state particles 
by $\tilde V_i$.  In some places we will need to refer to sets 
of these masses or speeds, and so we define: (i) the set 
consisting of the hypothesized masses of {\em all} invisible particles:
\bea 
\invismassset=\ourSet{{\tilde M_i} \mid {i\in\invisset } },
\label{eq:deftildemutot}
\eea
(ii) the set containing only the hypothesized masses of the 
invisible particles assocated with parent $\parent a$:
\bea
\invismassseti a=\ourSet{{\tilde M_i} \mid {i\in\invisassign{a}} },
\label{eq:deftildemua}
\eea
(iii) the set consisting of the hypothesized 3-speeds of {\em all} 
invisible particles: 
\bea 
\invisvelset=\ourSet{{\tilde V_i} \mid {i\in\invisset } },
\label{eq:deftildeveltot}
\eea  
and (iv) the set containing only the hypothesized 3-speeds 
of the invisible particles assocated with parent $\parent a$: 
\bea
\invisvelseti a=\ourSet{{\tilde V_i} \mid {i\in\invisassign{a}} }.
\label{eq:deftildevela}
\eea




We denote the missing transverse
{\em momentum} two-vector by the symbol\footnote{Note that due to its 
status as an experimentally measurable quantity, for the
missing transverse momentum $\mptvec$ we use the letter 
``p'' as opposed to ``q'', even though at high values $\mptvec$ 
is interpreted as the total transverse momentum of {\em invisible}
particles.} 
$\mptvec$ and its magnitude thus $\mpt$.  
Note that some authors use variants of the symbol ``$\met$'' 
to denote the missing transverse momentum,\footnote{By right, 
since its meaning is derived from conservation
of {\em momentum} in the transverse plane, the missing transverse
momentum ought universally to be known as $\mptvec$.  Alas, much of
the hadron-collider literature, especially that from the experimental
collaborations, calls the missing transverse momentum the ``missing
{\em energy}'' or ``missing transverse {\em energy}'' and denotes 
its magnitude ``$\met$'' and its two vector
by some variant of ``$\metvec$''.  This is perhaps a result of history
(a hang over from $e^+ e^-$ or LEP terminology where the collision of
point-particles from mono-energetic beams meant that one really {\em
could} talk about missing energy) and the fact that $\mptvec$ is
often reconstructed, at least in part, from calorimetric {\em energy}
deposits under the assumption they were produced by massless
\particles.} but the distinction is necessary in this paper 
as we shall (as others should)
make important distinctions between energy and momentum.  
\par
We wish to apply the constraint that the missing momentum 
in an event is entirely due to the $\numinvis$ invisible particles with 
momenta $Q^\mu_i$, rather than to jet mismeasurement, for example.   
In other words, we use the relationships expressed in:
\beq
 \sum_{i=1}^{\numinvis} \vec{q}_{iT} = \mptvec \equiv - \vec{u}_T - \sum_{i=1}^{\numvis}\vec{p}_{iT} .
\label{eq:theetmissassumption}
\eeq
in which the first equality represents our desire to constrain 
the momenta of the invisible particles (and only those particles) 
using $\mptvec$, while the second equality reminds us of our 
assumptions of how $\mptvec$ is constructed as an experimentally measurable quantity.
These relationships also remind us that we have assumed (i) that there 
are no sources of invisible momentum other than those coming from the 
parent decays, and (ii) that we have defined the ``Upstream visible momentum'' 
to contain all visible momentum deposits 
which did not originate from the decay of any parent.

\begin{figure}[t]
\begin{center}
\includegraphics[width=\linewidth]{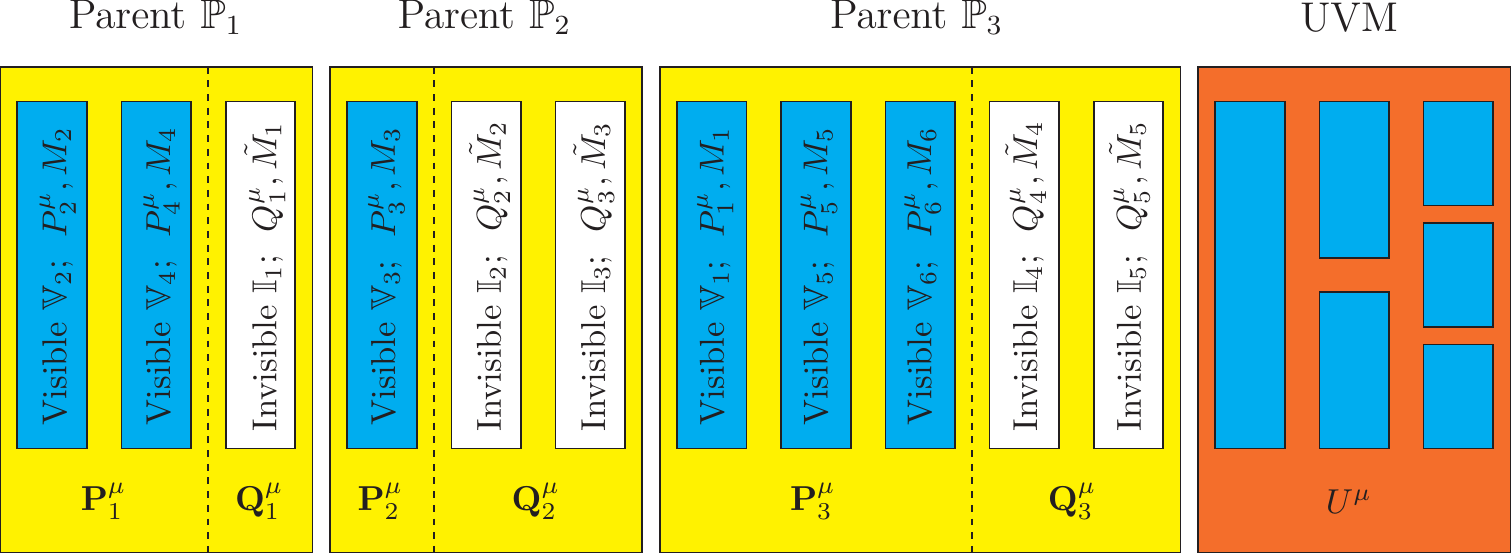}
\end{center}\vspace{-5mm}
\caption[Notation - example 1]{\label{fig:countingParticles} This figure illustrates the
notation used to label \particles\ and their assignments to parent hypotheses.
The figure
shows a hypothesis in which six ($\numvis=6$) visible \particles\ 
$\visset = \ourSet{\visible 1, \visible 2, \visible 3, \visible 4, \visible 5, \visible 6}$
and five ($\numinvis=5$) invisible \particles\ $\invisset = \ourSet{\invisible 1, \invisible 2, \invisible 3, \invisible 4,\invisible 5}$ have been assigned to three ($\numparents=3$) parents $\parentset=\ourSet{\parent 1, \parent 2, \parent 3}$ according to the assignments
$\visassign{1}=\ourSet{\visible 2,\visible 4}$,
$\visassign{2}=\ourSet{\visible 3}$,
$\visassign{3}=\ourSet{\visible 1,\visible 5,\visible 6}$,
$\invisassign{1}=\ourSet{\invisible 1}$,
$\invisassign{2}=\ourSet{\invisible 2,\invisible 3}$ and
$\invisassign{3}=\ourSet{ \invisible 4,\invisible 5}$.
The number of visible \particles\ assigned to each parent in turn are therefore
$\numvisi{1}=2$,
$\numvisi{2}=1$ and
$\numvisi{3}=3$ and the number of invisible \particles\ assigned to each parent in turn are
$\numinvisi{1}=1$,
$\numinvisi{2}=2$ and
$\numinvisi{3}=2$.}
\end{figure}

\begin{figure}[tbh]
\begin{center}
\includegraphics[width=0.65\linewidth]{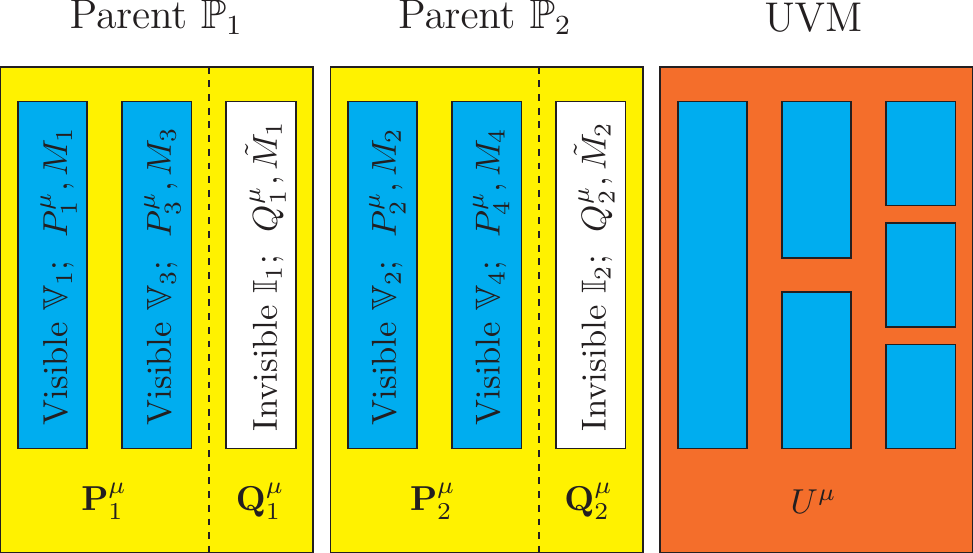}
\end{center}\vspace{-5mm}
\caption[Notation - example 2]{\label{fig:countingParticles2parents} This figure is provided for 
the benefit of readers unable to imagine a simpler version of 
Figure~\ref{fig:countingParticles}. (Readers finding this figure 
helpful need not admit this to close friends, relatives or colleagues.) 
The figure
shows a hypothesis in which four ($\numvis=4$) visible \particles\ 
$\visset = \ourSet{\visible 1, \visible 2, \visible 3, \visible 4}$
and two ($\numinvis=2$) invisible \particles\ $\invisset = \ourSet{\invisible 1, \invisible 2}$ have been assigned to two ($\numparents=2$) parents $\parentset=\ourSet{\parent 1, \parent 2}$ according to the assignments
$\visassign{1}=\ourSet{\visible 1,\visible 3}$,
$\visassign{2}=\ourSet{\visible 2,\visible 4}$,
$\invisassign{1}=\ourSet{\invisible 1}$ and
$\invisassign{2}=\ourSet{\invisible 2}$.
The number of visible \particles\ assigned to each parent in turn are therefore
$\numvisi{1}=2$,
$\numvisi{2}=2$
and the number of invisible \particles\ assigned to each parent in turn are
$\numinvisi{1}=1$,
$\numinvisi{2}=1$.
An explicit physics example corresponding to this 
figure is discussed in Section~\ref{sec:ttbar}.
}
\end{figure}

\begin{figure}[tbh]
\begin{center}
\includegraphics[width=0.5\linewidth]{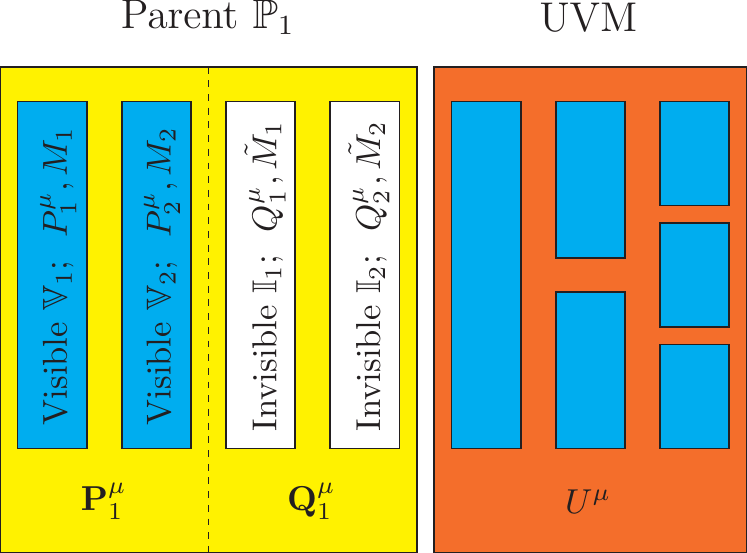}
\end{center}\vspace{-5mm}
\caption[Notation - example 3]{\label{fig:countingParticles1parent} This figure is provided for 
the benefit of readers unable to imagine an {\em even simpler} version of 
Figure~\ref{fig:countingParticles} than was shown in 
Figure~\ref{fig:countingParticles2parents}.  
(Readers finding this figure helpful are advised to seek 
gainful employment in some other field.)
The figure
shows a hypothesis in which two ($\numvis=2$) visible \particles\ 
$\visset = \ourSet{\visible 1, \visible 2}$
and two ($\numinvis=2$) invisible \particles\ $\invisset = \ourSet{\invisible 1, \invisible 2}$ have been assigned to one ($\numparents=1$) parent $\parentset=\ourSet{\parent 1}$ according to the assignments
$\visassign{1}=\ourSet{\visible 1,\visible 2}$ and
$\invisassign{1}=\ourSet{\invisible 1,\invisible 2}$.  For completeness we note $\numvisi{1}=\numinvisi{1}=2$.
An explicit physics example corresponding to this 
figure is discussed in Section~\ref{sec:higgs}.
}
\end{figure}

%

\label{sec:composite}

When considering the decay of a single parent $\parent a$ 
\beq
\parent a \longrightarrow \visassign a \cup \invisassign a.
\label{eq:Pa_decay}
\eeq
it is useful to have notation that can refer to composite 
quantities, e.g.~the total four momentum posessed by the visible 
daughters of $\parent a$, or the total invariant mass of that collection of visible daughters.
Accordingly, as illustrated in \theExampleFigs, 
we denote by $\comp{P}_a^\mu$ the total (1+3)-momentum of 
the visible daughters of parent $\parent a$:
\beq
\comp{P}_a^\mu \equiv 
\ourVec{\vphantom{\tilde{\comp{E}}_a } \comp{E}_a, \vec{\comp{p}}_{aT}, \comp{p}_{az} } 
\equiv \sum_{i\in \visassign{a}} P_i^\mu, 
\label{eq:defPa}
\eeq
or in components
\bea
\vec{\comp{p}}_{aT} &\equiv& \sum_{i\in \visassign{a}} \vec{p}_{iT},
\label{eq:paTdef} \\ [2mm]
\comp{p}_{az} &\equiv& \sum_{i\in \visassign{a}} p_{iz},
\label{eq:pazdef}
\eea
\begin{subequations}
\label{Eadef}
\bea
\comp{E}_a
 &=& \sum_{i\in \visassign{a}}
\sqrt{M_i^2 + \vec{p}_{iT}^{\, 2} + p_{iz}^{\, 2}}
\label{EadefM}
\\ [2mm]
&=& \sum_{i\in \visassign{a}} \frac{|\vec{P}_i|}{V_i}
= \sum_{i\in \visassign{a}} \frac{\sqrt{\vec{p}_{iT}^{\, 2} + p_{iz}^{\, 2}}}{V_i},
\label{EadefV}
\eea
\end{subequations}
where the former (latter) expression for $\comp{E}_a$
will be relevant later on for $\ourT$ ($\ourperp$) transverse 
projections since it is written in a form which depends explicitly 
on the masses (speeds) of the visible particles.

Similarly, we denote
the total hypothesized (1+3)-momentum of the invisible daughters  
of parent $\parent a$  by 
\beq
\comp{Q}_a^\mu \equiv \ourVec{\tilde{\comp{E}}_a, \vec{\comp{q}}_{aT}, \comp{q}_{az} }
\equiv \sum_{i\in \invisassign{a}} Q_i^\mu ,
\label{eq:defQa}
\eeq
or in components
\bea
\vec{\comp{q}}_{aT} &\equiv& \sum_{i\in \invisassign{a}} \vec{q}_{iT},
\label{eq:qaTdef} \\
\comp{q}_{az} &\equiv& \sum_{i\in \invisassign{a}} q_{iz},
\label{eq:qazdef}
\eea
\begin{subequations}
\label{tildeEadef}
\bea
\tilde{\comp{E}}_a
&=& \sum_{i\in \invisassign{a}}
\sqrt{\tilde M_i^2 + \vec{q}_{iT}^{\, 2} + q_{iz}^{\, 2}}\, ,
\label{tildeEadefM}
\\ [2mm]
&=&\sum_{i\in \invisassign{a}} \frac{|\vec{Q}_i|}{\tilde V_i}
= \sum_{i\in \invisassign{a}} \frac{\sqrt{\vec{q}_{iT}^{\, 2} + q_{iz}^{\, 2}}}{\tilde V_i},
\label{tildeEadefV}
\eea
\end{subequations}
where again the former (latter) expression for $\tilde{\comp{E}}_a$ 
will be relevant for $\ourT$ ($\ourperp$) transverse projections 
since it is written in a form which 
depends explicitly on the masses (speeds) of the invisible particles.

As already indicated in eqs.~(\ref{eq:defPa}-\ref{tildeEadef}), we
shall use {\bf bold-face} script to label ``composite'' momenta.
Each parent is thus also treated as a composite particle,
which has (1+3) momentum
\beq
\comp{P}_a^\mu+\comp{Q}_a^\mu
\label{eq:parentmomentum}
\eeq
with ({\it a priori} unknown) (1+3) dim. invariant mass
\beq
{\cal M}_a \equiv  
\sqrt{g_{\mu\nu}\, \left( \comp{P}_a+\comp{Q}_a \right)^\mu
\left( \comp{P}_a+\comp{Q}_a \right)^\nu}.
\label{eq:parentmass}
\eeq
The important distinction between the {\bf bold-face}
notation for composite momenta and the ordinary 
notation for the momenta of individual particles 
is pictorially illustrated in \theExampleFigs.

Note that $\tilde{\comp{E}}_a$ in \eqref{eq:defQa} (whose tilde is necessary to distinguish 
it from the energy $\comp{E}_a$ of the {\em visible} composite daughter of parent $a$) 
might legitimately be termed the missing {\em energy}\footnote{Really 
the missing {\em energy} rather than the missing {\em momentum}!} of the
parent $\parent a$.
We also introduce masses for the 
respective composite daughter objects as follows
\begin{eqnarray}
\comp{M}_a &\equiv& \sqrt{\comp{E}_a^2 - \vec{\comp{p}}_{aT}^2 - \comp{p}_{az}^2 },
\label{Macompvis} \\
\tilde{\comp{M}}_a (Q_i^\mu) &\equiv& 
\sqrt{ \tilde{\comp{E}}_a^2 -  \vec{\comp{q}}_{aT}^{\,2}- \comp{q}_{az}^{\,2}},
\label{Macompinv}
\end{eqnarray}
where again a tilde refers to the invisible object.
Note that the invariant mass (\ref{Macompinv})
of any composite invisible daughter 
is not ``constant'' or a fixed function of measured momenta. 
It depends on the hypothesized invisible momenta $Q_i^\mu $ and so is part of the event hypothesis.   

\subsection{Notation for ``early'' and ``late'' transverse projections}
\label{sec:earlylatenotation}

When forming transverse kinematic variables corresponding to
composite parent or daughter objects, 
one needs to construct the transverse 1+2 dim.~analogues of 
(\ref{eq:defPa}) and (\ref{eq:defQa}).
In doing so, one inevitably has to face the issue 
discussed in Section~\ref{sec:earlylate} ---
whether the agglomeration of individual particles into a 
composite object is done before or after 
projecting into the transverse plane.
As we already saw in Section~\ref{sec:earlylate},
the two outcomes are generally quite different, 
since the composite object is usually massive.
This is why we shall need to develop some additional notation
to help us keep track of the order in which those operations
are performed. Correspondingly, for the remainder of this paper
we shall adopt the following principle: {\em in forming
transverse quantities for composite objects, the order in which
the various operations of agglomeration and projection
are taken will be specified by the order (from left to right)
of the corresponding subscript indices}.

Let us illustrate this principle with a few relevant examples.
The ``late-projected'' (or ``early-partitioned'')
version of the composite visible momentum (\ref{eq:defPa})
is denoted by $\comp{p}_{a\genericT}^\alpha$ 
\beq
\comp{p}_{a\genericT}^\alpha \equiv \ourVec{ \comp{e}_{a\genericT},\vec{\comp{p}}_{a \genericT}}
\label{eq:defpaT}
\eeq
while the alternative ``early-projected'' (or ``late-partitioned'')
version is denoted by $\comp{p}_{\genericT a}^\alpha$:
\beq
\comp{p}_{\genericT a}^\alpha \equiv \ourVec{ \comp{e}_{\genericT a},\vec{\comp{p}}_{\genericT a}}.
\label{eq:defpTa}
\eeq
We remind the reader that the generic index ``$\genericT$''
in (\ref{eq:defpaT}) and (\ref{eq:defpTa})
stands for either ``$\ourT$'', ``$\ourperp$'' or ``$\massless$'', 
as discussed in Section~\ref{sec:projections}. 

We already saw in Section~\ref{sec:earlylate} (eqs.~(\ref{commuteourT}-\ref{commutemassless}))
that the space-like components of (\ref{eq:defpaT}) and (\ref{eq:defpTa})
are equivalent for any choice of ``$\genericT$'':
\beq
\vec{\comp{p}}_{a \genericT} \equiv \vec{\comp{p}}_{\genericT a} = \sum_{i\in \visassign{a}} \vec{p}_{iT},
\label{eq:defpaTpTa}
\eeq
but the time-like components $\comp{e}_{a\genericT}$ and
$\comp{e}_{\genericT a}$ are generally different. 
For example, in the case of $\genericT=\ourT$,
the late-projected (early-partitioned) 
transverse energy $\comp{e}_{a\ourT}$ is given by
\begin{subequations}
\label{eTMNT}
\bea
\hspace{-0.5cm}
\comp{e}_{a\ourT} &=& \sqrt{\comp{M}_a^2 + \vec{\comp{p}}_{aT}^{\, 2}}
= \sqrt{\comp{E}_a^2 - \comp{p}_{az}^{2}}    \label{eTMNT1} \\ [2mm]
&& \hspace{-1.0cm} = \sqrt{\left(\sum_{i\in \visassign{a}} \sqrt{M_i^2 + \vec{p}_{iT}^{\, 2} + p_{iz}^{\, 2}} \right)^2 
- \left(\sum_{i\in \visassign{a}} p_{iz} \right)^2} ,
\label{eTMNT2}
\eea
\end{subequations}
while the early-projected (late-partitioned)
transverse energy $\comp{e}_{\ourT a}$ is given by
\bea
\comp{e}_{\ourT a} &=&
\sum_{i\in \visassign{a}} e_{i\ourT} =
\sum_{i\in \visassign{a}} \sqrt{M_i^2+\vec{p}_{iT}^{\,2}}\,   .
\label{eTMTN} 
\eea
In the case of $\genericT=\ourperp$ projections, the corresponding
transverse energies are given by
\bea
\comp{e}_{a\ourperp} &=& 
\frac{\comp{p}_{aT}}{\sqrt{\comp{p}_{aT}^2+\comp{p}_{az}^2}}\, \comp{E}_a \, ,
\label{epMNperp} \\ [2mm]
\comp{e}_{\ourperp a} &=& \sum_{i\in \visassign{a}} e_{i\ourperp} =
\sum_{i\in \visassign{a}} \frac{p_{iT}}{V_i}\, .
\label{epMTN}
\eea
Finally, for $\genericT=\massless$, the two transverse energies are
\bea
\comp{e}_{a\massless} &=&
\left|\sum_{i\in \visassign{a}} \vec{p}_{iT}\right|,
\label{e0MNT}
\\ [2mm]
\comp{e}_{\massless a} &=&
\sum_{i\in \visassign{a}} p_{iT}.
\label{e0MTN}
\eea

The same conventions apply to the transverse projections of
the composite momentum of a collection of invisible daughter particles:
the ``late-projected'' (or ``early-partitioned'')
version of the composite invisible momentum (\ref{eq:defQa})
is denoted by $\comp{q}_{a\genericT}^\alpha$ 
\beq
\comp{q}_{a\genericT}^\alpha \equiv 
\ourVec{ \tilde{\comp{e}}_{a\genericT},\vec{\comp{q}}_{a \genericT}},
\label{eq:defqaT}
\eeq
while the alternative ``early-projected'' (or ``late-partitioned'')
version is denoted by $\comp{q}_{\genericT a}^\alpha$:
\beq
\comp{q}_{\genericT a}^\alpha \equiv 
\ourVec{ \tilde{\comp{e}}_{\genericT a},\vec{\comp{q}}_{\genericT a}}.
\label{eq:defqTa}
\eeq
Again, the space-like components of (\ref{eq:defqaT}) and (\ref{eq:defqTa})
are the same:
\beq
\vec{\comp{q}}_{a \genericT} \equiv \vec{\comp{q}}_{\genericT a} 
= \sum_{i\in \invisassign{a}} \vec{q}_{iT},
\label{eq:defqaTqTa}
\eeq
but the time-like components are not. Altogether, there are 
6 different possibilities:
\begin{subequations}
\label{tildeeTMNT}
\bea
\tilde{\comp{e}}_{a\ourT}
&=& \sqrt{\tilde{\comp{M}}_a^2 + \vec{\comp{q}}_{aT}^{\, 2}}
= \sqrt{\tilde{\comp{E}}_a^2 - \comp{q}_{az}^{\, 2}} 
\label{tildeeTMNT1}
\\ [2mm]
&& \hspace{-1.0cm} = \sqrt{\left(\sum_{i\in \invisassign{a}} 
\sqrt{\tilde M_i^2 + \vec{q}_{iT}^{\, 2} + q_{iz}^{\, 2}} \right)^2 
- \left(\sum_{i\in \invisassign{a}} q_{iz} \right)^2},~~~~
\label{tildeeTMNT2}
\eea
\end{subequations}
\bea
\tilde{\comp{e}}_{\ourT a} 
&=& \sum_{i\in \invisassign{a}} \tilde{e}_{i\ourT} 
=\sum_{i\in \invisassign{a}} \sqrt{\tilde M_i^2+\vec{q}_{iT}^{\,2}}\, ,
\label{tildeeTMTN}
\eea
\begin{subequations}
\label{tildeepMNperp}
\bea
\tilde{\comp{e}}_{a\ourperp} 
&=& \frac{\comp{q}_{aT}}{\sqrt{\comp{q}_{aT}^2+\comp{q}_{az}^2}}\, 
\tilde{\comp{E}}_a  \\ [2mm]
&=&
\frac{\left|\sum_{i\in \invisassign{a}} \vec{q}_{iT}\right| 
\sum_{i\in \invisassign{a}} \frac{ \sqrt{q_{iT}^2+q_{iz}^2 }}{\tilde V_i}}
{\sqrt{\left(\sum_{i\in \invisassign{a}} \vec{q}_{iT}\right)^2 
     + \left(\sum_{i\in \invisassign{a}} q_{iz}\right)^2 }} ,
\eea
\end{subequations}
\bea
\tilde{\comp{e}}_{\ourperp a} 
&=& \sum_{i\in \invisassign{a}} \tilde{e}_{i\ourperp} 
=\sum_{i\in \invisassign{a}} \frac{q_{iT}}{\tilde V_i}\, ,
\label{tildeepMTN}
\\ [2mm]
\tilde{\comp{e}}_{a\massless} 
&=& \left|\sum_{i\in \invisassign{a}} \vec{q}_{iT}\right|\, ,
\label{tildee0MNT}
\\ [2mm]
\tilde{\comp{e}}_{\massless a} 
&=&
\sum_{i\in \invisassign{a}} q_{iT}.
\label{tildee0MTN}
\eea

In general, our principle of index ordering will extend to
{\em any} transverse invariant mass or transverse energy variable.
For example, in analogy to (\ref{eTMNT}-\ref{e0MTN})
and (\ref{tildeeTMNT}-\ref{tildee0MTN})
there will be six different versions of the transverse masses 
of the composite parent particles and they will be denoted 
by ${\cal M}_{a\genericT}$ or ${\cal M}_{\genericT a}$, with 
$\genericT \in \{\ourT,\ourperp,\massless\}$.

\subsection{Comments on the characterization framework}

Note that we do not place
any a priori restrictions on the values of $N$, $N_{\cal I}$ or
on the way invisible particles are partitioned into the
subsets ${\cal I}_a$. In contrast, many studies on supersymmetry 
(SUSY) or Universal Extra Dimensions (UED) 
in the hadron collider literature are predicated on the following assumptions:
\begin{itemize}
\item $N=2$. This assumption is motivated if 
the new particles are charged under a conserved $Z_2$ parity, 
like $R$-parity in supersymmetry or KK-parity in UED.
However, other discrete symmetries are also possible,
e.g. $Z_3$ \cite{Castano:1994ec,Dreiner:2006xw}
and higher \cite{Dreiner:2005rd,Lee:2007qx},
which could in principle allow for $N>2$.
Even in models with a $Z_2$ parity one could still consider 
the production of any {\em even} number of parents, e.g.
$N=4$, $N=6$, etc.
\item $|{\cal I}_a|=1$ for all $a$.  
In the conventional models with conserved $Z_2$ parity,
this assumption implies that the decay of each 
parent generates one and only one massive invisible 
particle, excluding the possibility of any neutrinos
appearing among the invisible particles. However,
this assumption is not guaranteed ---
even in the conventional SUSY models with conserved 
$R$-parity, SM neutrinos can easily appear
among the decay products of charginos, sleptons, 
$W$-bosons, heavy flavor quarks (especially top), 
taus, etc. Furthermore,
$R$-parity conservation only guarantees that a 
given SUSY parent must decay into an {\em odd} 
(not necessarily 1) number of SUSY particles.
Finally, a $Z_3$ symmetry could allow {\em two}
massive invisible particles per parent, see e.g. \cite{Agashe:2010gt,Agashe:2010tu}.
\end{itemize}
Because of all these caveats, we prefer to keep 
our discussion as general as possible, and 
first define our invariant mass variables below in Sec.~\ref{sec:imass} for any
$\numparents$ and $\numinvis$, before specializing to 
$\numparents=1$ and $\numparents=2$ for illustration purposes only.  

One might ask whether the methods proposed here 
can be usefully applied to events with ``the wrong'' value of $N$.
The answer to this question is ``yes'', and we shall demonstrate this explicitly 
below in Section~\ref{sec:ttbar} (see in particular Figure~\ref{fig:ttb}) where we shall apply
$N=1$ variables in an example where not one, but two parents were produced in the hard scatter.
That study will show that one can sometimes obtain useful information from variables with 
``the wrong'' value of the parent number $N$.

\subsection{Choosing the partitioning}
\label{sec:choosepartition}

In conclusion of this section,
one more comment regarding the partitioning is in order.
One may wonder how one should decide whether a given visible
particle should be counted among the set of visible daughters
or whether it should be included in the ``Upstream visible momentum''
category. The answer to this question depends on the particular 
case at hand. There are simple cases of final states where 
the outgoing particles can be unambiguously associated with the particle 
sets \visassign{a} that match the expected decay products of an assumed parent. 
For example, a high $p_T$, isolated reconstructed lepton
is unlikely to have come from the typical sources of UVM such as
initial state radiation (ISR), multiple parton-parton interactions (MPI), 
multiple hadron-hadron interactions (pileup) etc., and can
probably be safely counted as a visible daughter.
On the other hand, there are also cases (typically involving jets of hadrons)
where the correct partitioning is not obvious at all.
In such cases, one possible approach is 
to consider all possible partitions, see e.g. \cite{Lester:2007fq}.
Another possible approach would be to devise a certain set of cuts,
using the generic differences between the kinematics of ISR jets
and jets from heavy parent decays \cite{Alwall:2009zu,Konar:2010ma,Nojiri:2010mk,Krohn:2011zp}.
Examples of choices for particular physical examples can be found in Section~\ref{sec:literature}.

\section{The mass-bound variables}
\label{sec:imass}

\subsection{Guiding principles}
\label{sec:principles}

The guiding principle we employ for creating useful hadron-collider event variables,
is that: {\em we should place the best possible bounds on any Lorentz invariants 
of interest, such as parent masses or the center-of-mass energy $\ROOTSHAT$, 
in any cases where it is not possible to determine the actual values of those 
Lorentz invariants due to incomplete event information}.  Such incomplete 
information could take the form of lack of knowledge of the longitudinal 
momentum of the primary collision, or lack of knowledge of the 4-momenta 
of individual invisible particles, or lack of knowledge of the number of 
invisible particles which were present, etc.

We contrast this principle with the alternative approach that is used to 
motivate event variables without any explicit regard to whether they have 
an interpretation as an optimal bound of a Lorentz invariant.  
This alternative approach tends to recommend the use of variables 
that are somewhat ad-hoc, but by construction possess useful 
invariances (such as invariance under longitudinal boosts) which are 
designed to remove sensitivity to quantities that are unknown.   
One example of this latter class of variables, which are usually 
considered to be simply ``made up'' without reference to our guiding 
principle, would include the missing transverse momentum $\mptvec$ 
(aready seen in \eqref{eq:theetmissassumption}) obtained by 
adding all transverse visible momenta vectorially. Another 
would be the so called $h_T$ variable\footnote{Note 
that the definition of $h_T$ in the literature is not well 
standardized.  Indeed even one LHC experiment has managed 
to define it in three different and inequivalent ways in 
the space of just a few years, and sometimes even inconsistently 
in a single document (see Section 2 of \cite{Barr:2010zj} 
for further details). The definition we adopt in 
equation~(\ref{eq:htdefbyus}) is the definition
which appears, at present, to be the most widely used in 
the literature.  We note that a conceivable consequence 
of this paper might be that purists will in the future settle 
on a definition in which $h_T$ is defined as a sum of transverse 
energies $e_T$ instead of transverse momenta, whereby three different
variants would be possible: $h_{\ourT}$, $h_{\ourperp}$ and 
$h_{\massless}$ (though these three definitions will be almost
equivalent under most practical experimental conditions, where
the visible particles are approximately massless).} which is 
defined as the {\em scalar} sum of the transverse momenta of 
some class of visible objects (typically jets) in the event: 
\beq
h_T \equiv \sum_{i=1}^{\numvis} p_{iT}.
\label{eq:htdefbyus}
\eeq
Another example is the sum of these two variables:
\beq
\MEFF \equiv h_T + \mpt,
\label{eq:meffdefbyus}
\eeq
a quantity which can be traced back to
the original literature \cite{Hinchliffe:1996iu}
and has become known as an ``effective mass'', 
even though it is not a mass.\footnote{In keeping with our conventions from Section~\ref{sec:vecnotation},
we use lowercase letters for both $h_T$ and $\MEFF$, since they are 
not 1+3 dimensional quantities.}

The main disadvantage of variables like $h_T$ and $\MEFF$, 
is that they do not utilize all the information available;
for example they are completely insensitive to all angles in the transverse plane.
This is why here we would like to construct a more optimal 
class of variables, to wit, those which bound the invariants of interest. 
These too must be invariant under global longitudinal 
boosts since a bound cannot depend on unknown quantities.
However by explict construction we can ensure that 
they also make best use of any available kinematic information. 

\subsection{Construction of mass-bounding variables}
\label{sec:imassdef}

\begin{table*}\begin{center}
\renewcommand\arraystretch{1.5}
\begin{tabular}{| c | c | c | c | c |}
\hline
Type of     & \multicolumn{3}{c|}{Operations } &   \\
\cline{2-4}
variables   & First & Second & Third &  Notation     \\
\hline
Unprojected & Partitioning & Minimization & --- & $M_N$ \\
\hline
\multirow{3}{*} {\begin{minipage}{3cm}{Early partitioned \\ (late projected) \\ $M_{NT}$ }\end{minipage}} 
    & Partitioning & $T=\ourT$ projection & Minimization & $M_{N\ourT}$ \\
    & Partitioning & $T=\ourperp$ projection & Minimization & $M_{N\ourperp}$ \\
    & Partitioning & $T=\massless$ projection & Minimization & $M_{N\massless}$ \\
\hline
\multirow{3}{*} {\begin{minipage}{3cm} Late partitioned \\ (early projected) \\ $M_{TN}$ \end{minipage}}
     & $T=\ourT$ projection & Partitioning & Minimization & $M_{\ourT N}$ \\
     & $T=\ourperp$ projection & Partitioning & Minimization & $M_{\ourperp N}$ \\
     & $T=\massless$ projection & Partitioning & Minimization & $M_{\massless N}$ \\
\hline
\end{tabular}
\caption[Basic mass-bound variables]{\label{tab:floxbridge_basic} 
Method of constructing the mass-bound variables and corresponding notation.
The Table lists the sequence of operations performed in the calculation of 
each variable. ``Partitioning'' refers to the operations discussed
in Sec.~\ref{sec:topology} and \ref{sec:composite} of partitioning 
the final state particles into daughter sets and then adding the momenta 
in each set to form corresponding composite daughter particles. 
``Minimization'' implies minimizing the largest (suitably projected) 
parent mass with respect to (the relevant components of) the
missing momenta of all invisible particles; while the remaining operations 
involve the different types of transverse projections defined and discussed
in Section~\ref{sec:projections}.}
\end{center}
\end{table*}

We are now ready to define the general procedure that can be used to 
construct the mass-bound variables.
In fact, we shall describe a broad class of such variables, 
where each individual variable $M_{\{indices\}}$ will be labelled by 
a certain set of indices $\{indices\}$ indicative of the 
way the particular variable was constructed, namely:
\begin{itemize}
\item Since we are targeting the general event topology of Fig.~\ref{fig:event}, 
where we imagine the inclusive production of $N$ parents, 
each one of our variables will necessarily carry a 
corresponding index $N$. In the process of constructing 
such a variable, we will have to partition (and then agglomerate) the observed
visible particles in the event into $N$ groups ${\cal V}_a$, 
$(a=1,2,\ldots,N)$, as already explained in Section~\ref{sec:topology}.
We will then form the 1+3 dimensional invariant mass of each parent 
$\parent a$
\beq
{\cal M}_{a} \equiv \sqrt{g_{\mu\nu}\, (\comp{P}_{a}^\mu+\comp{Q}_{a}^\mu)(\comp{P}^\nu_{a}+\comp{Q}^\nu_a)},
\label{defMa}
\eeq
which is constructed out of the 1+3 momenta $\comp{P}^\mu_{a}$
and $\comp{Q}^\mu_{a}$ of the respective composite daughter particles
(see Section~\ref{sec:composite}).
\item Optionally, instead of the 1+3 dimensional parent mass (\ref{defMa}),
we may choose to consider the corresponding early-partitioned
(late-projected) transverse mass
\beq
{\cal M}_{a\genericT}\equiv 
\sqrt{g_{\alpha\beta}\,(\comp{p}_{a\genericT}^\alpha+\comp{q}_{a\genericT}^\alpha)
      (\comp{p}^\beta_{a\genericT}+\comp{q}^\beta_{a\genericT})} ,
\label{MaTdef}
\eeq 
or the late-partitioned (early-projected) transverse mass 
\beq
{\cal M}_{\genericT a}\equiv 
\sqrt{g_{\alpha\beta}\,(\comp{p}_{\genericT a}^\alpha+\comp{q}_{\genericT a}^\alpha)
                       (\comp{p}^\beta_{\genericT a}+\comp{q}^\beta_{\genericT a})},
\label{MTadef}
\eeq 
where $\comp{p}^\alpha_{a\genericT}$, $\comp{p}^\alpha_{\genericT a}$,
$\comp{q}^\alpha_{a\genericT}$ and $\comp{q}^\alpha_{\genericT a}$
are the 1+2 dimensional momentum vectors defined in
(\ref{eq:defpaT}), (\ref{eq:defpTa}),
(\ref{eq:defqaT}) and (\ref{eq:defqTa}), correspondingly,
and the index $T$ takes values in $\ourSet{\ourT,\ourperp,\massless}$,
as explained in Section~\ref{sec:projections}.\footnote{We should 
point out that the projection specification $T\in\{\ourT,\ourperp,\massless\}$ refers to 
operations on the {\em visible} particles. One should keep in mind that the visible and 
the invisible composite particles are a priori independent and so could, in principle, 
be treated differently, both in terms 
of the order of the operations, as well as regarding the type of 
transverse projections. For example, consider the $M_{NT}$
class of variables, where one first forms composite visible particles
and transversifies later. In principle, for the
invisible particles, one could perform those 
operations in the opposite order and instead of
(\ref{MaTdef}) consider
$$
\sqrt{g_{\alpha\beta}\,(\comp{p}_{a\genericT}^\alpha+\comp{q}_{\genericT a}^\alpha)
      (\comp{p}^\beta_{a\genericT}+\comp{q}^\beta_{\genericT a})} 
$$
instead. Furthermore, one could choose a
different type of transverse projection 
for the invisibles than for the visible sector, e.g.
$$
\sqrt{g_{\alpha\beta}\,(\comp{p}_{a\ourT}^\alpha+\comp{q}_{a\ourperp}^\alpha)
      (\comp{p}^\beta_{a\ourT}+\comp{q}^\beta_{a\ourperp})} 
$$
and so on. 
One might therefore wonder whether
projected variables need to carry additional indices 
indicating how the {\em invisible} sector is being handled.
In the following, for simplicity we 
shall assume that the invisible particles 
are always projected in exactly the same way as the 
corresponding visible particles, so that the transversification
indices uniquely describe the transverse projections of both 
visible and invisible daughters.
Those readers who are curious about the remaining cases (when the visibles 
and the invisibles are projected differently) can easily
infer the corresponding results from the formulas given below.
}


\item The last step is to consider the {\em largest} hypothesized parent mass
($\max\ourMaxMinBracs{{\cal M}_{a}}$, $\max\ourMaxMinBracs{{\cal M}_{aT}}$ 
or $\max\ourMaxMinBracs{{\cal M}_{Ta}}$
as appropriate) and {\em minimize} it over all possible values of the 
unknown invisible momenta consistent with the constraints. This minimization 
is always a well-defined, unambiguous operation, 
which yields a unique numerical answer \cite{Konar:2008ei},
which we shall denote as
\bea
M_N &\equiv& \min_{\substack{
\sum \vec{q}_{iT} = \mptvec}} 
\ourMaxMinBracs{\max_a\ourMaxMinBracs{{\cal M}_{a}} },
\\ [2mm]
M_{N\genericT} &\equiv& \min_{\substack{
\sum \vec{q}_{iT} = \mptvec}} 
\ourMaxMinBracs{\max_a\ourMaxMinBracs{{\cal M}_{a\genericT}}},
\\ [2mm]
M_{\genericT N} &\equiv& \min_{\substack{
\sum \vec{q}_{iT} = \mptvec}} 
\ourMaxMinBracs{\max_a\ourMaxMinBracs{{\cal M}_{\genericT a}} },
\eea
as indicated in Table~\ref{tab:floxbridge_basic}.
The minimization over the unknown parameter is performed in order to guarentee that 
the resultant variable cannot be larger than the mass of the heaviest parent,
resulting in an event-by-event lower bound on the mass of the heaviest parent.
\end{itemize}

These are the basic steps, leading to the variables displayed 
in Table~\ref{tab:floxbridge_basic}. This basic set of variables 
will be further extended in Section~\ref{sec:doubleT} below, 
by considering a second level of projections {\em within} the transverse plane.
For the remainder of this section, however, we shall stick to the basic procedures above
and focus on the simplest classes of variables displayed in Table~\ref{tab:floxbridge_basic},
namely the ``unprojected'' $M_N$ and the ``singly projected'' $M_{NT}$ and $M_{TN}$ variables.

\subsection{The variables: $M_N$, $M_{NT}$ and $M_{TN}$}
\label{sec:TMNT}


%


In this subsection we provide analytic formulas
(where available) for calculating each of the basic mass-bound variables
from Table~\ref{tab:floxbridge_basic} on an event-by-event basis.

\subsubsection{The usual (``unprojected'') invariant mass: $M_\numparents$}
\label{MN}

Here we work directly with the usual (1+3)-dimensional
invariant masses ${\cal M}_a$ of the parent particles $\parent a$: 
\begin{subequations}
\label{eq:Madef}
\begin{eqnarray}
&& \hspace{-0.5cm}  {\cal M}_a^2(\comp{P}_a,\comp{Q}_{a},\invismassseti a)
 \equiv (\comp{P}_a+\comp{Q}_a)^2   \\ [2mm]
&& \hspace{-0.3cm}= \left(\comp{E}_a    + \tilde{\comp{E}}_a\right)^2
 -  \left(\vec{\comp{p}}_{aT} + \vec{\comp{q}}_{aT}\right)^2
 -  \left(\comp{p}_{az} + \comp{q}_{az}\right)^2.  ~~~~
\end{eqnarray}
\end{subequations}

The unprojected invariant mass variable $M_\numparents$ is defined by the right hand side of
\beq
M_\numparents(\chiM)\equiv
\min_{
\substack{
\sum \vec{q}_{iT} = \mptvec
}
}
\ourMaxMinBracs{ \max_a\ourMaxMinBracs{{\cal M}_{a}(\comp{P}_{a},\comp{Q}_{a},\invismassseti a)} },
\label{defMN}
\eeq
where the minimization needs to be performed over 
$3N_{\cal I}$ degrees of freedom ($\vec{q}_{iT}$ and $q_{iz}$ 
for $i=1,2,\ldots,\numinvis$), subject to the two scalar constraints
\eqref{eq:theetmissassumption} supplied by transverse momentum conservation.  
The invisible particle momenta $\vec{q}_{iT}$ and $q_{iz}$
are fixed by the minimization and $M_\numparents$ does not depend on 
them.

Note that we have emphasized in the left hand side of \eqref{defMN} 
that $M_\numparents$ turns out {\em not} to be
a function of the $\numinvis$ individual invisible mass hypotheses 
$\tilde M_i$ in $\invismassset = \bigcup_a \invismassseti a $,  
but instead turns out (see proof in Section~\ref{sec:proofoffloxdeponhypsets}) 
to be a function of the set 
\bea
\chiM=\ourSet{{\mmass_a}\mid{a\in\parentset}}.\label{eq:chimdefined}
\eea 
containing the $\numparents$ ``invisible mass-sum parameters, $\mmass_a$'' defined by
\beq
\mmass_a\equiv \sum_{i\in \invisassign{a}} \tilde M_i.
\label{Mslashadef}
\eeq  
These mass parameters are simple arithmetic sums 
of the hypothesized masses of the individual invisible particles 
associated with any given parent $\parent a$.  


Notice the simplification 
in going from the individual parent masses ${\cal M}_a$ to the 
variable $M_N$. The individual parent masses ${\cal M}_a$ collectively
depend on {\em all} invisible particle masses
$\tilde M_i$, (a total of $N_{\cal I}$ parameters),
while the invariant mass variable $M_N$ defined in 
(\ref{defMN}) only depends on 
the $N$ summed-invisible-mass parameters $\mmass_a$, $(a=1,2,\ldots,N)$,
which are simply related to the
individual particle masses $\tilde M_i$ via (\ref{Mslashadef}).
In the most common cases of $N=1$ or $N=2$,
we will therefore have to deal with only one or two unknown invisible mass-sum parameters.
A similar reduction in complexity will be found when we consider 
the $\ourperp$ projected variables, but there the mass bound will 
end up depending on a speed-related parameter for each parent. 
We see that from now on the index $N$ 
can be interpreted not only as the number of parents, but also
as the number of relevant independent mass inputs characterizing
the invisible sector.

The preceding discussion is best illustrated with a specific example.
Let us consider the simplest case of $N=1$.
The minimization of the corresponding variable $M_1$ 
with respect to $\vec{q}_{iT}$ and $q_{iz}$
is straightforward. One finds that the minimum 
is located at \cite{Konar:2008ei}
\begin{eqnarray}
\vec{q}_{iT} &=& \mptvec \frac{\tilde M_i}{\mmass_{1}} ,      \label{eq:qiT}    \\
q_{iz} &=& \comp{p}_{1z}\, \frac{\tilde M_i}{\mmass_{1}}\,
\sqrt{\frac{\mmass_{1}^2+\mpt^2}{\comp{M}_1^2+\comp{p}_{1T}^2}}\, ,
\label{eq:qiz}   
\end{eqnarray}
and its value (see \cite{Konar:2010ma}) is given by 
\beq
M_1^2(\mmass_1)\equiv \left( \sqrt{\comp{M}_1^2+\comp{p}_{1T}^2} 
+\sqrt{\mmass_{1}^2+\mpt^2}\right)^2-u_T^2
\label{eq:M1formula}
\eeq
in which, to save space, we have slightly abused our notation by 
writing $M_1^2(\mmass_1)$ in place of $M_1^2(\ourSet{\mmass_1})$ --- a convention 
we will adopt throughout this document wherever $\numparents=1$.  
We remind the reader that $\comp{M}_1$ is the measured (1+3)-mass of the (single) visible 
composite daughter (see also eq.~(\ref{Macompvis}))
\beq
\comp{M}_1 \equiv \sqrt{\comp{E}_1^2 - \vec{\comp{p}}_{1T}^2 - \comp{p}_{1z}^2 },
\eeq
while $\mmass_1$ is the only invisible mass
parameter needed\footnote{Note the analogy between $\mptvec$ and $\mmass_1$.
$\mptvec$ measures the {\em total} transverse momentum of 
the whole collection of missing particles. 
Similarly, $\mmass_1$ measures the {\em total} mass of the 
whole collection of missing particles. 
Both $\mptvec$ and $\mmass_1$ are given by simple 
sums of the corresponding quantities $\vec{q}_{iT}$
and $\tilde M_i$ of the individual missing particles, 
compare (\ref{eq:theetmissassumption}) and (\ref{eq:mismass1def}).}
defined in (\ref{Mslashadef})
\beq
\mmass_1 \equiv \sum_{i=1}^{\numinvis} \tilde M_i.
\label{eq:mismass1def}
\eeq
In Ref.~\cite{Konar:2010ma}, the quantity $M_1(\mmass_1)$
defined in (\ref{eq:M1formula}) was labelled $\sqrt{\hat{s}}_{\rm min}^{(\rm sub)}$:
\beq
M_1(\mmass_1) \equiv \sqrt{\hat{s}}_{\rm min}^{(\rm sub)}(\mmass_1),
\label{M1eqsminsub}
\eeq
since it provides a lower bound on the parton-level 
center-of-mass energy of the parent subsystem 
${\cal V}_1\oplus {\cal I}_1$, not counting
the uninteresting upstream visible momentum $U^\mu$.
In the special case of a vanishing upstream momentum ($u_T=0$), 
$M_1(\mmass_1)$ reduces to the global variable
$\sqrt{\hat{s}}_{\rm min}$ from \cite{Konar:2008ei}:
\beq
\lim_{u_T\to 0} M_1(\mmass_1) = \sqrt{\hat{s}}_{\rm min}(\mmass_1).
\eeq

We will not consider the next simplest example ($M_2$) until 
Section~\ref{sec:mt2}, as simple analytic (as opposed to numerical 
or iterative) formulae for it are only known to exist 
in certain special cases \cite{Lester:2011nj}, 
such as when $\mmass_1 = \mmass_2 = {\bf M}_1 = {\bf M}_2 =0$, 
or when the upstream visible momentum $\vec{u}_T$ is either 
zero or (anti-)parallel to the missing transverse momentum $\mptvec$.

\subsubsection{The early partitioned,  $\ourT$-projected invariant mass: $M_{\numparents\ourT}$}
\label{sec:MNT}

Here the momenta $\comp{P}_a^\mu$ and $\comp{Q}_a^\mu$ of the composite 
particles are first formed in 1+3 dimensions, as in (\ref{eq:defPa})
and (\ref{eq:defQa}), then {\em afterwards} are projected on the 
transverse plane according to the mass-preserving $\ourT$ method 
defined in eq.~(\ref{defpT}) of Sec.~\ref{sec:TBGformalism}.
This results in {\em transverse} masses of the parents given by
\begin{subequations}
\label{eq:MaourTdef} 
\begin{eqnarray}
{\cal M}^2_{a\ourT}(\comp{p}^\alpha_{a\ourT},\comp{q}^\alpha_{a\ourT}, \invismassseti a) 
&\equiv& 
\left( \comp{p}_{a\ourT}+ \comp{q}_{a\ourT} \right)^2
\\ [2mm]
&& \hspace{-2.5cm}\equiv (\comp{e}_{a\ourT}+\tilde{\comp{e}}_{a\ourT})^2 
-(\vec{\comp{p}}_{aT}+\vec{\comp{q}}_{aT})^2,   
\end{eqnarray}
\end{subequations}
where the transverse momenta $\vec{\comp{p}}_{aT}$ and $\vec{\comp{q}}_{aT}$ are given by
(\ref{eq:paTdef}) and (\ref{eq:qaTdef}), while the transverse
energies $\comp{e}_{a\ourT}$ and $\tilde{\comp{e}}_{a\ourT}$ 
are given by (\ref{eTMNT}) and (\ref{tildeeTMNT}).

Then the ``early partitioned, $\ourT$-projected'' variable
$M_{N\ourT}$ is defined in a manner very similar to
(\ref{defMN})
\beq
M_{\numparents\ourT}(\chiM)\equiv 
\min_{\substack{
\sum \vec{q}_{iT} = \mptvec}} 
\ourMaxMinBracs{\max_a \ourMaxMinBracs{{\cal M}_{a\ourT}
(\comp{p}_{a\ourT}^\alpha,\comp{q}_{a\ourT}^\alpha,\invismassseti a)
} }\ .
\label{defMNourT}
\eeq
Just like $M_{\numparents}$, this variable also
depends only\footnote{At this point 
readers who are familiar with the Cambridge $m_{T2}$
variable \cite{Lester:1999tx,Barr:2003rg} have probably recognized 
that for the special case of $N=2$, the $M_{\numparents \ourT}$
variable (\ref{defMNourT}) recovers the Cambridge $m_{T2}$.
Note that the original literature \cite{Barr:2003rg}
on the Cambridge $m_{T2}$ variable also defined more 
general variables $m_{TX}$, e.g. $m_{T3}$, $m_{T4}$, etc.
However, we caution readers to make the distinction between
the index ``$\numparents$'' in $M_{\numparents \ourT}$,
which refers to the number of hypothesized {\em parents},
and the index ``$X$'' in the Cambridge $m_{TX}$,
which stood for the total number of {\em invisible particles} 
(in this paper denoted by $N_{\cal I}$).
For example, the index ``2'' in the Cambridge 
$m_{T2}$ notation implies the presence of exactly two
{\em invisible} particles, the number of parents 
already being implicitly assumed to be two.
In contrast, the variable $M_{2\ourT}$
defined in (\ref{defMNourT}) does not imply any 
particular number of invisible particles, and in this sense
is equivalent to the whole class of $m_{TX}$ for any $X$.
}
on the $\numparents$ summed-invisible-mass parameters 
$\mmass_a$ within $\chiM$ as opposed to the $N_{\cal I}$ 
individual invisible masses $\tilde M_i$ within $\invismassset$. 
Eq.~(\ref{defMNourT}) again represents a constrained 
minimization problem for the $3N_{\cal I}$ variables
$\vec{q}_{iT}$ and $q_{iz}$. Note that in spite of its 
``transverse'' index, $M_{\numparents\ourT}$ still depends on the
longitudinal momenta $q_{iz}$ through the transverse energy
$\tilde{\comp{e}}_{a\ourT}$, see (\ref{tildeeTMNT2}).

In order to gain some intuition, let us again consider the 
simplest case of $\numparents=1$. The minimization of (\ref{defMNourT})
is once again straightforward and the minimum is found at
\begin{eqnarray}
\vec{q}_{iT} &=& \mptvec\, \frac{\tilde M_i}{\mmass_{1}} , \label{eq:qiTmin}         \\
q_{iz} &=& \comp{q}_{1z}\, \frac{\tilde M_i}{\mmass_{1}} ,
\end{eqnarray}
with an arbitrary choice of $\comp{q}_{1z}$. This leads to
\beq
M_{1\ourT}^2(\mmass_1)\equiv \left( \sqrt{\comp{M}_1^2+\comp{p}_{1T}^2} 
+\sqrt{\mmass_{1}^2+\mpt^2}\right)^2-u_T^2\, .
\label{eq:M1ourTformula}
\eeq
Comparing (\ref{eq:M1ourTformula}) to (\ref{eq:M1formula}), we see that
\beq
M_{1\ourT} = M_{1}.
\label{M1eqM1T}
\eeq
This is in fact a special case of the more general mathematical 
identity
\beq
M_{\numparents\ourT} = M_{\numparents},
\label{MNeqMNT}
\eeq 
for which a proof is provided in the appendix --- see equation~(\ref{eq:proofofMNtisMNT}).
This identity reveals that ``transverse'' quantities
do not necessarily ``forget'' about relative longitudinal 
momenta. In particular, (\ref{MNeqMNT}) teaches us that
whenever the composite particles are formed {\em before}
the transverse projection, the information about 
the relative longitudinal momenta is retained, 
and the result is the same as if everything 
was done in 1+3 dimensions throughout. As a result,
$M_{\numparents\ourT}$ automatically inherits all the advantages 
and disadvantages of its 1+3 cousin $M_{\numparents}$.

\subsubsection{The late partitioned, $\ourT$-projected invariant mass: $M_{\ourT\numparents}$}
\label{sec:MTN}

This is the first example of an ``early projected'', ``late partitioned'' variable. 
We follow the procedure of the previous subsection \ref{sec:MNT},
only this time we switch the order of the operations, and we first
$\ourT$-project the momentum of each individual particle on the 
transverse plane, before forming composite particles.
The transverse invariant mass of each composite parent 
is then given by
\begin{subequations}
\label{eq:MaourTdef2}
\begin{eqnarray}
{\cal M}^2_{\ourT a}(\comp{p}^\alpha_{\ourT a},\comp{q}^\alpha_{\ourT a},\invismassseti a) 
&\equiv& 
\left( \comp{p}_{\ourT a}+ \comp{q}_{\ourT a} \right)^2
\\ [2mm]
&& \hspace{-2cm}\equiv (\comp{e}_{\ourT a}+\tilde{\comp{e}}_{\ourT a})^2 
-(\vec{\comp{p}}_{aT}+\vec{\comp{q}}_{aT})^2,  ~~~ 
\end{eqnarray}
\end{subequations}
with $\vec{\comp{p}}_{aT}$ and $\vec{\comp{q}}_{aT}$ still given by
(\ref{eq:paTdef}) and (\ref{eq:qaTdef}), while
the composite transverse energies 
$\comp{e}_{\ourT a}$ and $\tilde{\comp{e}}_{\ourT a}$
are given by (\ref{eTMTN}) and (\ref{tildeeTMTN}), correspondingly.
Notice that these expressions do not contain the longitudinal momenta
$p_{iz}$ and $q_{iz}$. This is in contrast to the ``early partitioned''
case represented by (\ref{eTMNT}) and (\ref{tildeeTMNT}), where
the longitudinal momenta appear explicitly. The comparison between
(\ref{eTMNT}) and (\ref{tildeeTMNT}) on the one hand, and
(\ref{eTMTN}) and (\ref{tildeeTMTN}) on the other, nicely
illustrates the main point of Section~\ref{sec:earlylate} ---
that by adding the momenta {\em before} the projection,
one retains sensitivity to the relative longitudinal momenta.
Conversely, when the operations are performed in reverse order
and the transverse projection is done first, the
longitudinal momenta completely drop out of the game.

Now we are ready to apply the usual definition and obtain
\beq
M_{\ourT\numparents}(\chiM)\equiv 
\min_{\substack{
\sum \vec{q}_{iT} = \mptvec}} 
\ourMaxMinBracs{\max_a \ourMaxMinBracs{{\cal M}_{\ourT a}
(\comp{p}_{\ourT a}^\alpha,\comp{q}_{\ourT a}^\alpha,\invismassseti a)
} }.
\label{defMourTN}
\eeq

Let us again investigate the simplest case of $\numparents=1$.
With the help of the transverse momentum conservation constraint 
(\ref{eq:theetmissassumption}), eq.~(\ref{defMourTN}) reduces to
\beas
M_{\ourT1}^2
&=& \min_{\substack{
\sum \vec{q}_{iT} = \mptvec}}  
\ourMaxMinBracs{ {\left( \sum_{i=1}^{\numvis}{
e_{i\ourT}}+\sum_{i=1}^{\numinvis}{\tilde e_{i\ourT}} \right)}^2  - u_T^{2} } \\
&= & {\left(   \sum_{i=1}^{\numvis} {
e_{i\ourT}}+\min_{\substack{
\sum \vec{q}_{iT} = \mptvec}} 
\ourMaxMinBracs{ \sum_{i=1}^{\numinvis} {\tilde e_{i\ourT}}} \right)}^2  - u_T^{2}. 
\eeas
The minimum is once again found at (\ref{eq:qiTmin})
and we get
\beq
M_{\ourT1}^2(\mmass_1) = 
\left( \sum_{i=1}^{\numvis} \sqrt{M_i^2+\vec{p}_{iT}^{\,2}}
+\sqrt{\mmass_1^2 + \mpt^2 } \right)^2
 -  u_T^{2}.
\label{eq:MourT1formula}
\eeq
As expected, this result differs from (\ref{eq:M1ourTformula}),
although the two formulas follow a similar pattern.
The difference is only in the term corresponding to the 
visible sector, where the transverse energy of the 
composite visible particle is computed differently,
compare (\ref{eTMNT1}) and (\ref{eTMTN}).

An interesting result emerges if we consider the 
further simplification that all visible particles are
massless, i.e. $M_i=0, \forall\, i$. This, in fact, 
is a very good approximation for the leptons and 
quarks/gluons of the SM, whose masses can be safely 
neglected. Setting $M_i=0$ in (\ref{eq:MourT1formula}) 
and using (\ref{eq:htdefbyus}), we get
\beq
\lim_{M_i\to 0}M_{\ourT1}^2(\mmass_1) = \left( h_T
+\sqrt{\mmass_1^2 + \mpt^2 } \right)^2
 -  u_T^{2}.
\label{eq:MT1eqht}
\eeq
This result is quite interesting. It allows us to reinterpret 
the usual $h_T$ variable in terms of a bona fide invariant mass
variable like $M_{\ourT1}$, properly accounting 
for the effects of upstream visible momentum
$u_T$ {\em and} the total mass $\mmass_1$ of the invisible 
particles present in the event. We shall return to this 
point in the next Section~\ref{sec:math}.

Another interesting result follows from eq.~(\ref{eq:MT1eqht})
in the special case when we set $\mmass_1=0$. 
Using (\ref{eq:meffdefbyus}), we get
\beq
\lim_{M_i\to 0}M_{\ourT1}^2(\mmass_1=0) = 
\left( h_T + \mpt \right)^2
 -  u_T^{2}
= \MEFF^2 -  u_T^{2},
\label{eq:MT1eqmeff}
\eeq
providing a connection between the ``effective mass''
$\MEFF$ and $M_{\ourT1}(0)$.

\subsubsection{The late partitioned, $\ourperp$-projected mass: $M_{\ourperp \numparents}$}
\label{sec:MperpN}

This is the second example of an ``early projected'' variable, only
this time we use the speed-preserving $\ourperp$ projection 
described in Section~\ref{sec:perpmalism}. Correspondingly, 
the individual visible (invisible) particles will be characterized 
by their 3-speeds $V_i$ ($\tilde V_i$) instead of their masses
$M_i$ ($\tilde M_i$) and so we remind the reader of the notation 
introduced in \eqref{eq:deftildeveltot} and \eqref{eq:deftildevela}.

The 1+2 momentum vectors of the individual particles after the 
$\ourperp$ projection are obtained from (\ref{defpperp})
\bea
p_{i\ourperp}^\alpha &\equiv& \ourVec{ e_{i\ourperp},\vec{p}_{i\ourperp}} = 
\ourVec{ \frac{p_{iT}}{V_i}, \vec{p}_{iT}}, \\
q_{i\ourperp}^\alpha &\equiv& \ourVec{ \tilde e_{i\ourperp},\vec{q}_{i\ourperp}} = 
\ourVec{ \frac{q_{iT}}{\tilde V_i}, \vec{q}_{iT}}.
\eea
Then we form composite particles with $\ourperp$ projected 1+2 momenta
$\comp{p}_{\ourperp a}^\alpha$ and $\comp{q}_{\ourperp a}^\alpha$ 
given by (\ref{eq:defpTa}) and (\ref{eq:defqTa}), respectively.

The transverse parent masses are now formed in terms of 
$\comp{p}_{\ourperp a}^\alpha$ and $\comp{q}_{\ourperp a}^\alpha$
as follows
\begin{subequations}
\label{eq:Mperpadef}
\begin{eqnarray}
{\cal M}^2_{\ourperp a}(\comp{p}^\alpha_{\ourperp a},\comp{q}^\alpha_{\ourperp a},\invisvelseti a) 
&\equiv& 
\left( \comp{p}_{\ourperp a}+ \comp{q}_{\ourperp a} \right)^2
\label{eq:Mperpadef1} \\ [2mm]
&& \hspace{-2cm}\equiv (\comp{e}_{\ourperp a}+\tilde{\comp{e}}_{\ourperp a})^2 
-(\vec{\comp{p}}_{aT}+\vec{\comp{q}}_{aT})^2,   
\label{eq:Mperpadef2}
\end{eqnarray}
\end{subequations}
where the transverse energies $\comp{e}_{\ourperp a}$ 
and $\tilde{\comp{e}}_{\ourperp a}$ are
specified by (\ref{epMTN}) and (\ref{tildeepMTN})
and the transverse momenta $\vec{\comp{p}}_{aT}$
and $\vec{\comp{q}}_{aT}$ are given by (\ref{eq:paTdef}) and (\ref{eq:qaTdef}).

This is a convenient place to introduce another two small pieces of 
notation.\footnote{Contrast with the definition of $\mmass_a$ 
in equation~(\ref{Mslashadef}) and the definition of $\chiM$ 
in equation~(\ref{eq:chimdefined}).}  Firstly we will need to 
define a ``maximum invisible velocity parameter'' $\mv_a$ for 
each parent $\parent a$ according to 
\beq
\mv_a \equiv \max_{i\in \invisassign{a}}  \ourMaxMinBracs{ \tilde V_i }.
\label{eq:chivadefined}
\eeq
Then we would like to denote by  $\chiV$ the set of all 
the above velocity parameters, i.e.
\bea
\chiV=\ourSet{{\mv_a}\mid{a\in\parentset}}\label{eq:chivdefined}.
\eea

Now we are in a position to state (see proof in Section~\ref{sec:proofoffloxdeponhypsets}) 
that the only dependence of the ``late partitioned'', 
$\ourperp$-projected mass variable $M_{\ourperp \numparents}$ 
on the velocity parameters of the invisible particles is through $\chiV$, i.e.: 
\beq
M_{\ourperp \numparents}(\chiV)\equiv 
\min_{\substack{
\sum \vec{q}_{iT} = \mptvec}} 
\ourMaxMinBracs{\max_a \ourMaxMinBracs{{\cal M}_{\ourperp a}
(\comp{p}_{\ourperp a}^\alpha,\comp{q}_{\ourperp a}^\alpha,\invisvelseti a)
}}.
\label{defMourperpN}
\eeq

Once again, it is instructive to consider the special case of $N=1$.
With the help of (\ref{eq:theetmissassumption}), eq.~(\ref{defMourperpN}) becomes 
\beas
M_{\ourperp 1}^2 (\mv_1)
&=& \min_{\sum \vec{q}_{iT} = \mptvec}
 \ourMaxMinBracs{ {\left( 
\sum_{i=1}^{\numvis}
e_{i\ourperp}+\sum_{i=1}^{\numinvis} \tilde e_{i\ourperp} \right)}^2  
- u_T^2 } \\ [2mm]
 &= & {\left(   \sum_{i=1}^{\numvis} {e_{i\ourperp}}
+\min_{\sum \vec{q}_{iT} = \mptvec}
 \ourMaxMinBracs{ \sum_{i=1}^{\numinvis} {\tilde e_{i\ourperp}}} 
\right)}^2  - u_T^2  \\ [2mm]
 &= & {\left(   \sum_{i=1}^{\numvis} {e_{i\ourperp}}
+\min_{\sum \vec{q}_{iT} = \mptvec}
 \ourMaxMinBracs{ \sum_{i=1}^{\numinvis} \frac{q_{iT}}{\tilde V_i}} 
\right)}^2  - u_T^2.  
\eeas
The minimization selects the invisible particle with 
the largest speed, whose transverse momentum becomes $\mptvec$,
while all other invisible particles have $q_{iT}=0$.
This configuration leads to the final answer
\beq
M_{\ourperp 1}^2 (\mv_1)
 = {\left(   \sum_{i=1}^{\numvis} \frac{p_{iT}}{V_i}
+\frac{\mpt}{\mv_1} \right)}^2  - u_T^2.
\label{Mperp1formula}
\eeq

When we make the approximation that all visible particles 
are massless ($V_i=1$), we again obtain a relation to $h_T$:
\beq
\lim_{V_i\to 1}M_{\ourperp 1}^2 (\mv_1)
= \left( h_{T} +\frac{\mpt}{\mv_1} \right)^2  - u_T^2\, ,
\label{eq:Mperp1eqht}
\eeq
which is the analogue of (\ref{eq:MT1eqht}) for the
case of $\ourperp$ transverse projections. But note that
unlike (\ref{eq:MT1eqht}), here the 
unknown parameter characterizing the invisible sector is the
maximum {\em speed} parameter $\mv_1$ 
instead of the summed-invisible-mass parameter $\mmass_1$.

Finally, if in addition we also assume that all invisible 
particles are massless as well, then 
$$\tilde V_i=1, \, \forall i\quad \Longrightarrow\quad \mv_1=1,$$ 
so that
\beq
\lim_{V_i\to 1}M_{\ourperp 1}^2 (\mv_1=1)
= \left( h_{T} + \mpt \right)^2  - u_T^2
= \MEFF^2 - u_T^2\, ,
\label{eq:Mperp1eqmeff}
\eeq
which is the analogue of (\ref{eq:MT1eqmeff}).
The fact that (\ref{eq:MT1eqmeff}) and
(\ref{eq:Mperp1eqmeff}) are the same should not come as a surprise:
recall from Sec.~\ref{sec:masslesscomments} that the two 
transverse projections $\ourT$ and $\ourperp$ are equivalent 
in the massless limit.

\subsubsection{The early partitioned, $\ourperp$-projected mass: $M_{\numparents\ourperp}$}
\label{sec:MNperp}

Here we follow a procedure analogous to that of Sec.~\ref{sec:MNT},
where the composite momenta $\comp{P}_a^\mu$ and $\comp{Q}_a^\mu$
are first formed in 1+3 dimensions, before being projected on 
the transverse plane, only this time we use the $\ourperp$ 
projection for this purpose:
\bea
\sum_{i\in \visassign{a}} P_i^\mu 
&\longrightarrow& \comp{P}_a^\mu 
\ \overset{\ourperp}{\longrightarrow} \ 
\comp{p}_{a\ourperp}^\alpha = \ourVec{\comp{e}_{a\ourperp}, \vec{\comp{p}}_{a\ourperp}}, \\ [2mm]
\sum_{i\in \invisassign{a}} Q_i^\mu 
&\longrightarrow& \comp{Q}_a^\mu 
\ \overset{\ourperp}{\longrightarrow} \ 
\comp{q}_{a\ourperp}^\alpha = \ourVec{\tilde{\comp{e}}_{a\ourperp}, \vec{\comp{q}}_{a\ourperp}}.
\eea

The transverse parent masses are now formed in terms of 
$\comp{p}_{a\ourperp}^\alpha$ and $\comp{q}_{a\ourperp}^\alpha$
as usual
\begin{subequations}
\label{eq:Maperpdef}
\begin{eqnarray}
{\cal M}^2_{a\ourperp}(\comp{p}^\alpha_{a\ourperp},\comp{q}^\alpha_{a\ourperp},\invisvelseti a) 
&\equiv& 
\left( \comp{p}_{a\ourperp}+ \comp{q}_{a\ourperp} \right)^2
\label{eq:Maperpdef1} \\ [2mm]
&& \hspace{-1.5cm}\equiv (\comp{e}_{a\ourperp}+\tilde{\comp{e}}_{a\ourperp})^2 
-(\vec{\comp{p}}_{aT}+\vec{\comp{q}}_{aT})^2.  
\label{eq:Maperpdef2}
\end{eqnarray}
\end{subequations}
Here the composite transverse momenta $\vec{\comp{p}}_{a\ourperp}$
and $\vec{\comp{q}}_{a\ourperp}$ are still given by 
(\ref{eq:paTdef}) and (\ref{eq:qaTdef}), while the transverse energies 
$\comp{e}_{a\ourperp}$ and $\tilde{\comp{e}}_{a\ourperp}$
are given by (\ref{epMNperp}) and (\ref{tildeepMNperp}), correspondingly.

Then the early-partitioned, $\ourperp$-projected variable is defined as usual:
\beq
M_{\numparents \ourperp}(\chiV)\equiv 
\min_{\substack{
\sum \vec{q}_{iT} = \mptvec}} 
\ourMaxMinBracs{\max_a \ourMaxMinBracs{{\cal M}_{a\ourperp}
(\comp{p}_{a\ourperp}^\alpha,\comp{q}_{a\ourperp}^\alpha,\invisvelseti a)
}}\ .
\label{defMNourperp}
\eeq

Once again, let us specify this to the case of $\numparents=1$.
Using (\ref{eq:theetmissassumption}), we get
\begin{subequations}
\bea
M_{1\ourperp} &=& \min_{\sum \vec{q}_{iT} = \mptvec}
\ourMaxMinBracs{
\left(\comp{e}_{1\ourperp}+\tilde{\comp{e}}_{1\ourperp}\right)^2-u_T^2
}
\\ [2mm]
&=& 
\left(\comp{e}_{1\ourperp}
+\min_{\sum \vec{q}_{iT} = \mptvec}\ourMaxMinBracs{\tilde{\comp{e}}_{1\ourperp}}\right)^2-u_T^2.
\eea
\end{subequations}
The minimization is performed over the $3N_{\cal I}$ variables
$\vec{q}_{iT}$ and $q_{iz}$, $i=1,2,\ldots,N_{\cal I}$ and the result is
\begin{subequations}
\label{eq:M1perpformula}
\bea
\hspace{-0.5cm}M_{1\ourperp} (\mv_1) &=& \left(\comp{e}_{1\ourperp}
+\frac{\mpt}{\mv_1}\right)^2-u_T^2
\label{eq:M1perpformula1} \\ [2mm]
&=&\left(\frac{\comp{p}_{1T}}{\sqrt{\comp{p}_{1T}^2+\comp{p}_{1z}^2}}\, \comp{E}_1 
+\frac{\mpt}{\mv_1}\right)^2-u_T^2,
\label{eq:M1perpformula2}
\eea
\end{subequations}
which is similar, but not equivalent to (\ref{Mperp1formula}).

\subsubsection{The late partitioned, $\massless$-projected mass: $M_{\massless\numparents}$}
\label{sec:M0N}

Here we follow the procedure of Secs.~\ref{sec:MTN} and \ref{sec:MperpN},
only this time we use the $\massless$ transverse projection
from Sec.~\ref{sec:masslessism}. One first forms the 1+2 momenta of the
individual particles
\bea
p_{i\massless}^\alpha &\equiv& \ourVec{ e_{i\massless},\vec{p}_{i\massless}} = 
\ourVec{ p_{iT}, \vec{p}_{iT}}, \\ [2mm]
q_{i\massless}^\alpha &\equiv& \ourVec{ \tilde e_{i\massless},\vec{q}_{i\massless}} = 
\ourVec{ q_{iT}, \vec{q}_{iT}},
\eea
then the composite momenta
\bea
\comp{p}_{\massless a}^\alpha &\equiv& \ourVec{ \comp{e}_{\massless a}, \vec{\comp{p}}_{\massless a}  }
= \ourVec{
\sum_{i\in \visassign{a}} p_{iT}, \sum_{i\in \visassign{a}} \vec{p}_{iT}
}\, , \\ [2mm]
\comp{q}_{\massless a}^\alpha &\equiv& \ourVec{ \tilde{\comp{e}}_{\massless a}, \vec{\comp{q}}_{\massless a}  }
= \ourVec{
\sum_{i\in \visassign{a}} q_{iT}, \sum_{i\in \visassign{a}} \vec{q}_{iT}
}\, .
\eea

The transverse parent masses are now formed in terms of 
$\comp{p}_{\massless a}^\alpha$ and $\comp{q}_{\massless a}^\alpha$
as usual:
\begin{subequations}
\label{eq:Mmasslessadef}
\begin{eqnarray}
{\cal M}^2_{\massless a}(\comp{p}^\alpha_{\massless a},\comp{q}^\alpha_{\massless a}) 
&\equiv& 
\left( \comp{p}_{\massless a}+ \comp{q}_{\massless a} \right)^2
\label{eq:Mmasslessadef1} \\ [2mm]
&& \hspace{-1.5cm}\equiv (\comp{e}_{\massless a}+\tilde{\comp{e}}_{\massless a})^2 
-(\vec{\comp{p}}_{aT}+\vec{\comp{q}}_{aT})^2, 
\end{eqnarray}
\label{eq:Mmasslessadef2}
\end{subequations}
and the ``late partitioned'', $\massless$-projected mass variable 
$M_{\massless \numparents}$ is defined as before:
\beq
M_{\massless \numparents}\equiv 
\min_{\sum \vec{q}_{iT} = \mptvec}
\ourMaxMinBracs{\max_a \ourMaxMinBracs{{\cal M}_{\massless a}
(\comp{p}_{\massless a}^\alpha,\comp{q}_{\massless a}^\alpha)
}}\ .
\label{defMmasslessN}
\eeq
Notice that the $M_{\massless N}$ variables do not depend on any 
unknown parameters related to the invisible sector (i.e.~we need no ``$\chiO$'' 
where previously we needed an $\chiM$ or a $\chiV$) and so can be 
uniquely computed in terms of the measured momenta of 
the visible particles and the missing transverse momentum alone.

Specializing (\ref{defMmasslessN})  
to the simplest case of $\numparents=1$, we get
\beas
M_{\massless 1}^2
&=& \min_{\sum \vec{q}_{iT} = \mptvec}
 \ourMaxMinBracs{ {\left( 
\sum_{i=1}^{\numvis}
e_{i\massless}+\sum_{i=1}^{\numinvis} \tilde e_{i\massless} \right)}^2  
- u_T^2 } \\ [2mm]
 &= & {\left(   \sum_{i=1}^{\numvis} {e_{i\massless}}
+\min_{\sum \vec{q}_{iT} = \mptvec}
 \ourMaxMinBracs{ \sum_{i=1}^{\numinvis} {\tilde e_{i\massless}}} 
\right)}^2  - u_T^2   \\ [2mm]
 &= & {\left(   \sum_{i=1}^{\numvis} {p_{iT}}
+\min_{\sum \vec{q}_{iT} = \mptvec}
 \ourMaxMinBracs{ \sum_{i=1}^{\numinvis} q_{iT}} 
\right)}^2  - u_T^2.   
\eeas
The minimization over the $2N_{\cal I}$
variables $\vec{q}_{iT}$ is straightforward and
we obtain several equivalent expressions for the answer
\begin{subequations}
\label{Mmassless1formula}
\bea
M_{\massless 1}^2
&=& \left(\sum_{i=1}^{\numvis} p_{iT} + \mpt \right)^2  - u_T^2 
\label{Mmassless1formula_general} \\ [2mm]
&=& \left( h_T + \mpt \right)^2  - u_T^2, \\ [2mm]
\label{Mmassless1formula_ht}
&=&  \MEFF^2 - u_T^2.
\label{Mmassless1formula_meff}
\eea
\end{subequations}
showing the close connection between $M_{\massless 1}$ and the usual 
$h_T$ and $\MEFF$ variables. We see that in the absence of
any upstream visible momentum ($\vec{u}_T=0$), the variable
$M_{\massless 1}$ itself is nothing but the effective mass $\MEFF$.
However, these two variables differ if (as is typically the case) 
the event also has some nonzero upstream momentum $u_T$.
The importance of the result (\ref{Mmassless1formula_meff}) 
is that it teaches us how to properly account for the 
presence of UVM in such cases: $u_T$ should be {\em subtracted in quadratures}
from $\MEFF$ in order to obtain the proper invariant mass variable 
(in this case $M_{\massless 1}$). Furthermore, it also 
reveals the physical meaning of the widely used $\MEFF$
variable (see also Sec.~\ref{sec:meff} below): 
it is the minimum allowed transverse mass 
constructed out of ``$\massless$''-projected momenta,
for a semi-invisibly decaying parent, whenever 
that parent is produced exclusively with $u_T=0$
(i.e.~with no additional upstream momentum in the event).

\subsubsection{The early partitioned, $\massless$-projected mass: $M_{\numparents\massless}$}
\label{sec:MN0}

Finally, we discuss the early partitioned, $\massless$-projected 
version $M_{\numparents\massless}$, where the composite momenta are first formed 
in 1+3 dimensions, then transversified via the ``$\massless$'' projection:
\bea
\sum_{i\in \visassign{a}} P_i^\mu 
&\longrightarrow& \comp{P}_a^\mu 
\ \overset{\massless}{\longrightarrow} \ 
\comp{p}_{a\massless}^\alpha = \ourVec{\comp{e}_{a\massless}, \vec{\comp{p}}_{a\massless}}, \\ [2mm]
\sum_{i\in \invisassign{a}} Q_i^\mu 
&\longrightarrow& \comp{Q}_a^\mu 
\ \overset{\massless}{\longrightarrow} \ 
\comp{q}_{a\massless}^\alpha = \ourVec{\tilde{\comp{e}}_{a\massless}, \vec{\comp{q}}_{a\massless}},
\eea
where in light of (\ref{e0MNT}) and (\ref{tildee0MNT})
\bea
\comp{p}_{a\massless}^\alpha &=& \ourVec{ \comp{e}_{a\massless}, \vec{\comp{p}}_{a\massless}  }
= \ourVec{
\left|\sum_{i\in \visassign{a}} \vec{p}_{iT}\right|, \sum_{i\in \visassign{a}} \vec{p}_{iT}
}\, , \\ [2mm]
\comp{q}_{a\massless}^\alpha &=& \ourVec{\tilde{\comp{e}}_{a\massless}, \vec{\comp{q}}_{a\massless}  }
= \ourVec{
\left|\sum_{i\in \invisassign{a}} \vec{q}_{iT}\right|, \sum_{i\in \invisassign{a}} \vec{q}_{iT}
}\, .
\eea
These (1+2) composite momenta are now used to form the 
corresponding transverse parent masses 
\begin{subequations}
\label{eq:Mamasslessdef}
\begin{eqnarray}
{\cal M}^2_{a\massless}(\comp{p}^\alpha_{a\massless},\comp{q}^\alpha_{a\massless}) 
&\equiv& 
\left( \comp{p}_{a\massless}+ \comp{q}_{a\massless} \right)^2
\label{eq:Mamasslessdef1} \\ [2mm]
&& \hspace{-1.5cm}\equiv (\comp{e}_{a\massless}+\tilde{\comp{e}}_{a\massless})^2 
-(\vec{\comp{p}}_{aT}+\vec{\comp{q}}_{aT})^2.   
\label{eq:Mamasslessdef2}
\end{eqnarray}
\end{subequations}

Now the ``early partitioned'', $\massless$-projected mass variable 
$M_{\numparents\massless}$ is defined as before:
\beq
M_{\numparents\massless}\equiv 
\min_{\sum \vec{q}_{iT} = \mptvec}
\ourMaxMinBracs{\max_a \ourMaxMinBracs{{\cal M}_{a\massless}
(\comp{p}_{a\massless}^\alpha,\comp{q}_{a\massless}^\alpha)
}}.
\label{defMNmassless}
\eeq
Just like its cousin $M_{\massless \numparents}$ defined in 
(\ref{defMmasslessN}), $M_{\numparents\massless}$ does not 
depend on any unknown parameters like $\mmass_a$ or $\mv_a$.

Specifying (\ref{defMNmassless})  
to the simplest case of $\numparents=1$, we get
\beas
M_{1\massless}^2
&=& \min_{\sum \vec{q}_{iT} = \mptvec}
 \ourMaxMinBracs{ {\left( 
\comp{e}_{1\massless}+\tilde{\comp{e}}_{1\massless} \right)}^2  
- u_T^2 } \\ [2mm]
 &= & {\left( \comp{e}_{1\massless}
+\min_{\sum \vec{q}_{iT} = \mptvec}
 \ourMaxMinBracs{ \tilde{\comp{e}}_{1\massless} } 
\right)}^2  - u_T^2  \\ [2mm]
 &= & {\left( \comp{e}_{1\massless}
+\min_{\sum \vec{q}_{iT} = \mptvec}
 \ourMaxMinBracs{\left| \sum_{i=1}^{\numinvis} \vec{q}_{iT}\right|} 
\right)}^2  - u_T^2.  
\eeas
The minimization over the $2N_{\cal I}$
variables $\vec{q}_{iT}$ gives
\begin{subequations}
\label{M1masslessformula}
\bea
M_{1 \massless}^2
&=& \left(\left|\sum_{i=1}^{\numvis} \vec{p}_{iT}\right| 
+ \mpt \right)^2  - u_T^2  \label{M1masslessformula_general} \\ [2mm]
&=& \left( \left| \mptvec + \vec{u}_T \right|
+ \mpt \right)^2  - u_T^2   
\label{M1masslessformula1}
\\ [2mm]
&=& 2 \left( 
   \mptvec\cdot\left(\mptvec + \vec{u}_T\right) 
+  \mpt        \left|\mptvec + \vec{u}_T\right|
\right),
\label{M1masslessformula2}
\eea
\end{subequations}
providing a connection between our $M_{1 \massless}$
variable and the usual missing transverse momentum $\mpt$.
In order to see the physical meaning of $\mpt$, let us take
the ``no upstream momentum'' limit $u_T\to 0$ in 
(\ref{M1masslessformula1}) or (\ref{M1masslessformula2}), 
resulting in
\beq
\lim_{u_T\to 0}M_{1 \massless}^2 = 4 \mpt^2.
\eeq
One can thus interpret the variable $2\mpt$ 
(and not just the $\mpt$!) as the 
minimum allowed ``$\massless$''-projected transverse mass 
of a semi-invisibly decaying parent, whenever 
the parent is produced exclusively with $u_T=0$, 
i.e. with no additional upstream momentum in the event.
However, in situations when the parent is produced inclusively, 
with $u_T\ne 0$, the relevant variable to consider would be 
$M_{1 \massless}$ as given by (\ref{M1masslessformula1}) or 
(\ref{M1masslessformula2}),
which properly accounts for the $u_T$ effect
(see also Sec.~\ref{sec:met} below).

\section{Additionally projected variables}
\label{sec:doubleT}


\subsection{Momentum decompositions with respect to $\vec{u}_T$}

\begin{figure}[t]
\begin{center}
\includegraphics[width=0.99\columnwidth]{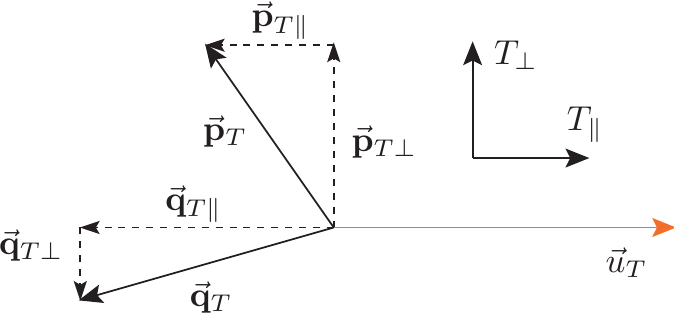}
\caption[Transverse vector decomposition onto parallel and perpendicular axes]{\label{fig:Uprojection} 
Transverse vector decomposition onto the direction $T_\myL$
specified by the UVM transverse momentum vector $\vec{u}_T$
and the direction $T_\myT$ orthogonal to it \cite{Matchev:2009ad,Konar:2009wn}.
All vectors shown are in the plane perpendicular to the beam axis.
}
\end{center}
\end{figure}

An additional level of projection {\em within} the plane
transverse to the beam has been shown to be
useful in certain circumstances~\cite{Matchev:2009ad,Konar:2009wn}.
To orient such projections we note that the total transverse momentum $\vec{u}_T$
of the UVM category breaks the rotational symmetry 
of the transverse plane and selects two preferred directions
$T_\myL$ (along $\vec{u}_T$) and $T_\myT$ (transverse to $\vec{u}_T$),
as shown in Fig.~\ref{fig:Uprojection}.
Having projected the 1+3 momentum vectors onto the transverse 
plane as in Fig.~\ref{fig:Tprojection}, one may then
additionally project the resulting 1+2 transverse momentum 
vectors onto these special directions, as illustrated in 
Fig.~\ref{fig:Uprojection}. 
The corresponding momentum components resulting from such 
``double transverse'' projections will carry a ``double transverse'' 
index: ``$T$$\myT$'' for components along $T_\myT$ and 
``$T$$\myL$'' for components along $T_\myL$
(see Fig.~\ref{fig:Uprojection}). For example, the $\mptvec$ vector
can be decomposed into a $T_\myT$ component $\slashed{\vec p}_{T\perp}$ 
\beq
\slashed{\vec p}_{T\perp}  = \frac{1}{u_T^2} \vec{u}_T 
\times \Big (  \vec{p}_{T}   \times \vec{u}_T   \Big ) \, ,
\label{defuTperp}
\eeq
and a $T_\myL$ component $\slashed{\vec p}_{T \parallel}$
\beq
\slashed{\vec p}_{T \parallel}  =  \slashed{\vec p}_{T}  - \slashed{\vec p}_{T\perp} 
= \frac{1}{u_T^2} \Big (\slashed{\vec p}_{T} \cdot \vec{u}_T \Big ) \vec{u}_T \, .
\label{defuTparallel}
\eeq
By definition, the upstream transverse momentum vector $\vec{u}_T$ 
has only a $T_\myL$ component, i.e. 
\bea
\vec{u}_{T\myL}&=&\vec{u}_{T}, \label{uTparallel} \\ [2mm]
\vec{u}_{T\myT}&=&0.
\label{uTperp}
\eea
In view of (\ref{eq:theetmissassumption}) and (\ref{uTperp}),
momentum conservation in the $T_\myT$ direction reads
\begin{equation}
      \sum_{i=1}^{\numinvis} \vec{q}_{iT\perp}  
=    \slashed{\vec p}_{T\perp}  \label{METconstraint}
= -  \sum_{i=1}^{\numvis}  \vec{p}_{iT\perp} \, .
\end{equation}
It is precisely the absence of a $\vec{u}_{T\myT}$ term in this
equation which allows one to derive exact analytical formulas
for the $T_\myT$ doubly projected variables defined next in 
Section \ref{sec:doubly-projected}.

\subsection{Doubly-projected mass bound variables}
\label{sec:doubly-projected}

\subsubsection{Homogeneously-doubly-projected mass bound variables}

For our purposes, the additional projections in Fig.~\ref{fig:Uprojection}
allow us to extend the original set of mass-bound variables
from Table~\ref{tab:floxbridge_basic} by considering the 
``doubly projected'' variables shown in
Table~\ref{tab:floxbridge_double}\footnote{To save space,
Table~\ref{tab:floxbridge_double} lists only $T_\myT$ projected
variables. An analogous set of $T_\myL$ projected variables is obtained 
by replacing the $T_\myT$ projection in Table~\ref{tab:floxbridge_double}
with a $T_\myL$ projection.}.
\begin{table*}
\begin{center}
\renewcommand\arraystretch{1.75}
\begin{tabular}{| c | c | c | c | c | c |}
\hline
Type of     & \multicolumn{4}{c|}{Operations } &   \\
\cline{2-5}
variables   &     First  &  Second   & Third  & Fourth &  Notation     \\
\hline
Early partitioned  & Partitioning & $T=\ourT$ projection & $\myT=\ourT$ projection on $T_\myT$ & Minimization & $M_{N\ourT \myT}$ \\
doubly projected   & Partitioning & $T=\ourperp$ projection & $\myT=\ourperp$ projection on $T_\myT$  & Minimization & $M_{N\ourperp \myT}$ \\
      $M_{NT\myT}$    & Partitioning & $T=\massless$ projection & $\myT=\massless$ projection on $T_\myT$  & Minimization & $M_{N\massless \myT}$ \\
\hline
Late partitioned,  & $T=\ourT$ projection & $\myT=\ourT$ projection on $T_\myT$ & Partitioning  & Minimization & $M_{\ourT \myT N}$ \\
doubly projected  & $T=\ourperp$ projection & $\myT=\ourperp$ projection on $T_\myT$  & Partitioning & Minimization & $M_{\ourperp \myT N}$ \\
     $M_{T\myT N}$    & $T=\massless$ projection & $\myT=\massless$ projection on $T_\myT$  & Partitioning & Minimization & $M_{\massless \myT N}$ \\
\hline
In-between partitioned,  & $T=\ourT$ projection & Partitioning & $\myT=\ourT$ projection on $T_\myT$ & Minimization & $M_{\ourT N\myT}$ \\
doubly projected & $T=\ourperp$ projection & Partitioning & $\myT=\ourperp$ projection on $T_\myT$  & Minimization & $M_{\ourperp N\myT}$ \\
    $M_{TN\myT}$    & $T=\massless$ projection & Partitioning & $\myT=\massless$ projection on $T_\myT$  & Minimization & $M_{\massless N\myT}$ \\
\hline
\end{tabular}
\caption[``Doubly projected'' mass-bounds]{\label{tab:floxbridge_double} 
An extended version of Table~\ref{tab:floxbridge_basic}, 
containing the additional variables found by including the option of 
a $T_\myT$ projection shown in Fig.~\ref{fig:Uprojection}.
An analogous set of variables is obtained by considering a $T_\myL$ projection instead.
}
\end{center}
\end{table*}
The benefit of such additionally projected varibles has been noted and 
discussed in \cite{Matchev:2009ad,Konar:2009wn}. For example,
the shapes and the kinematic endpoints of the distributions of
$T_\myT$-projected variables can be independent of the value of $u_T$. 
Therefore, such distributions can be constructed from the whole event 
sample, without any loss in statistics due to a specific $u_T$ selection.
Furthermore, the relation (\ref{uTperp}) leads to significant simplifications
in the analytical treatment of $T_\myT$ doubly projected variables.
For example, for singly projected variables, the case of $\numparents =2$ 
is untractable by analytical means, and (apart from some special cases \cite{Lester:2011nj})
has to be treated numerically \cite{oxbridgeStransverseMassLibrary,zenuhanStransverseMassLibrary}. 
In contrast, one can derive exact analytical formulas for 
calculating $\numparents=2$, $T_\myT$ doubly projected mass bound variables on 
an event-per-event basis, without any need for numerical minimizations
\cite{Matchev:2009ad,Konar:2009wn}.

In general, the variables in Table~\ref{tab:floxbridge_double}
are independent, with one exception:
\beq
M_{N\massless \myT} \equiv M_{\massless N\myT}.
\label{eq:N00eq0N0}
\eeq
Later on in Section~\ref{sec:ttbar} (see in particular Fig.~\ref{fig:ttb2}(b)),
we shall consider a specific example illustrating some of the 
homogeneously-doubly-projected variables from Table~\ref{tab:floxbridge_double}.

\subsubsection{Heterogeneously-doubly-projected mass bound variables}

Notice that in defining the mass bound variables in Table~\ref{tab:floxbridge_double},
we have chosen the second level of projection (along $T_\myT$) to be performed with
{\em the same} type of transverse projection (``\ourT'', ``\ourperp'' or ``\massless'')
which was used to project into the transverse plane.
Of course, this does not have to be the case --- and by
allowing for {\em different} types of transverse projections for 
$T$ and for $T_\myT$, one would obtain 18 additional variables 
with ``mixed'' transverse projections. These heterogeneously-doubly-projected
variables are listed in Table~\ref{tab:floxbridge_super_extended}, where
the additional subindex on $\myT$ specifies the type of $T_\myT$ projection
as being of the ``\ourT'', ``\ourperp'' or ``\massless'' type.\footnote{Another 
set of 18 additional variables can be trivially obtained from Table~\ref{tab:floxbridge_super_extended}
by considering a $T_\parallel$ type of projection at the second level instead.}
\begin{table}[htb]
\renewcommand\arraystretch{1.75}
\begin{tabular}{| c | c | c |}
\hline
Early partition & Hedged partition & Late partition \\
\hline
$M_{N \ourT \perp_\ourperp},$ $M_{N \ourT \perp_\massless}$ & 
$M_{\ourT N \perp_\ourperp},$ $M_{\ourT N \perp_\massless}$ & 
$M_{\ourT \perp_\ourperp N},$ $M_{\ourT \perp_\massless N}$\\
$M_{N \ourperp \perp_\ourT},$ $M_{N \ourperp \perp_\massless}$ & 
$M_{\ourperp N \perp_\ourT},$ $M_{\ourperp N \perp_\massless}$ & 
$M_{\ourperp \perp_\ourT N},$ $M_{\ourperp \perp_\massless N}$\\
$M_{N \massless \perp_\ourT},$ $M_{N \massless \perp_\ourperp}$ & 
$M_{\massless N \perp_\ourT},$ $M_{\massless N \perp_\ourperp}$ & 
$M_{\massless \perp_\ourT N},$ $M_{\massless \perp_\ourperp N}$\\
\hline
\end{tabular}
\caption[The 18 heterogeneously-doubly-projected transverse mass variables]{\label{tab:floxbridge_super_extended}
The 18 additional heterogeneously-doubly-projected transverse mass variables for each $N$,
where the additional subindex on $\myT$ specifies the type of $T_\myT$ projection
as being of the ``\ourT'', ``\ourperp'' or ``\massless'' type.
As was the case in Tables~\ref{tab:floxbridge_basic} and \ref{tab:floxbridge_double}, 
``partition'' implies the combined operation of partitioning the objects and 
agglomerating them by summation into composite objects.}
\end{table}
As usual, the sequence of indices in both Tables~\ref{tab:floxbridge_double}
and \ref{tab:floxbridge_super_extended} represents the order in which the operations 
are to be performed. For example, $M_{\massless N \myT_{\ourT} }$ means
\begin{itemize}
\item project all objects using the massless `$\massless$' projection, 
then
\item partition and agglomerate into $N$ parents, 
then 
\item project into the direction perpendendicular to $\vec{u}_T$ 
using the mass-preserving `\ourT' projection, then, as ever, 
\item minimize over all values of the unknown momenta that satisfy the constraints.
\end{itemize}

Interestingly, most of the ``$\myT_\massless$'' heterogeneously-doubly-projected variables
turn out to be related to each other and to the corresponding homogeneously-doubly-projected 
variables from Table~\ref{tab:floxbridge_double}. For example:
\bea
M_{N\massless \myT}&\equiv& M_{N\massless \myT_\massless}
= M_{N\ourT \myT_\massless}
= M_{N\massless \myT_\ourT}
= M_{\ourT N \myT_\massless},~~~~~~
\label{eq:N00identities}
\\ [2mm]
M_{\massless \myT N}&\equiv& M_{\massless \myT_\massless N}
= M_{\massless \myT_\ourT N}
= M_{\ourT \myT_\massless N},
\\ [2mm]
M_{\massless N\myT}&\equiv& M_{\massless N\myT_\massless}
= M_{N\ourT \myT_\massless}
= M_{N\massless \myT_\ourT}
= M_{\ourT N \myT_\massless},
\label{eq:0N0identities}
\eea
where the last line (\ref{eq:0N0identities}) follows from (\ref{eq:N00eq0N0})
and (\ref{eq:N00identities}).
The one remaining variable $M_{\massless N\myT_\ourT}$
is rather similar to $M_{\ourT N\myT_\ourT}\equiv M_{\ourT N\myT}$, 
since the difference between them may arise only due to nonzero 
masses ($M_i\ne 0$) of the individual visible particles.

\section{Properties of the mass-bound variables}
\label{sec:math}

We should stress that proliferating the number of 
kinematic variables in the literature is certainly not
among the goals of this paper --- on the contrary, 
we emphasise that these variables are different implementations of  
the general principle described in Section~\ref{sec:principles}.
What's more we will soon begin to reveal further connections
of these variables to each other (in Section~\ref{sec:relations})
and to existing proposals (in Section~\ref{sec:literature}).
But before we proceed, perhaps now is a good time to
summarize what we have accomplished so far.

The previous discussion has hopefully convinced 
the reader that, once the decision 
on the targeted event topology (Fig.~\ref{fig:event})
is made, the choice of relevant invariant mass variables is 
straightforward and rather unambiguous. 
Following the general recipe outlined in 
Section~\ref{sec:imassdef}, one is able to overcome
the two main obstacles in any analysis involving missing momentum:
\begin{itemize}
\item {\em The fact that the momenta of the invisible particles are unknown.}
To construct a bound, this problem is solved by performing a {\em minimization}
over all possible values of the invisible momenta, 
consistent with the measured $\mptvec$.
The minimization fixes the values of the 
invisible momenta (e.g.~as in (\ref{eq:qiT}) and (\ref{eq:qiz}))
and from that point on, one works with 
fully specified kinematics in the event.
Of course, the momenta found in the 
process of minimization, are not equal to
the {\em actual} momenta of the invisible 
particles in the event, although in some
cases they can be close, see \cite{Cho:2008tj}.
\item {\em The fact that the total number and the masses of the
invisible particles are unknown.}
This problem is also resolved through the minimization ---
as we have seen in the explicit $\numparents=1$ examples discussed in 
Section~\ref{sec:TMNT}, the mass bound resulting from the minimization 
turned out to be a function which depends only on a set of $\numparents$ summed
mass parameters (\ref{eq:chimdefined}) or a set of $\numparents$ 
3-speeds \eqref{eq:chivdefined}, and is insensitive to the
number of invisible particles or to the fine structure of the 
individual masses $\tilde M_a$ or 3-speeds $\tilde V_a$ connected 
to parent $\parent a$.  We set out a general proof for general 
$\numparents$ in Section~\ref{sec:proofoffloxdeponhypsets} below.
\end{itemize}

 
It should be recognized that for any practical application, 
there is no need to consider every one of the
variables in Tables~\ref{tab:floxbridge_basic}
and \ref{tab:floxbridge_double}, 
since some will be better suited than others
to the particular task at hand. 

For example, we have seen that the ``$\ourperp$''-projected quantities 
assume knowledge of the $\vec{p}_T$ and the speed of the particle,
but leave the mass and $p_z$ undetermined.
This means that all of the ``$\ourperp$''-projected variables in
Tables~\ref{tab:floxbridge_basic} and \ref{tab:floxbridge_double}
should be considered appropriate only for experimental situations 
in which the $p_T$ and speed of the particles are known, 
but nothing is known about their masses or longitudinal momenta. 
Such situations may exist -- for example if $p_T$ can be determined from 
the particles' bending radii in a solenoidal magnetic field 
and speed can be inferred from time-of-flight information 
or from the characteristic angle of any emitted \v Cerenkov radiation. 
However such cases are the exception, rather than
the norm in current experiments.\footnote{Even in those cases, it is usual that
$\vec{p}_T$ is determined from a measured track, so one would 
also expect to be able to reconstruct the polar angle $\theta$, 
from that track, which would permit $p_z$ and hence the full 1+3-vector of the 
particle to be determined. There would be no need to then restrict oneself
to the subset of that information held by the corresponding ``$\ourperp$''-projected quantities.}  
In what follows we shall therefore give greater attention 
to the remaining three classes of variables: ``unprojected'', 
``$\ourT$''-projected, and ``$\massless$''-projected.

We shall denote a generic mass-bound variable as $M_\flox$,
where the composite index $\flox$ is made up from (any number of)
objects taken from the set $\ourSet{\numparents, \ourT, \ourperp, \massless}$.
There are seven such possibilities\footnote{The number of possibilities increases 
to 17 if one allows a second level of projection, as discussed in Section~\ref{sec:doubleT}.}:
\bea
\flox &\in& \ourSet{
{\scriptstyle \numparents},
{\scriptstyle \numparents\ourT},
{\scriptstyle \ourT\numparents},
{\scriptstyle \numparents\massless},
{\scriptstyle \massless\numparents},
{\scriptstyle \numparents\ourperp},
{\scriptstyle \ourperp\numparents}
}.
\label{floxlabel}
\eea
For later convenience, we also introduce the 
generic notation $\flox_\ourT$ for the
$\ourT$-projected variables:
\beq
\flox_\ourT \in \ourSet{
{\scriptstyle \numparents\ourT},
{\scriptstyle \ourT\numparents}
}
\eeq
$\flox_\ourperp$ for their two
``$\ourperp$''-projected counterparts:
\beq
\flox_\ourperp \in \ourSet{
{\scriptstyle \numparents\ourperp},
{\scriptstyle \ourperp\numparents}
},
\eeq
and $\flox_\massless$ for the two
``$\massless$''-projected equivalents:
\beq
\flox_\massless \in \ourSet{
{\scriptstyle \numparents\massless},
{\scriptstyle \massless\numparents}
}.
\eeq

The large multiplicity is partially
due to the different possible ways to transversify the
energy-momenta of the composite daughter particles 
whose masses $\comp{M}_a$ and $\tilde{\comp{M}}_a$ are typically nonzero.
First, one can choose whether or not to project, 
and then those projections can be of type
$\ourT$, $\ourperp$ or $\massless$
(see Section~\ref{sec:projections}).
In addition, as emphasized in Section~\ref{sec:earlylate},
the operations of partitioning into composite particles
and transversifying do not commute, so that in general
we obtain non-equivalent variables simply by switching 
the order of those operations. As illustrated in 
Tables~\ref{tab:floxbridge_basic}-\ref{tab:floxbridge_super_extended},
we use the ordering of indices on each variable (from left to right)
to indicate the order of the corresponding operations. 
For example $M_{2\ourT}$ means that we add the 1+3 vectors
first to form two composite visible daughter particles 
and transversely project later,
while $M_{\ourT 2}$ implies the opposite --- make 
a $\ourT$ transverse projection before forming the 
composite daughter particles. 


\subsection{Dependence of mass bounds $M_\flox$ on $\chiM$, $\chiV$, etc.}
\label{sec:proofoffloxdeponhypsets}

We have stated that the dependence of the mass bound variables, $M_\flox$, 
on parameters of the hypothesis is always confined to a set of $\numparents$ 
parameters contained within $\chiM$ or $\chiV$ etc.  We have not yet proved 
this statement for general values of $\numparents$, or indicated whether we 
can demonstrate this to be true for other classes of projection not already 
discussed.  All we have proved, so far, are the following statements, which 
are specific to $N=1$ and consider at most one projection:
\begin{itemize}
\item  That   $M_{1}$ depends only on $\chiM = \ourSet{\chiM_1}$ (see \eqref{eq:M1formula})
\item  That  $M_{1\ourT}$ depends only on $\chiM= \ourSet{\chiM_1}$ (see \eqref{eq:M1ourTformula})
\item  That  $M_{\ourT 1}$ depends only on $\chiM= \ourSet{\chiM_1}$ (see \eqref{eq:MourT1formula})
\item  That  $M_{\ourperp 1}$ depends only on $\chiV= \ourSet{\chiV_1}$ (see \eqref{Mperp1formula})
\item  That  $M_{1 \ourperp}$ depends only on $\chiV= \ourSet{\chiV_1}$ (see \eqref{eq:M1perpformula})
\item  That  $M_{1 \massless}$ depends on no hypothesis parameters\footnote{Note that depending 
on ``no hypothesis parameters'' is a special case of depending on a very dull set of 
parameters $\chiO=\ourSet{\chiO_1}$ which contain no information.},
\item  That  $M_{\massless 1}$ depends on no hypothesis parameters.
\end{itemize}
We now seek to generalize the proofs of the above to all other values of $\numparents$, 
in a manner that does not make specific requirements on $\flox$.  Specifically, 
we would like to prove that:
\bea
\mbox{``$M_\flox$ depends only on 
$\chiS_\flox = \ourSet{\chiS^\flox_1,\ldots,\chiS^\flox_\numparents}$''}\label{eq:WHATWEwouldliketoprovebutcannot}
\eea
where $\chiS_\flox$ is a set of $\numparents$ parameters, of which there 
is one ($\chiS^\flox_a$) for each parent $\parent a$, and where the nature 
of $\chiS^\flox_a$ depends on the type of projection in $\flox$ 
(which may be arbitrary), and on $a$, but not on the number $\numparents$ 
of parents in total.  In particular, we have already seen to 
expect $\chiS_\flox=\ourSet{\chiM_1}$ when $\flox \in \ourSet{\scriptstyle 1,1\ourT,\ourT 1}$, 
and to expect $\chiS_\flox=\ourSet{\chiV_1}$ when 
$\flox \in \ourSet{\scriptstyle 1\ourperp,\ourperp 1}$, 
and we are now seeking to generalize these to results like 
``$\chiS_\flox=\ourSet{\chiM_1, \chiM_2}$ when $\flox \in \ourSet{\scriptstyle 2,2\ourT,\ourT 2}$'' 
or ``$\chiS_\flox=\ourSet{\chiV_1, \chiV_2,\chiV_3}$ when $\flox \in \ourSet{\scriptstyle 3\ourperp,\ourperp 3}$''etc.


What we will actually succeed in proving is the marginally less general statement that:
\begin{align}
\framebox[0.85\columnwidth]{\parbox{0.83\columnwidth}{
$M_{\flox}$ depends only on $\chiS_{\flox}=\ourSet{\chiS_1^\flox,\chiS_2^\flox,\ldots,\chiS_n^\flox}$, 
when $\numparents=n$, provided that $M_{\flox}$ depends only 
on $\chiS_{\flox}=\ourSet{\chiS_1^\flox}$ when $\numparents=1$}}\label{eq:whatwecanactuallyshow}
\end{align} 
which reminds us that the generality of the desired (but unattainable) result 
\eqref{eq:WHATWEwouldliketoprovebutcannot} is constrained 
(for any particular projection $\flox$) by the need to prove the 
result for the $\numparents=1$ case.  In other words, though the 
proof of \eqref{eq:whatwecanactuallyshow} found below will be valid 
for any projection, the desired result \eqref{eq:WHATWEwouldliketoprovebutcannot} 
will only be true for projections that experience simplification in 
the $\numparents=1$ case.

The proof of \eqref{eq:whatwecanactuallyshow} is astonishingly simple. 
Consider an arbitrary mass-bound variable
\beq
M_{\flox} \equiv \min_{\substack{
\sum \vec{q}_{iT} = \mptvec}} 
\ourMaxMinBracs{\max_a\ourMaxMinBracs{{\cal M}_{a}
\ourVec{{\cal S}^\flox_a}}}.
\label{Mfloxdefgeneral}
\eeq
where each parent mass ${\cal M}_a$ depends on a corresponding
set of invisible parameters $S^\flox_i$
\beq
{\cal S}^\flox_a \equiv \ourSet{ S^\flox_i \mid {i\in\invisassign{a}}}
\eeq
(compare this to the analogous relations (\ref{eq:deftildemua}) and (\ref{eq:deftildevela})).
Let us now perform the minimization in (\ref{Mfloxdefgeneral})
in two steps. In the first step, for each parent, 
we hold the sum of the invisible daughters' momenta 
constant, and minimize over the internal partition of invisible 
momentum between those daughters:
\beq
M_{\flox} =
\min_{\substack{\sum \vec{q}_{aT} = \mptvec}} 
\ourMaxMinBracs{\min_{\substack{\sum\limits_{i\in \invisassign{a}} \vec{q}_{iT} = \vec{q}_{aT}}} 
\ourMaxMinBracs{\max_a
  \ourMaxMinBracs{{\cal M}_{a}
    \ourVec{
      {\cal S}^\flox_a
    }
  }
}
}.
\label{Mflox2steps}
\eeq
Since the internal partitions over the invisible momenta are done independently 
for each parent, (\ref{Mflox2steps}) can be equivalently rewritten as
\beq
M_{\flox} =
\min_{\substack{\sum \vec{q}_{aT} = \mptvec}} 
\ourMaxMinBracs{
\max_a\ourMaxMinBracs{
\min_{\substack{\sum\limits_{i\in \invisassign{a}} \vec{q}_{iT} = \vec{q}_{aT}}} 
\ourMaxMinBracs{
{\cal M}_{a}
    \ourVec{
      {\cal S}^\flox_a
    }
  }
}
}.
\label{Mfloxswitch}
\eeq
Now we use the assertion that for $N=1$ (i.e. for any individual parent)
the minimization over internal partitions yields a function of
a single parameter $\chiS^\flox_a$ as opposed to the whole 
set of parameters ${\cal S}^\flox_a$:
\beq
\min_{\substack{
\sum\limits_{i\in \invisassign{a}} \vec{q}_{iT} = \vec{q}_{aT}}
} 
\ourMaxMinBracs{
{\cal M}_{a}
    \ourVec{
      {\cal S}^\flox_a
}
}
\equiv
{\cal M}_{a}
\ourVec{\chiS^\flox_a}.
\label{eq:assertion}
\eeq
Substituting (\ref{eq:assertion}) into (\ref{Mfloxswitch}),
we obtain the desired result
\beq
M_{\flox}\ourVec{\chiS_\flox} =
\min_{\substack{\sum \vec{q}_{aT} = \mptvec}} 
\ourMaxMinBracs{
\max_a
\ourMaxMinBracs{
{\cal M}_{a}
    \ourVec{
      \chiS^\flox_a      
    }
  }
},
\label{MfloxQED}
\eeq
which makes it obvious that $M_{\flox}$ can only be a function of the 
set of parameters 
$\chiS_{\flox}=\ourSet{\chiS_1^\flox,\chiS_2^\flox,\ldots,\chiS_n^\flox}$.

\subsection{Parental masses and upper kinematic endpoints}
\label{sec:endpoints}

By construction, the mass-bound variables of Table~\ref{tab:floxbridge_basic}
are designed to provide an event-by-event lower bound on the true invariant mass 
$\trueparentmass_{\cal P}^{\rm max}$ of the heaviest parent
\beq
\trueparentmass_{\cal P}^{\rm max} \equiv 
\max_{a\in\parentset}\ourMaxMinBracs{M_{a}}.
\label{MPmaxdef}
\eeq
Such bounding properties are contingent on us being able to make
appropriate choices when analyzing the events.
We initially restrict our discussion of the bound to the case where
the set of momentum configurations $\assumed$ permitted under our assumptions
is equal to the set of momenta $\allevents$ sampled by nature. 
We observe that to ensure $\assumed=\allevents$
we must (a) correctly reconstruct the event topology 
(the number of parents, the number and types of daughters,
and the association of daughters to parents)
and (b) employ the true values of the parameters $\chiS$
used in constructing of any $m_\flox$ variable --- 
i.e. $\chiV^{\rm true}$ for $\flox_\ourperp$,
and $\chiM^{\rm true}$ for $\flox_\ourT$ or $\flox_N$. 
Under $\assumed=\allevents$ conditions all $M_\flox$ variables
are designed to return values smaller than the mass of the heaviest parent
\beq
M_{\flox} \le \trueparentmass_{\cal P}^{\rm max}
\quad(\assumed=\allevents).
\label{eq:Mbound}
\eeq
From eq.~(\ref{eq:Mbound}), it follows directly that,
if we were to consider the differential distribution
of the same variable $M_{\flox}$ over all events,
the upper kinematic endpoint $M^{\rm max}_{\flox}$
of this distribution also satisfies
\beq
M^{\rm max}_{\flox}
\equiv \max_{\rm all\ events} \ourMaxMinBracs{ M_{\flox}} 
\le {\trueparentmass}_{\cal P}^{\rm max}\quad (\assumed=\allevents),
\label{eq:endpoint}
\eeq
where we make explicit the requirement that
the true values of the $\chiS$ parameters are used.

There remains the important question as to the circumstances
under which the inequalities in \eqref{eq:Mbound} and \eqref{eq:endpoint}
are saturated -- i.e. the conditions for which a measurement of the $M_\flox$
kinematic endpoint will provide a determination of (rather than simply a lower bound on) 
the largest parent mass ${\trueparentmass}_{\cal P}^{\rm max}$.


We observe that when $\assumed=\allevents$:
(i) that for any selected event $\event\in\allevents$ 
the minimization picks out some non-empty subset of 
momenta $K_\event$ that satisfy the global minimum; 
(ii) that $K_{\event} \subset \assumed$;
(iii) we may define for convenience ${\cal M}_\flox^{\rm max}(c) = 
\max_{a\in\parentset}\ourMaxMinBracs{{\cal M}_{a\flox}(c)}$ for any configuration 
$c\in\assumed$;
(iv) that since we minimize over any unknown momentum components, 
the value of ${\cal M}_{\flox}^{\rm max}$ evaluated for some minimum 
configuration $k\in K$ cannot exceed the value that would be obtained 
elsewhere in $\assumed$ (and therefore in $\allevents$);
(v)  that projections $\genericT\in\ourSet{\ourT,\ourperp,\massless}$ 
do not increase the invariant mass \eqref{masshierarchy};
(vi) that $M_{\flox}$ can therefore not exceed the largest parent's 
invariant mass since 
\begin{equation}M_\flox \equiv {\cal M}_{\flox}^{\rm max} (K_{\event}) 
\leq {\cal M}_{\flox}^{\rm max} (\event) \leq {\trueparentmass}_{\cal P}^{\rm max} 
\quad\left(\assumed=\allevents\right).
\label{eq:saturation}
\end{equation}

The necessary and sufficient condition for saturation of \eqref{eq:endpoint}
is therefore that there exist some event 
$\event$ for which two inequalities in \eqref{eq:saturation}
simultaneously become equalities.

Given that $M_\flox^{\rm max} \equiv \max_{a\in\parentset}\ourMaxMinBracs{{\cal M}_a(K_\event)}$ 
it follows from \eqref{eq:saturation} that a necessary condition for saturation is that
\begin{equation}\label{eq:saturationa}
\exists(\event\in\allevents,a\in\parentset)\left[
{\cal M}_a(\event)=M_\parentset^{\rm max}
\right].
\end{equation}

There are cases for which \eqref{eq:saturationa} is {\em not} satisfied, 
and for which the inequality in \eqref{eq:endpoint} must therefore remain unsaturated. 
The ``${\massless}$'' projection discards all previous information about
the mass of the 1+3 vector being projected,
and so calculation of $M_{\massless N}$ will return the same value
that would be obtained if one were to set both $M_i=0\,(\forall i\in\visset)$ 
and $\chiM_a=0\,(\forall a\in\parentset)$.
If all events contain
massive invisibles (or indeed massive visibles) in all daughter sets 
--- as would be the case for models like $R$-parity conserving supersymmetry and UED ---
then \eqref{eq:saturationa} cannot be true for any ${\cal M}_{\massless a}$ 
and so $M_{\massless N}$ can only bound from below, rather than determine, 
the mass of the heaviest parent.
In Appendix~\ref{sec:hierproofs} we prove the event-by-event inequality
$M_{N\massless}\leq M_{\massless N}$, so 
any bound that is unsaturated for $M_{\massless N}$ must
also be unsaturated for $M_{N\massless}$.

Compared to the $\flox_\massless$ variables, 
the $\flox_\ourT$ and $\flox=N$ variables are subject to less stringent
conditions for saturation, because they
retain mass information during the process of (absence of) projection.
Some of the necessary conditions can be inferred from the results of
in App.~\ref{sec:boundsonpairs} and \ref{sec:boundsonsets}. As an example
of these less-stringent conditions, 
if $\left|\invisassign{a}\right|\ne 0 \,\forall a$ then
a necessary condition for saturation will include the existence of 
events $\event\in\allevents$ with vanishing relative rapidity
between $\comp{P}^\mu_a$ and $\comp{Q}^\mu_a$ for some $a\in\parentset$.

As discussed in Section~\ref{sec:TMNT}, some of the widely 
used collider variables like $h_T$, $\MEFF$ and $\mpt$
belong to the ${\flox_\massless}$ class of mass bound
variables, and as such can generally only place a lower bound on 
the parent mass if nature produces heavy invisibles. In order to 
really measure the mass scale of the new particles when massive invisible particles
are pervasive, one must work with variables which retain the dependence on the 
missing mass parameters and therefore belong to either 
the $M_\numparents$ or the $M_{\flox_\ourT}$
class of mass-bound variables.
Other than the full 1+3 dim. invariant mass, other common examples of such variables
include $\sqrt{\hat{s}}_{\rm min}$~\cite{Konar:2008ei} (discussed below in Sec.~\ref{sec:smin})
the transverse mass $m_T$ in the form \cite{Nakamura:2010zzi} that 
accounts for the mass of all daughters (shown in Sec.~\ref{sec:MT}) 
and the `stransverse mass' $m_{T2}$~\cite{Lester:1999tx} (Sec.~\ref{sec:mt2}).

Before concluding this subsection, we note that 
the $M_\flox$ variables are still useful
even when the true values of the mass $\chiM$ (or speed $\chiV$) parameters are not known.
The most conservative procedure in these situations of uncertainty is to 
minimize $M_\flox$ over the complete physically relevant range of 
any unknown parameter. 
This leads to $\chiM_a\rightarrow 0$ for $M_N,\,M_{\flox_\ourT}$, 
and to $\chiV_a\rightarrow 1$ for $M_{\flox_\massless}$.
The resulting, conservative, $M_\flox$ variables still provide lower bounds on the mass
of the heaviest parent -- though those bounds will generally not be saturated.
\par
A more sophisticated treatment is also possible.
For example if the ``true'' value
of the summed-mass parameter set $\chiM^\mathrm{true}$ for the
calculation of $M_{\numparents}(\chiM)$ or
$M_{\flox_\ourT}(\chiM)$ were not known --- then one
could still view the set of endpoint measurements for all possible
values of $\mmass_a$ as one constraint among the $\numparents+1$ 
unknowns ${\trueparentmass}_{\cal P}^{\rm max}$ and $\mmass_a$,
$(a=1,2,\ldots,\numparents)$.  
Not only is this valuable information on its own, the derived functional relationship
${\trueparentmass}_{\cal P}^{\rm max}(\chiM)$ 
is, in addition, often sufficient for determining the 
{\em individual} mass parameters $\mmass_a$.
The function ${\trueparentmass}_{\cal P}^{\rm max}(\chiM)$,
when viewed as an $\numparents$-dimensional hyper-surface 
in the $(\numparents+1)$-dimensional space
spanned by ${\trueparentmass}_{\cal P}^{\rm max}$ and $\chiM$,
exhibits certain ridge or crease features,
which commonly originate from the point 
marking the set of true values of $\chiM^\mathrm{true}$
\cite{Barr:2009jv,Konar:2009qr}.
(The one-dimensional version of this phenomenon
was originally discussed in 
\cite{Cho:2007qv,Gripaios:2007is,Barr:2007hy,Cho:2007dh,Burns:2008va}
and is known as the $m_{T2}$ ``kink''. 
Also see \cite{Kim:2009si} for algebraic singularity in relation to the kink.)

\subsection{Relations among the mass-bound variables}
\label{sec:relations}

Some of the variables in Table~\ref{tab:floxbridge_basic}
are related to each other, either in general\footnote{Previously 
in (\ref{M1eqM1T}) we already encountered the $\numparents=1$ version 
of eq.~(\ref{eq:MNeqMNTagain}). The general proof for 
arbitrary $\numparents$ is provided in
the appendix in equation~(\ref{eq:proofofMNtisMNT}).}
\beq
M_{\numparents\ourT}(\chiM) = M_{\numparents}(\chiM),
\label{eq:MNeqMNTagain}
\eeq
\beq
M_{N\massless \myT}= M_{\massless N\myT} 
= M_{N\ourT \myT_\massless}
= M_{N\massless \myT_\ourT}
= M_{\ourT N \myT_\massless},
\eeq
\beq
M_{\massless \myT N}
= M_{\massless \myT_\ourT N}
= M_{\ourT \myT_\massless N},
\eeq
or under some special circumstances, e.g. massless particles:
\bea
M_{\ourT\numparents}( \ourSet{\mmass_a=0},\ourSet{M_i=0}) &=& M_{\massless\numparents},
\label{eq:MTNeqM0N}
\\ [2mm]
M_{\ourT\numparents\myT}( \ourSet{\mmass_a=0},\ourSet{M_i=0}) &=& M_{\massless\numparents\myT_\ourT}.
\label{eq:MTNperpeqM0Nperp}
\eea
Given such exact identities like (\ref{eq:MNeqMNTagain}), 
the reader may wonder why we even bothered
to introduce separately variables like
$M_{\numparents\ourT}$ and $M_{\numparents}$.
In our view, such redundancy is a virtue, since it 
offers deeper intuitive understanding of these kinematic
variables, and allows one to think about the same fundamental 
quantity in different contexts, e.g. in (1+3)-dimensions or 
in (1+2)-dimensions.

We additionally find (see proof terminating in \eqref{eq:thehierarchyisprovedforgeneraln} 
in Appendix~\ref{sec:hierproofs}) that the mass-bounds 
from Table~\ref{tab:floxbridge_basic} obey a hierarchy:
\beq
M_{\numparents} =  M_{\numparents\ourT} \ge M_{\ourT\numparents} 
\ge  M_{\massless\numparents} \ge M_{\numparents\massless}. \label{eq:sneakpreviewofhierarchy}
\eeq
Similarly, the doubly-projected mass-bounds from Table~\ref{tab:floxbridge_double}
obey the hierarchy
\beq
M_{N\ourT \myT} \ge M_{\ourT N\myT} \ge M_{\ourT \myT N} \ge M_{\massless \myT N} 
\ge M_{\massless N\myT} = M_{N\massless \myT}. 
\eeq
From these hierarchies, it becomes apparent that the \ourT{}-projected, late-partitioned
variables bear a cost associated with the insensitivity to the longitudinal momenta. By
dropping this information we necessarily weaken the bound relative to the early-partitioned
versions. Interestingly enough, the order of projection and partition has the opposite effect
with the \massless{}-projection, since both longitudinal and transverse information is 
contained in the masses of the agglomerates, and hence by throwing away the masses at a
later stage, we in fact throw away maximal information and are forced to produce the worst
possible bound!

\section{Connections to other variables in the literature}
\label{sec:literature}

The existing literature is abundant with a number of 
(transverse) invariant mass variables which were suggested 
(at various times and for a variety of reasons) for
the study of missing momentum event topologies 
(see \cite{Barr:2010zj} for a recent review). 
At the same time, the mass-bound variables which we defined earlier
in Table~\ref{tab:floxbridge_basic}, were meant to be very 
general, since they target the rather generic event topology of 
Fig.~\ref{fig:event}, and are intended to have as few hidden 
assumptions as possible. It follows that we should be able to 
correlate the most useful mass-scale 
variables in the literature to one of our mass-bound
variables from Table~\ref{tab:floxbridge_basic}.\footnote{A 
corollary from this statement is that
invariant mass variables which make similar sorts of assumptions  but do not fit into 
the classification of Table~\ref{tab:floxbridge_basic},
are often both poorly motivated and sub-optimal.}
The purpose of this subsection is to demonstrate 
that this is indeed the case.

\subsection{Missing transverse momentum $\mpt$}
\label{sec:met}

The defining feature of any ``missing particle'' event
is the presence of missing momentum
(more precisely, missing transverse momentum) $\mpt$.
This is due to the production and escape of 
a certain number of ``invisible'' particles,
which are either sterile, or very weakly interacting, 
so that they are not seen in the detector.
The $\mpt$ distribution\footnote{The missing transverse momentum 
is often labelled called ``missing transverse energy'' and labelled 
$\met{}$ or $E_{\rm T}^{\rm miss}$ in experimental papers.
As previously discussed we prefer to recognize the imporant 
distinction between energy and momentum, so use the symbol $\mpt$.}
is perhaps the most widely studied distribution in relation to new physics
searches, especially in models with WIMP dark matter 
candidates like supersymmetry, UED and so on.
Eq.~(\ref{M1masslessformula}) allows us to correlate
the $\mpt$ variable to our $M_{1 \massless}$ variable as
\beq
M_{1 \massless} \stackrel{u_T\to 0}{\longrightarrow} 2\mpt.
\label{eq:mptinterpretation}
\eeq
We see that as $M_{1 \massless}$ is defined more and more inclusively,
it eventually becomes equal to twice the missing transverse momentum.
Thus in the case of a singly produced parent,
eq.~(\ref{eq:mptinterpretation}) allows us to interpret 
the usual $\mpt$ variable (more precisely, the variable $2\mpt$) 
as a suitably constructed (in the $M_{1 \massless}$ sense) 
transverse invariant mass of the parent
(see also the discussion at the end of Sec.~\ref{sec:MN0}).
In accordance with (\ref{eq:Mbound}),
in the $u_T\rightarrow 0$ limit the upper kinematic 
endpoint of the $2\mpt$ distribution 
gives a {\em lower bound} on the parent mass 
in events interpreted as single-parent ($\numparents=1$) production.

\subsection{Effective mass $\MEFF$}
\label{sec:meff}

The ``effective mass'' variable defined in
(\ref{eq:meffdefbyus}) can be also directly related to 
one of our variables, namely the late-partitioned,
``$\massless$''-projected variable $M_{\massless 1}$
discussed in Sec.~\ref{sec:M0N}. The previously derived
eq.~(\ref{Mmassless1formula_meff}) reads
\beq
M_{\massless 1}^2
=  \MEFF^2 - u_T^2.
\eeq
Therefore, we obtain the correspondence
\beq
M_{\massless 1} \stackrel{u_T\to 0}{\longrightarrow} \MEFF,
\label{eq:meffinterpretation} 
\eeq
allowing us to interpret $\MEFF$ as
a suitably constructed (in the $M_{\massless 1}$ sense) 
transverse invariant mass of a singly produced,
semi-invisibly decaying parent
(see also the discussion at the end of Sec.~\ref{sec:M0N}).

The comparison between eq.~(\ref{eq:mptinterpretation})
and eq.~(\ref{eq:meffinterpretation}) rather nicely illustrates 
the main point of Sec.~\ref{sec:earlylate}:
that when it comes to transverse projections and 
forming composite particles, performing these
operations in different order yields
different results. In the case at hand, 
when forming composite particles {\em before} 
the ``$\massless$'' transverse projection, 
one obtains $2\mpt$, while by forming
composite particles {\em after} the ``$\massless$'' 
transverse projection, one obtains $\MEFF$.

\subsection{Florida $\sqrt{\hat{s}}_{\rm min}$ and $\sqrt{\hat{s}}_{\rm min}^{(\rm sub)}$ variables}
\label{sec:smin}

As already seen in eq.~(\ref{M1eqsminsub}), in
the special case of $\numparents=1$, the unprojected 
mass-bound variable $M_\numparents=M_1$ is nothing but 
the subsystem $\sqrt{\hat{s}}_{\rm min}^{(\rm sub)}$ variable from
\cite{Konar:2010ma}:
\begin{align}
M_1(\mmass_1) & \equiv \sqrt{\hat{s}}_{\rm min}^{(\rm sub)}(\mmass_1)
\nonumber \\
 & =
\left[\left( \sqrt{\comp{M}_1^2+\comp{p}_{1T}^2} 
+\sqrt{\mmass_{1}^2+\mpt^2}\right)^2-u_T^2   \right]^{1/2}  .
\label{M1eqsminsubagain}
\end{align}
Restricting to events with vanishing upstream momentum ($u_T=0$),
one gets the inclusive $\sqrt{\hat{s}}_{\rm min}$ variable from \cite{Konar:2008ei}:
\bea
\lim_{u_T\to 0}M_1(\mmass_1) &\equiv& \sqrt{\hat{s}}_{\rm min}(\mmass_1)
\nonumber \\
&=&
 \sqrt{\comp{M}_1^2+\comp{p}_{1T}^2} 
+\sqrt{\mmass_{1}^2+\mpt^2}.~~
\label{M1eqsminagain}
\eea
As advocated in Refs.~\cite{Konar:2008ei,Konar:2010ma},
practical applications of $M_1(\mmass_1)$ need not be limited
to events in which the {\em actual} number of parents was
$\numparents=1$. The work of \cite{Konar:2008ei,Konar:2010ma}
showed that in events with $\numparents=2$, the peak in
the $M_1(\mmass_1)$ distribution is correlated with 
the parent mass threshold $\sum_{a=1,2} M_{\parent a}$, even 
if the two parent particles $\parent 1$ and $\parent 2$ are different.

Note that the mathematical identity (\ref{eq:MNeqMNTagain}) 
also allows us to write
\bea
\sqrt{\hat{s}}_{\rm min}^{(\rm sub)}(\mmass_1) &=& M_{1\ourT}(\mmass_1),
\label{sminsubeqM1}
\\ [2mm]
\sqrt{\hat{s}}_{\rm min}(\mmass_1)  &=& \lim_{u_T\to 0}M_{1\ourT}(\mmass_1),
\eea
relating the $\sqrt{\hat{s}}_{\rm min}^{(\rm sub)}$ and $\sqrt{\hat{s}}_{\rm min}$
variables to the {\em transverse} invariant mass quantity $M_{1\ourT}$,
which is simply the total transverse invariant mass in the event
(after accounting for the potential presence of any transverse 
upstream momentum $u_T$).

\subsection{Transverse mass}
\label{sec:MT}

Perhaps the most popular variable which specifically targets 
a semi-invisibly decaying resonance, is the transverse mass $m_{Te\nu}$,
which, as suggested by our notation, was first applied in
searches for a leptonically decaying $W$-boson
(see, e.g.~\cite{Barger:1983wf}):
\begin{subequations}
\label{eq:MWT}
\bea
m_{Te\nu}^2 &\equiv& (e_{eT}+e_{\nu T})^2 - (\vec{p}_{eT}+\vec{q}_{\nu T})^2 
\label{eq:MWTdef}
\\ [2mm]
&\approx& 2( |\vec{p}_{eT}||\vec{q}_{\nu T}| - \vec{p}_{eT}\cdot\vec{q}_{\nu T}),
\label{eq:MWTapprox}
\eea
\end{subequations}
where $\vec{p}_{eT}$ ($\vec{q}_{eT}$) is the transverse momentum of the 
lepton (neutrino), and in the second line one makes
the approximation that the lepton and the neutrino are approximately 
massless. Assuming that the $W$ boson is produced singly, 
with zero recoil (i.e. $u_T=0$ in our language), 
the neutrino transverse momentum $\vec{q}_{eT}$
can be identified with the measured missing transverse momentum 
$\mptvec$, and (\ref{eq:MWTapprox}) becomes
\beq
m_{Te\nu}^2 \approx 2 p_{eT} \mpt \left(1-\cos\phi_{e\nu}\right),
\eeq
where $\phi_{e\nu}$ is the measured opening angle between 
the transverse vectors $\vec{p}_{eT}$ and $\mptvec$.

In this simple example of a $W$-decay, 
the two daughter particles are massless, but the same idea can be
easily generalized to the case of massive daughters as
\cite{Nakamura:2010zzi}
\beq
m_{T e\nu}^2(M_e,M_\nu)  =
M_e^2 + M_\nu^2 + 2( e_{eT} e_{\nu T} - \vec{p}_{eT}\cdot\vec{q}_{\nu T}),
\label{eq:mTenumassive}
\eeq
where $M_e$ and $M_\nu$ are the electron and neutrino masses, respectively,
and
\begin{subequations}
\label{electronneutrinoet}
\bea
e_{eT} &\equiv& \sqrt{M_e^2 + \vec{p}_{eT}^{\, 2}}\, ,
\label{electroneT} \\ [2mm]
e_{\nu T} &\equiv& \sqrt{M_\nu^2 + \vec{q}_{\nu T}^{\, 2}}\, .
\label{neutrinoeT}
\eea
\end{subequations}

Now let us obtain these results with our formalism.
In general, we have a singly produced ($\numparents=1$) 
parent resonance, which decays to a single ($N_{\cal V}=1$)
visible daughter particle and a single ($N_{\cal I}=1$)
invisible daughter particle. Since there is only one particle in each 
daughter set, ${\cal V}_1=\ourSet{e}$ and  ${\cal I}_1=\ourSet{\nu}$,
there is no need to form composite particles, so the order of
the operations becomes unimportant. However, if the 
daughter particles are massive, the two different types
of transverse projections give two different versions
of the transverse mass variable:
\begin{align}
M_{1\ourT}^2(M_{\nu})&=M_{\ourT 1}^2(M_{\nu})   \nonumber   \\
&=\left( \sqrt{M_e^2+\vec{p}_{eT}^{\,2}} 
+\sqrt{M_\nu^2+\vec{q}_{\nu T}^{\,2}}\right)^2-\vec{u}_T^{\,2}, 
\label{eq:transversemassT}  \\ 
M_{1\massless}^2&=M_{\massless 1}^2= 
\left(p_{eT}+q_{\nu T}\right)^2-\vec{u}_T^{\,2}.  \label{eq:transversemass0}
\end{align}
Here eq.~(\ref{eq:transversemassT}) follows simply from
the general formulas 
(\ref{eq:M1ourTformula}) or (\ref{eq:MourT1formula}) 
with the identifications $\comp{M}_1=M_1=M_e$,
$\vec{\comp{p}}_{1T}=\vec{p}_{1T}=\vec{p}_{eT}$,
$\mmass_1 = M_\nu$ and $\mptvec=\vec{q}_{\nu T}$.
Similarly, eq.~(\ref{eq:transversemass0})
is obtained from either (\ref{Mmassless1formula_general}) 
or (\ref{M1masslessformula_general}).

Now it is trivial to eliminate $\vec{u}_T$ using
the transverse momentum relation (\ref{eq:theetmissassumption})
$\vec{u}_T=-\vec{p}_{eT}-\vec{q}_{\nu T}$, 
and show that eq.~(\ref{eq:transversemassT})
is equivalent to (\ref{eq:mTenumassive}):
\beq
M_{1\ourT}(M_{\nu})=M_{\ourT 1}(M_{\nu})= m_{T e\nu}(M_e,M_\nu)
\label{eq:M1TeqMT1eqMTenu}
\eeq
while eq.~(\ref{eq:transversemass0})
is equivalent to (\ref{eq:MWTapprox}):
\beq
M_{1\massless}=M_{\massless 1}=m_{Te\nu}(M_e=0,M_\nu=0).
\eeq

\subsection{Cluster transverse mass variables}
\label{sec:cluster}

Next we consider a couple of more complicated single resonance 
processes. The first example is $h\to ZZ\to e^+e^- \nu\bar{\nu}$
where each $Z$-boson is assumed to be on-shell, one decaying invisibly, 
the other decaying visibly to a pair of leptons. 
For this particular scenario, Ref.~\cite{Barger:1987du} 
suggested the cluster transverse mass variable
\begin{subequations}
\begin{align}
M^2_{T,ZZ}&=
 \left(E_{T,Z_1}+E_{T,Z_2}\right)^2
-\left(\vec{p}_{T,Z_1}+\vec{p}_{T,Z_2}\right)^2   \\
\label{MTZZ}
\begin{split}
&= 
\left(\sqrt{M_Z^2+p_{T,e^+e^-}^2}+\sqrt{M_Z^2+\mpt^2}\right)^2
\\
&\qquad\qquad - \left(\vec{p}_{T,e^+e^-}+\mptvec\right)^2.
\end{split}
\end{align}
\end{subequations}

Note that the 1+3-dimenstional invariant mass of
the visible and that of the invisible systems
have each been constrained to be equal to $M_Z^2$.
These two constraints reflect the on-shell hypothesis we 
have chosen to assume for each of the two $Z$ bosons.\footnote{We note that if one wishes to relax the assumption 
of an on-shell $Z$ leading to the visible $e^+e^-$ system one may do so by treating the 
electron and positron vectors as separate inputs to the visible system {\cal V}.
Similarly one may relax the assumption that the invisible system
is the result of the decay of an on-shell $Z$ by treating the neutrinos
as independent invisible inputs.
}

Once again, we can obtain this variable from our
$M_{1\ourT}$ or $M_{\ourT 1}$. In analogy to
the case of $m_{Te\nu}$, we have a single parent 
resonance, the Higgs boson $h$, decaying to a single
massive visible daughter, the first $Z$ boson 
and a single massive invisible particle, the other $Z$-boson.
This corresponds to $\numparents=1$, 
$N_{\cal V}=1$, $N_{\cal I}=1$, 
${\cal V}_1=\ourSet{Z\to e^+e^-}$ and 
${\cal I}_1=\ourSet{Z\to\nu\bar{\nu}}$.
Correspondingly, we identify
$\comp{M}_1=M_1=M_Z$,
$\vec{\comp{p}}_{1T}=\vec{p}_{1T}=\vec{p}_{T,e^+e^-}$,
$\mmass_1 = M_Z$ and $\mptvec=\vec{q}_{T,\nu\bar{\nu}}$.
Then (\ref{eq:M1ourTformula}) and (\ref{eq:MourT1formula}) 
simply give
\begin{align}
M_{1\ourT}^2(M_{Z})&=M_{\ourT 1}^2(M_{Z})
\nonumber \\
&= \left( \sqrt{M_Z^2+\vec{p}_{T,e^+e^-}^{\,2}} 
+\sqrt{M_Z^2+\mpt}\right)^2 -\vec{u}_T^{\,2},
\label{eq:MTZZours}
\end{align}
which is equivalent to (\ref{MTZZ}) in light of the 
momentum conservation relation 
$\vec{u}_T=-\vec{p}_{T,e^+e^-}-\mptvec$.
Thus we have proved
\beq
M_{1\ourT}(M_{Z}) = M_{\ourT 1}(M_{Z}) = M_{T,ZZ}.
\label{eq:M1TeqMT1eqMTZZ}
\eeq

Another interesting example is provided by the process
$h\to W^+W^-\to e^+e^-\nu\bar{\nu}$, for which 
Ref.~\cite{Barger:1987re} proposed the
cluster transverse mass variable 
\beq
M_{C,WW}^2\equiv
\left(\sqrt{M_{e^+e^-}^2+\vec{p}_{T,e^+e^-}^{\, 2}}+\mpt\right)^2-
\left(\vec{p}_{T,e^+e^-} + \mptvec \right)^2.
\label{MCWW}
\eeq
Here the two leptons are clustered together (even though 
they originate from different $W$-bosons, they have a common 
parent in $h$) and their total transverse momentum is 
$\vec{p}_{T,e^+e^-}$. The definition
(\ref{MCWW}) is similar to (\ref{MTZZ}), 
the difference now being that the two leptons are not correlated,
and their invariant mass does not have to be consistent with $M_Z$.
In addition, the invisible mass parameter $\mmass_1$ 
is now set to zero (as opposed to $M_Z$), because 
the invisible particles (the two neutrinos) are massless.

The cluster variable (\ref{MCWW}) can be readily 
obtained from $M_{N\ourT}$
with the following interpretation: $\numparents=1$,
$N_{\cal V}=2$, $N_{\cal I}=2$, 
${\cal V}_1=\ourSet{e^+,e^-}$ and 
${\cal I}_1=\ourSet{\nu,\bar{\nu}}$.
Correspondingly, we identify
$\comp{M}_1=M_{e^+e^-}$,
$\vec{\comp{p}}_{1T}=\vec{p}_{T,e^+e^-}$
and $\mmass_1 = 2M_\nu=0$.
Then the general formula (\ref{eq:M1ourTformula})
reduces to
\beq
M_{1\ourT}^2(0)= 
\left( \sqrt{M_{e^+e^-}^2+\vec{p}_{T,e^+e^-}^{\,2}} 
+\mpt\right)^2 -\vec{u}_T^{\,2},
\label{eq:MCWWours}
\eeq
which is the same as (\ref{MCWW}), so that
\beq
M_{1\ourT}(0)=M_{C,WW}.
\label{eq:M1TeqMCWW}
\eeq
Notice that in this example we are clustering two visible particles,
and the order of operations becomes important.
Therefore, here $M_{1\ourT}$ and $M_{\ourT 1}$ in general 
lead to distinct variables, unlike the case of 
(\ref{eq:M1TeqMT1eqMTenu}) and (\ref{eq:M1TeqMT1eqMTZZ}).

\subsection{The $m_T^{\rm true}$ transverse mass variable}

Concerning the same $h\to W^+W^-\to e^+e^-\nu\bar{\nu}$ example,
Ref.~\cite{Barr:2009mx} advertized the variable
(assuming massless neutrinos)
\begin{equation}
\begin{split}
\left(m_T^{\rm true}\right)^2
\equiv
M_{e^+e^-}^2 
+ 2
\Bigl(
\mpt \sqrt{M_{e^+e^-}^2+\vec{p}_{T,e^+e^-}^{\,2}}    
\Bigr.
\\ 
\Bigl.
-\mptvec\cdot \vec{p}_{T,e^+e^-} 
\Bigr)
,
\end{split}
\label{eq:mTtruedef}
\end{equation}
which can be rewritten as
\bea
&&\left(m_T^{\rm true}\right)^2
=
M_{e^+e^-}^2 + \vec{p}_{T,e^+e^-}^{\,2} + \mptvec^{\,2}
             - \vec{p}_{T,e^+e^-}^{\,2} - \mptvec^{\,2}
\nonumber \\ [2mm]
&&\qquad\qquad  +2\mpt\sqrt{M_{e^+e^-}^2+\vec{p}_{T,e^+e^-}^{\,2}} 
- 2 \mptvec\cdot \vec{p}_{T,e^+e^-}
\nonumber \\ [2mm]
&&\quad = 
\left(\sqrt{M_{e^+e^-}^2+\vec{p}_{T,e^+e^-}^{\, 2}}+\mpt\right)^2-
\left(\vec{p}_{T,e^+e^-} + \mptvec \right)^2
\nonumber \\ [2mm]
&&\quad \equiv M_{C,WW}^2 \, .
\label{eq:MTtrueeqMCWW}
\eea
From (\ref{eq:M1TeqMCWW}) and (\ref{eq:MTtrueeqMCWW}) 
it now follows that
\beq
M_{1\ourT}(0) = m_T^{\rm true}.
\eeq
This connection in fact was the primary motivation for
introducing the $m_T^{\rm true}$ variable in the first place
\cite{Barr:2009mx}.

\subsection{The $m_{T Z'}^{reco}$ transverse mass variable}
\label{sec:mTZprime}

Our final single resonance example will be taken from 
a new physics scenario, namely a generic model with 
a new $Z'$ gauge boson which decays to a SM Higgs boson $h$ 
and a SM $Z$-boson as $Z'\to hZ\to b\bar{b}\nu\bar{\nu}$.
For this particular topology, Ref.~\cite{Katz:2010mr}
considered the transverse mass variable
\beq
m_{TZ'}^{reco}\equiv \sqrt{M_h^2+p^2_{Th}}+\sqrt{M_Z^2+\mpt^{\,2}},
\label{eq:mTZprimedef}
\eeq
where $M_h$ ($p_{Th}$) is the measured invariant mass 
(transverse momentum) of the $b\bar{b}$ jet pair resulting from the
decay $h\to b\bar{b}$.

In our language, the event topology $Z'\to hZ\to b\bar{b}\nu\bar{\nu}$
corresponds to a single parent, the $Z'$ boson, thus $\numparents=1$.
There is a single ($N_{\cal V}=1$) visible daughter particle,
which is the reconstructed Higgs boson: ${\cal V}_1=\ourSet{h\to b\bar{b}}$.
There is also a single ($N_{\cal I}=1$) invisible daughter particle, 
which is the invisibly decaying $Z$-boson: ${\cal I}_1=\ourSet{Z\to \nu\bar{\nu}}$.
Thus we identify $\comp{M}_1=M_1=M_h$,
$\vec{\comp{p}}_{1T}=\vec{p}_{1T}=\vec{p}_{Th}$ and
$\mmass_1 = M_Z$.

Again, we have chosen to make assumptions
about the 1+3-dimensional invariant masses of the visible and the invisible systems, 
requiring the former to be equal to $M_h$, and the latter
to be equal to $M_Z$, reflecting our assumptions about the decay topology.
As before it would be possible to independently relax either of both those assumptions
by treating the  $b\bar{b}$ and $\nu\bar{\nu}$ as independent inputs to the 
visible and invisible systems respectively.

If we retain the mass-shell constraint for both the $h$ boson and the $Z$ boson then
(\ref{eq:M1ourTformula}) and (\ref{eq:MourT1formula}) 
give
\begin{align}
M_{1\ourT}^2(M_{Z})&=M_{\ourT 1}^2(M_{Z})
\nonumber \\ 
&= \left( \sqrt{M_h^2+\vec{p}_{Th}^{\,2}} 
+\sqrt{M_Z^2+\mpt}\right)^2 -\vec{u}_T^{\,2}.
\label{eq:mTZprimeours}
\end{align}
Comparing to (\ref{eq:mTZprimedef}), we see that
\beq
\lim_{u_T\to 0} M_{1\ourT}(M_{Z})
= \lim_{u_T\to 0} M_{\ourT 1}(M_{Z})
= m_{TZ'}^{reco}.
\label{eq:M1TismTZprime}
\eeq
Since $m_{TZ'}^{reco}$ was properly defined as a transverse mass
variable, it is not surprising that it can be obtained
as a special case of the mass-bounding variables shown in Table~\ref{tab:floxbridge_basic}.
The importance of eq.~(\ref{eq:mTZprimeours})
is that it shows the proper way to generalize $m_{TZ'}^{reco}$
to the case where the $Z'$ is produced {\em inclusively},
with some non-vanishing UVM $u_T$ in the event.

This concludes our discussion of singly produced resonances.
We are hopeful that after all these examples, the reader
is prepared to handle any assumed event topology, and
will be able to construct the proper transverse invariant 
mass variable for the case at hand.

\subsection{Cambridge $m_{T2}$ variable}
\label{sec:mt2}

The variables considered in our previous examples 
referred either to the event as a whole 
(as in Secs.~\ref{sec:met}--\ref{sec:smin})
or to the production of a single resonance
(as in Secs.~\ref{sec:MT}--\ref{sec:mTZprime}).
We now move on to discussing variables 
intended to handle the production of more than one
parent resonance ($\numparents>1$).
Such cases are very common in new physics scenarios, 
especially if the new model contains a dark matter candidate, 
whose lifetime is protected by some discrete symmetry 
(typically a $Z_2$). In such models (e.g. supersymmetry, 
extra dimensions, little Higgs theories etc.)
the main production mechanisms usually involve 
the {\em pair-production} of new particles, thus
the case of $\numparents=2$ has received the most 
attention so far in the literature, although
$\numparents=3,4,\ldots$ cannot be ruled out, and
in principle deserve attention as well.

A popular variable of this type is the
Cambridge $m_{T2}$ variable defined as \cite{Lester:1999tx}
\beq
m_{T2}\equiv 
\min_{\substack{
\sum \vec{q}_{iT} = \mptvec}} 
\ourMaxMinBracs{\max \ourMaxMinBracs{{\cal M}_{1\ourT},{\cal M}_{2\ourT} 
}}\ ,
\label{eq:mt2def}
\eeq
where ${\cal M}_{1\ourT}$ and ${\cal M}_{2\ourT}$ are 
the transverse masses of the two parent particles,
and the minimization is done over all possible partitions 
of the transverse momenta of the invisible particles, 
consistent with the measured $\mpt$.

We note that if there is only one visible particle belonging to each parent, 
we can immediately identify the Cambridge variable $m_{T2}$ with both 
$M_{2\ourT}$ and $M_{\ourT 2}$ (and even with $M_2$ using (\ref{MNeqMNT})) 
since partitioning and projection commute for single particles.  If, however, 
we intend to apply $m_{T2}$ to events in which one or either parent has two or 
more physical daughters (e.g.~when doing top quark mass measurements in the 
di-leptonic $t\bar t \rightarrow b\bar b l^+ l^- \nu \bar \nu $ events) then  
$M_{2\ourT}$  will become inequivalent to $M_{\ourT 2}$ and we should decide 
which of these is the right thing to use. The answer to this question is subtle.
The original $m_{T2}$ paper \cite{Lester:1999tx} does not explicitly state 
how parent momenta should be constructed in the event that they have come 
from compound objects, so it is left up to users to decide which inputs to supply. 
It was certainly in the minds of the authors of \cite{Lester:1999tx} that 
users ought always to supply the maximal amount of trustworthy information 
to any analysis of any kind.  In the context of $m_{T2}$ this maxim would 
imply projecting only {\em after} combining the primary (1+3) momenta of 
any constituents of parents, provided that those constituents could be ``trusted''.\footnote{This 
might include the case where an experiment that records di-leptonic top-pairs 
with good (signed) $b$-tagging could allow the $l^+$ to be associated 
unambiguously with the $b$ and the $l^-$ to be associated with the $\bar b$.}  
Only when defined in this manner (i.e.~as $M_{2\ourT}$) can the maximum 
amount of information be squeezed from the variable in ``clean'' events.  
However, there can be benefits from using $M_{\ourT 2}$ (see for example 
Ref~\cite{Lester:2007fq})
in high-multiplicity or inclusive situations in which the indiviual momenta 
making up each parent have dubiuos provenance or poorly measured longitudinal 
momenta.  In such cases, one can benefit from using $M_{\ourT 2}$, even though 
its end point is less sharp for the signal, simply because it is less 
sensitive to longitudinal momenta and momenta at high rapidities. 

One might ask which of the two $m_{T2}$ choices --- $M_{2\ourT}$ or $M_{\ourT 2}$ ---
is ``better''. Unfortunately,
this question does not allow a ``one size fits all'' answer.
Each of the two $m_{T2}$ implementations has its unique 
advantages and disadvantages. 
The longitudinal correlations among the visible particles which are preserved by $M_{2 \ourT}$
result in
steeper, better defined endpoint structures --- 
see Figs.~\ref{fig:hWW}(a) and \ref{fig:ttb}(a) below.
On the other hand, $M_{\ourT 2}$ dampens the effects 
of any longitudinal momenta, which would be beneficial
in circumstances where forward jet activity due to 
ISR may be a problem.

We hope that the current paper will serve as a reminder that studies
using $m_{T2}$ in which either parent is built from two or more reconstructed
momenta should think carefully about the advantages and disadvantages
of both approaches before choosing the option that is best for them.  Both versions of
$m_{T2}$, namely $M_{2 \ourT}$ and $M_{\ourT 2}$, may prove to be
useful, and it can be important to make the
distinction between them.

In conclusion of this subsection, we 
highlight the analogy between $m_{T2}$ 
and $\sqrt{\hat{s}}_{\rm min}$, two variables 
which are more closely related than one might think.
We have shown that the (1+3) dimensional version of $m_{T2}$, together with
the mathematical identity (\ref{eq:MNeqMNTagain}) implies
\beq
m_{T2}^{(1+3)}(
\chiM
) \equiv M_{2 \ourT}(
\chiM
) = M_{2}(
\chiM
).
\label{mt2eqM2}
\eeq
The second equality here emphasizes that, in spite of the transverse index ``T'',
the $m_{T2}$ variable is a bona fide
1+3 dimensional quantity. In other words, 
the apparent transverse projection in the definition
(\ref{eq:mt2def}) does not lead to any loss of useful information.\footnote{This 
fact is found to be suprising to people who view $m_{T2}$ and similar 
variables as acting on ``projected'' quantities.  On the other hand it 
is no surprise to those who have always viewed  $m_{T2}$ as a variable 
insensitive to relative rapidity differences beteen the (total) invisible 
and (total) visible decay products of each parent --- the line taken 
in \cite{Lester:2007fq}.} 
Of course, the same cannot be said about the (1+2)-dimensional
version $m_{T2}^{(1+2)}\equiv M_{\ourT 2}$.

Now compare (\ref{mt2eqM2}) to the analogous equation following from
(\ref{M1eqsminsubagain}) and (\ref{sminsubeqM1})
\beq
\sqrt{\hat{s}}_{\rm min}^{(\rm sub)}(\mmass_1)
= M_{1\ourT}(\mmass_1)
= M_1(\mmass_1).
\label{smineqM1eqM1T}
\eeq
We are reminded that $\sqrt{\hat{s}}_{\rm min}^{(\rm sub)}$ and $m_{T2}^{(1+3)}$ have
essentially the same physical meaning: they both give a lower bound on a mass {\bf in (1+3) dimensions}
as a function of the corresponding invisible mass parameters.  
In the case of $\sqrt{\hat{s}}_{\rm min}^{(\rm sub)}$ that mass is the center
of mass energy of the collision since it views the whole collision as 
a ``single parent'', while $m_{T2}^{(1+3)}$ hypothesizes that the collision 
was a $2\rightarrow 2$ process, and therefore bounds the mass of the 
heavier of the two outgoing particles. 

\subsection{The doubly projected variables $m_{T2\perp}$ and $m_{T2\parallel}$}

The doubly projected variables $m_{T2\perp}$ and $m_{T2\parallel}$ 
introduced in \cite{Konar:2009wn} are nothing but the one-dimensional 
analogues of the Cambridge variable $m_{T2}$ (\ref{eq:mt2def}),
where one performs an additional projection on the directions 
$T_\myT$ and $T_\myL$, correspondingly.
As was the case for $m_{T2}$ discussed above,
the transverse projection should be interpreted in the $T=\ourT$ 
sense, and there are three possible versions of each variable, 
depending on the order of operations:
\beq
m_{T2\perp}(\chiM) = 
\left\{
\begin{array}{l}
M_{2 \ourT \myT}(\chiM) \quad {\rm in\ 1+3\ dims;} 
\\ [3mm]
M_{\ourT 2 \myT}(\chiM) \quad {\rm in\ 1+2\ dims;} 
\\ [3mm]
M_{\ourT \myT 2}(\chiM) \quad {\rm in\ 1+1\ dims,} 
\end{array}
\right.
\label{eq:mt2perpinterpretation}
\eeq
and similarly for $m_{T2\parallel}(\chiM)$.
The example considered in \cite{Konar:2009wn}
was inclusive chargino production, where each chargino parent
decays to a visible lepton and an invisibly decaying sneutrino.
In this case, each visible daughter partition 
${\cal V}_a$ has a single massless visible particle
(a lepton) and the distinction between 
the early partitioned version $m_{T2\perp}^{(1+3)}\equiv M_{2 \ourT \myT}$, 
the in-between partitioned version $m_{T2\perp}^{(1+2)}\equiv M_{\ourT 2 \myT}$ and
the late partitioned version $m_{T2\perp}^{(1+1)}\equiv M_{\ourT \myT 2}$
does not become manifest. However, in more complicated 
scenarios with multiple visible daughter particles, 
one would in principle obtain different results from
$m_{T2\perp}^{(1+3)}$, $m_{T2\perp}^{(1+2)}$ and
$m_{T2\perp}^{(1+1)}$, which is why one is advised to
carefully define which particular version 
of $m_{T2\perp}$ (and similarly for $m_{T2\parallel}$) is being used.

\subsection{Additionally constrained variables}

So far we have been discussing very general variables, 
which target the most general event topology
of Fig.~\ref{fig:event}. Notice that we have made very few
assumptions on how the decays (\ref{eq:Pa_decay})
actually take place,
and where such assumptions have been made (such as in Section~\ref{sec:cluster}),
it is always possible to relax those constraints if desired.
Also we did not use any additional information which may be available 
from the preliminary studies of other variables
related to our events, for example the invariant 
mass distributions of the visible daughter collections
${\cal V}_a$. Armed with such additional information, 
one may in principle further constrain the minimization 
over the unknown momenta, and obtain new, more specialized 
versions of our variables. However, the downside is that
such additional information typically comes at a cost:
the need to make additional assumptions about the 
event topology.

As an example, consider the Oxford $M_{2C}$ variable
\cite{Ross:2007rm,Barr:2008ba}, which is a variant of $m_{T2}$,
subject to the following additional assumptions:
\begin{enumerate}
\item The two parents $\parent 1$ and $\parent 2$ are identical,
with {\it a priori} unknown mass $M_{\parent 1}=M_{\parent 2}$,
therefore the minimization over invisible momenta is performed subject 
to the additional constraint
\beq
\left(\comp{P}_1+\comp{Q}_1 \right)^2 = \left(\comp{P}_2+\comp{Q}_2 \right)^2 
\label{eq:equalparents}
\eeq
with $\comp{P}_a$ and $\comp{Q}_a$ given by (\ref{eq:defPa}) and 
(\ref{eq:defQa}) correspondingly.
\item There is only one and only one invisible particle in each
invisible daughter set, i.e. that
$$
\left|{\cal I}_1\right|=\left|{\cal I}_2\right|=1.
$$
\item There is more than one visible particle in each visible 
daughter set, i.e. that
$$
\left|{\cal V}_1\right|=\left|{\cal V}_2\right|\ge 2.
$$
\item There are no intermediate on-shell resonances, so that
the decay (\ref{eq:Pa_decay}) is effectively
$(|{\cal V}_a|+|{\cal I}_a|)$-body for each $a=1,2$.
\end{enumerate}
Given a large sample of events $\ourevents$ that satisfy these conditions
one can study the distribution of the invariant mass $M_{{\cal V}_a}$ of the 
visible particles in each set ${\cal V}_a$.
If presented with some sufficiently large sample of events 
$\ourevents=\ourSet{\event\mid\event\in\assumed}$
one can measure the upper bound for the $M_{{\cal V}_a}$
distribution which will be found at the mass difference 
between the parent and the single invisible daughter:
\beq
\Delta^{\rm max} \equiv \max_\mathrm{events}\ourMaxMinBracs{ M_{{\cal V}_a}} 
= M_{\parent a} - \mmass_a 
\label{eq:thatisperhapswrong}
\eeq
One can then reuse this measurement for any event $\event\in\ourevents$
by asserting the constraint
\beq
\Delta^{\rm max} = \sqrt{(\comp{P}_a+\comp{Q}_a)^2} - \sqrt{\comp{Q}_a^2},
\eeq
during the process of minimization over $\comp{Q}_a$. 

The advantage of such additionally constrained variables is that they are
clearly better adapted for the study of the corresponding class
of more restricted event topologies. 
And additional constraints can bring qualitatively new features, including 
otherwise unobtained {\em upper} bounds
on parental masses~\cite{Barr:2008ba,Barr:2008hv}.
Their disadvantage is that they are better adapted 
(perhaps only suitable) for the study of those restricted topologies, 
and it is not clear how to interpret them once some of the 
assumptions hardwired in their definitions cease to be valid.

\subsection{Other variables}

In the literature one may sometimes encounter variables
which have the appearance of a (transverse) invariant mass, 
but cannot be related to any of our variables in 
Table~\ref{tab:floxbridge_basic}. As an illustrative example, 
consider the transverse mass variable 
\begin{align}
\begin{split}
M_{T_{WW}}^2 & \equiv 
\left(\sqrt{M_{e^+e^-}^2+\vec{p}_{T,e^+e^-}^{ 2}} \right.\\
&{}\ \ +\left.\sqrt{M_{e^+e^-}^2+\mpt^{ 2}}    \right)^2
-
\left(\vec{p}_{T,e^+e^-} + \mptvec \right)^2,
\end{split}
\label{MTWW}
\end{align}
proposed in Ref.~\cite{Rainwater:1999sd} 
in relation to the $h\to W^+W^-\to e^+e^-\nu\bar{\nu}$
process discussed in Sec.~\ref{sec:cluster}.
Comparing to the definition (\ref{MCWW}) of $M_{C,WW}$
and to the identity (\ref{eq:MCWWours}),
we see that $M_{T_{WW}}$ can be formally obtained from
$M_{C,WW}=M_{1\ourT}(0)$ with the rather ad hoc replacement 
\beq
\mmass_1=0 \quad \to \quad \mmass_1 = M_{e^+e^-}.
\eeq
However, there is no good physics justification for this 
conjecture and as a result, $M_{T_{WW}}$ cannot be related to 
any of the variables in Table~\ref{tab:floxbridge_basic}.
Not surprisingly, subsequent studies \cite{Barr:2009mx} 
found that $M_{C,WW}=M_{1\ourT}(0)$ outperforms $M_{T_{WW}}$.

\section{Simulation: physics examples}
\label{sec:examples}

In this section we provide an illustration of our previous discussion
with two specific physics examples from the Standard Model:
\begin{itemize}
\item {\em A case with $N=1$.} Here we consider the inclusive 
(single) production of a SM Higgs boson (mostly from gluon fusion), 
followed by the decay of the Higgs to a leptonic $W$-pair:
\begin{align}
\begin{split}
p p &\to h+X \\ &\to
W^+W^-+X \\&\to 
\ell^+\ell^-+\mpt+X,
\end{split}
\label{eq:hdecays}
\end{align}
where $X$ plays the role of UVM and stands for
jets from initial state radiation, 
unclustered hadronic energy, etc. 
In terms of our previous notation, this case 
involves one parent ($N=1$), two visible particles 
($N_{\cal V}=2$) and two invisible particles 
($N_{\cal I}=2$).
\item {\em A case with $N=2$.} Here we consider dilepton 
events from inclusive $t\bar{t}$ pair production, where 
both $W$'s decay leptonically:
\begin{align}
\begin{split}
p p &\to t\bar{t}+X \\ &\to
b\bar{b}W^+W^-+X \\ &\to 
b\bar{b}\ell^+\ell^-+\mpt+X.
\end{split}
\label{eq:ttdecays}
\end{align}
This case corresponds to two parents ($N=2$), four visible particles 
($N_{\cal V}=4$) and two invisible particles 
($N_{\cal I}=2$).
\end{itemize}
In both of those two cases, the events very closely 
resemble the typical SUSY-like events, in which there 
are two missing dark matter particles. Parton-level
event simulation is performed with PYTHIA \cite{Sjostrand:2006za}
at an LHC of 7 TeV, including the effects from the underlying event
(using PYTHIA's default model for it).

\subsection{An $\numparents=1$ example: a Higgs resonance}
\label{sec:higgs}

\begin{figure*}[t]
\begin{center}
\includegraphics[width=0.45\linewidth]{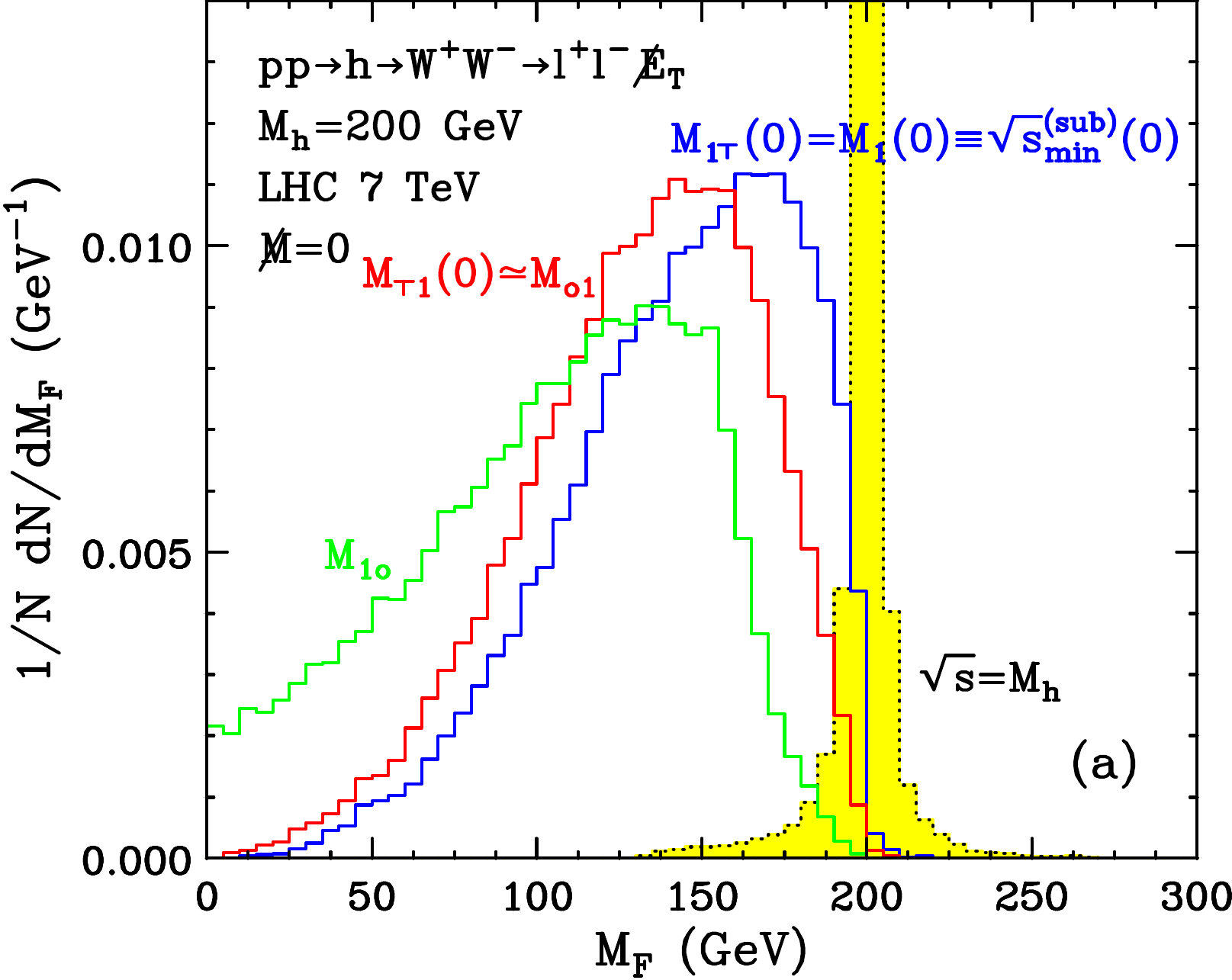} 
\hspace{0.05\linewidth}
\includegraphics[width=0.45\linewidth]{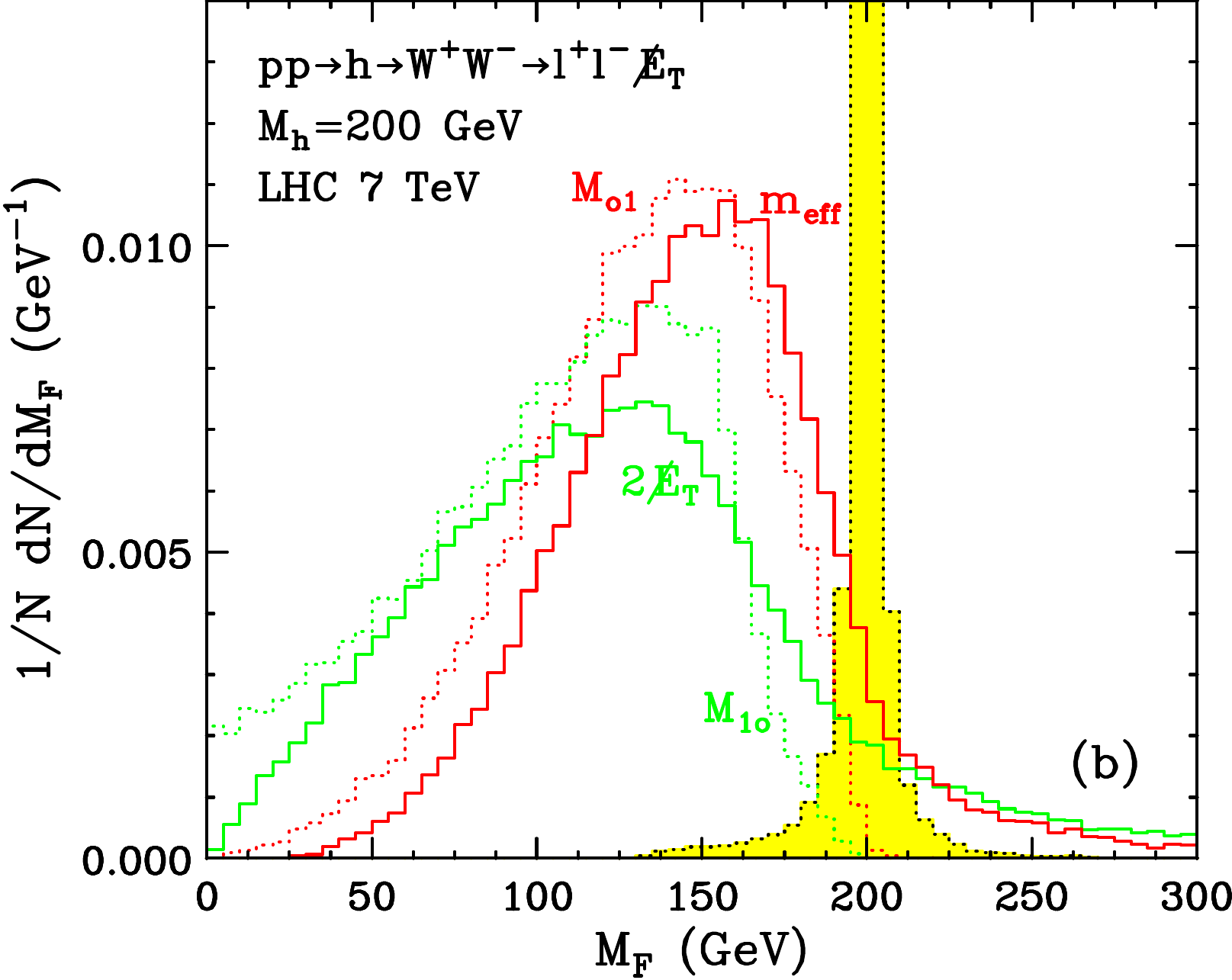} 
\end{center}
\caption[Distributions of mass variables for H->WW]{\label{fig:hWW}
(a) Unit-normalized distribution of the five $\numparents=1$ mass-bound variables
$M_{\flox}$, ${\flox}\in \{1, 1\ourT, \ourT 1, 1\massless, \massless 1\}$
for the inclusive Higgs production process
$h\to W^+W^-\to \ell^+\ell^-+\mpt$ at a 7 TeV LHC, with $m_h=200$ GeV
and $\mmass=0$.
The dotted (yellow-shaded) histogram gives the true $\sqrt{\hat{s}}$
distribution, which in this case is given by the Breit-Wigner 
$h$ resonance. (b) Unit normalized distributions of the 
variables $\MEFF$ and $2\mpt=2\mpt$ (solid lines), contrasted with 
$M_{\massless 1}$ and $M_{1\massless}$ (dotted lines).
}
\end{figure*}

We start with the Higgs production process (\ref{eq:hdecays})
for a Higgs boson mass $M_h=200$ GeV. In the language of Fig.~\ref{fig:event},
the Higgs resonance is treated as the only heavy parent particle ($\numparents=1$)
and the event is partitioned as
\beas
{\cal V}_1 &=& \ourSet{\ell^+,\ell^- },  \\ [2mm]
{\cal I}_1 &=& \ourSet{\nu_\ell,\bar{\nu}_\ell }.  
\eeas
This partitioning is pictorially represented in Fig.~\ref{fig:countingParticles1parent}.
We now concentrate on the five unprojected or singly projected 
variables which are of interest to us,
namely $M_{\flox}$ with ${\flox}\in \ourSet{1, 1\ourT, \ourT 1, 1\massless, \massless 1}$. 
Their distributions are shown in Fig.~\ref{fig:hWW}(a),
and for proper comparison, we use the correct 
value of the missing mass parameter $\mmass_1=0$ where necessary.
In that case, according to the general property (\ref{eq:endpoint}),
all $M_{\flox}$ variables are bounded from above by the
parent mass, in this case $M_h$.
For reference, Fig.~\ref{fig:hWW}(a) also shows
the Breit-Wigner distribution of the Higgs resonance
(yellow-shaded histogram). Fig.~\ref{fig:hWW}(a)
confirms that the distributions obey the bound of eq.~(\ref{eq:endpoint}).
Furthermore, it also shows that each
of the five distributions 
appears to be saturated -- i.e. that each has a kinematic 
endpoint at the value of the Higgs boson mass $M_h$
(only a very tiny fraction of events is observed to exceed
the bound, but this is due to the finite width of the Higgs parent).

\begin{figure}[tbh]
\begin{center}
\setlength{\unitlength}{1mm}
\begin{picture}(30,16)(-15,-10)
\linethickness{1pt}
\put(0,4){\makebox(0,0){$h$}}
\put(0,0){\circle{3}}
\put(0,0){\vector(1,0){5}}
\put(5,-5){\circle{2}}
\put(-5,-9){\circle{2}}
\put(8,1){\makebox(1,0){$W^+$}}
\put(0,0){\vector(-1,0){5}}
\put(-8,1){\makebox(1,0){$W^-$}}
\put(6,-5){\vector(-1,0){8}}
\put(16,-5){\makebox(1,0){$\ell^+$}}
\put(6,-5){\vector(1,0){8}}
\put(-4,-5){\makebox(1,0){$\nu$}}
\put(-6,-9){\vector(-1,0){8}}
\put(-16,-9){\makebox(1,0){$\bar{\nu}$}}
\put(-6,-9){\vector(1,0){8}}
\put(4,-9){\makebox(1,0){$\ell^-$}}
\end{picture}
\caption{\label{fig:configuration1}
Illustration of an example extremal configuration for the $N=1$ variables
when applied to the $h$ example.}
\end{center}
\end{figure}
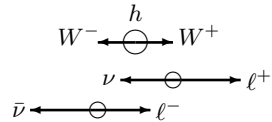

We can confirm the endpoint is saturated for each of the variables
$M_\flox\in\ourSet{1, 1\ourT, \ourT 1, 1\massless, \massless 1}$
by explicitly constructing an extremal event.
We do so under the approximation that $M_\ell = M_\nu = 0$.
Following the arguments of Sec.~\ref{sec:endpoints} 
we should construct an extremal event $\event$ from
the subset $\allevents$ sampled by nature of the total set of 
momentum configurations $\assumed$ that satisfy our general $N=1$, $M_{\parent 1}=M_H$ topology.
In the case of the decay of interest~\eqref{eq:hdecays}
nature obliges us to impose an additional on-mass-shell condition 
for the intermediate $W^\pm$ bosons
$\allevents=\ourSet{a\in\assumed \mid (P_i+Q_i)^2=M_W^2\ \left(i\in\{1,2\}\right)}$.\footnote{We 
assume the $W$ to have narrow widths. In fact since we can construct an 
extremal event while imposing a strict on-shell requirement for the 
intermediate $W^\pm$ then we can certainly also do so when that requirement 
is relaxed by allowing the $W$ to sample from its natural width distribution.}
An example of an extremal event $\event\in\allevents$
that satisfies the constraints and that also saturates 
the two inequalities of \eqref{eq:saturation} is (see also Fig.~\ref{fig:configuration1})
\beas
P_{\ell^+}&=&(E_1,E_1,0,0)\\
P_{\ell^-}&=&(E_2,E_2,0,0)\\
Q_{\nu}&=&(E_2,-E_2,0,0)\\
Q_{\bar\nu}&=&(E_1,-E_1,0,0),
\eeas
where $E_{1,2}=\tfrac{M_h}{4} \pm \tfrac{1}{2}\sqrt{M_h^2/4 - M_W^2}$.

We note that in the cases of 
$M_{1\massless}$ and $M_{\massless 1}$,
the kinematic endpoint coincides with the 
mass of the parent only because the final state objects
in this example happened to be massless.
In more general scenarios with massive particles
the endpoints of $M_{1\massless}$ and $M_{\massless 1}$
will provide only an unsaturated lower bound on the parent mass, 
in line with \eqref{masshierarchy}.

Fig.~\ref{fig:hWW}(a) also allows us to
compare the different $M_{\flox}$ distributions to each other.
As expected from the general property (\ref{eq:MNeqMNTagain}),
the distributions of $M_1$ and $M_{1\ourT}$ (given in blue)
are identical. As discussed in Section~\ref{sec:smin}
and shown in  eqs.~(\ref{M1eqsminsubagain}) and (\ref{sminsubeqM1}), 
they also coincide with the distribution of
the $\sqrt{\hat{s}}_{\rm min}^{(\rm sub)}$ variable from \cite{Konar:2010ma}.
Similarly, in line with eq.~(\ref{eq:MTNeqM0N}),
the distributions of $M_{\ourT 1}(\mmass_1=0)$ and $M_{\massless 1}$
(shown in red) are practically indistinguishable, since the 
lepton masses are so tiny.
Notice that this is only true when $M_{\ourT 1}$ is calculated 
with $\mmass_1=0$, as was done here, otherwise the distributions of
$M_{\ourT 1}$ and $M_{\massless 1}$ would generally be different.
Finally, the distribution of $M_{1\massless}$ (shown in green) 
is distinct, as this variable is not related to any of the others.

Upon inspection of the shapes of different distributions in
Fig.~\ref{fig:hWW}(a), one observes that 
$M_1$ and $M_{1\ourT}$ appear to peak closest to the parent mass $M_h$,
and consequently, have the best defined endpoint structures.
On the other hand, $M_{1\massless}$ peaks much farther from $M_h$, 
and has a rather low event population in the vicinity of its endpoint.
Finally, the case of $M_{\ourT 1}(0)\simeq M_{\massless 1}$ represents 
an intermediate situation --- the peak is found in between the peaks of 
$M_1=M_{1\ourT}$ and $M_{1\massless}$; and
the endpoint structure is more pronounced than 
the case of $M_{1\massless}$, but not as sharp 
as the case of $M_1=M_{1\ourT}$.
This is an inevitable consequence of the hierarchy 
\eqref{eq:thehierarchyisprovedforgeneraln} among the 
mass bounds which is present in every event.

Next, in Fig.~\ref{fig:hWW}(b) we compare the distributions of the 
standard variables $\MEFF$ (red solid line) and $2\mpt$ 
(green solid line) to their mass-bound counterparts 
$M_{\massless 1}$ (red dotted line) and $M_{1\massless}$ (green dotted line).
Recall from the discussion in Sec.~\ref{sec:meff}
(and in particular eq.~(\ref{eq:meffinterpretation})) 
that $M_{\massless 1}$ is the analogue of $\MEFF$,
and in the limit of no upstream momentum the two 
variables become identical. This is confirmed in Fig.~\ref{fig:hWW}(b),
which shows rather similar distributions for $M_{\massless 1}$ 
and $\MEFF$. However, the analogy is not perfect
and the $\MEFF$ distribution is slightly shifted to the right. 
The only\footnote{We checked 
that when one restricts the plot only to events with $u_T=0$,
the distributions of $\MEFF$ and $M_{\massless 1}$
become identical, as required by eq.~(\ref{eq:meffinterpretation}).} 
reason for this effect is the fact that 
we allow for initial state radiation in our sample, 
so that the Higgs parent is typically produced with 
some recoil and $u_T\ne0$. This is why the 
the $\MEFF$ distribution does {\em not} terminate at $M_h$, 
but shows a long tail extending to $\MEFF>M_h$.
In contrast, the $M_{\massless 1}$ distribution has
an exact endpoint at $M_h$.

A similar analysis holds for the other pair of distributions 
(color coded in green) which are shown in Fig.~\ref{fig:hWW}(b).
As explained in Sec.~\ref{sec:met} and seen from
eq.~(\ref{eq:mptinterpretation}), the variable
$M_{1\massless}$ is the analogue of $2\mpt$,
since the two become identical in the limit of 
no upstream momentum ($u_T\to 0$). However, in the presence of
upstream momentum, the proper behavior 
(an endpoint located at the parent mass) is retained
only by the $M_{1\massless}$ distribution, while
the $2\mpt$ distribution picks up a long tail 
extending beyond the true value of $M_h$.

This concludes our discussion of the $N=1$ 
unprojected and singly projected variables 
in relation to Higgs production (\ref{eq:hdecays}). 
We note that one could also apply $N=2$ variables
to this example, this time considering the two 
$W$ bosons as the two heavy parent particles,
and partitioning as
\beas
{\cal V}_1 &=& \ourSet{\ell^+ },\quad {\cal I}_1 = \ourSet{\nu_\ell }, \\ [2mm]
{\cal V}_2 &=& \ourSet{\ell^- },\quad {\cal I}_2 = \ourSet{\bar{\nu}_\ell }.  
\eeas
The upper kinematic endpoints of the resulting distributions will 
be found at the corresponding parent mass, in this case the mass
$M_W$ of the $W$-bosons. 

\subsection{An $\numparents=2$ example: top quark pair production}
\label{sec:ttbar}

As our next example, we consider dilepton events from the top 
quark pair production process (\ref{eq:ttdecays}). We assume that
the two $b$-jets from the top quark decays have been tagged, 
which distinguishes them from QCD jets from initial state radiation.
Correspondingly, the tagged $b$-jets will be included among the
set of visible particles, while any remaining QCD jets 
will contribute to the UVM category.

\begin{figure*}[t]
\begin{center}
\includegraphics[width=0.45\linewidth]{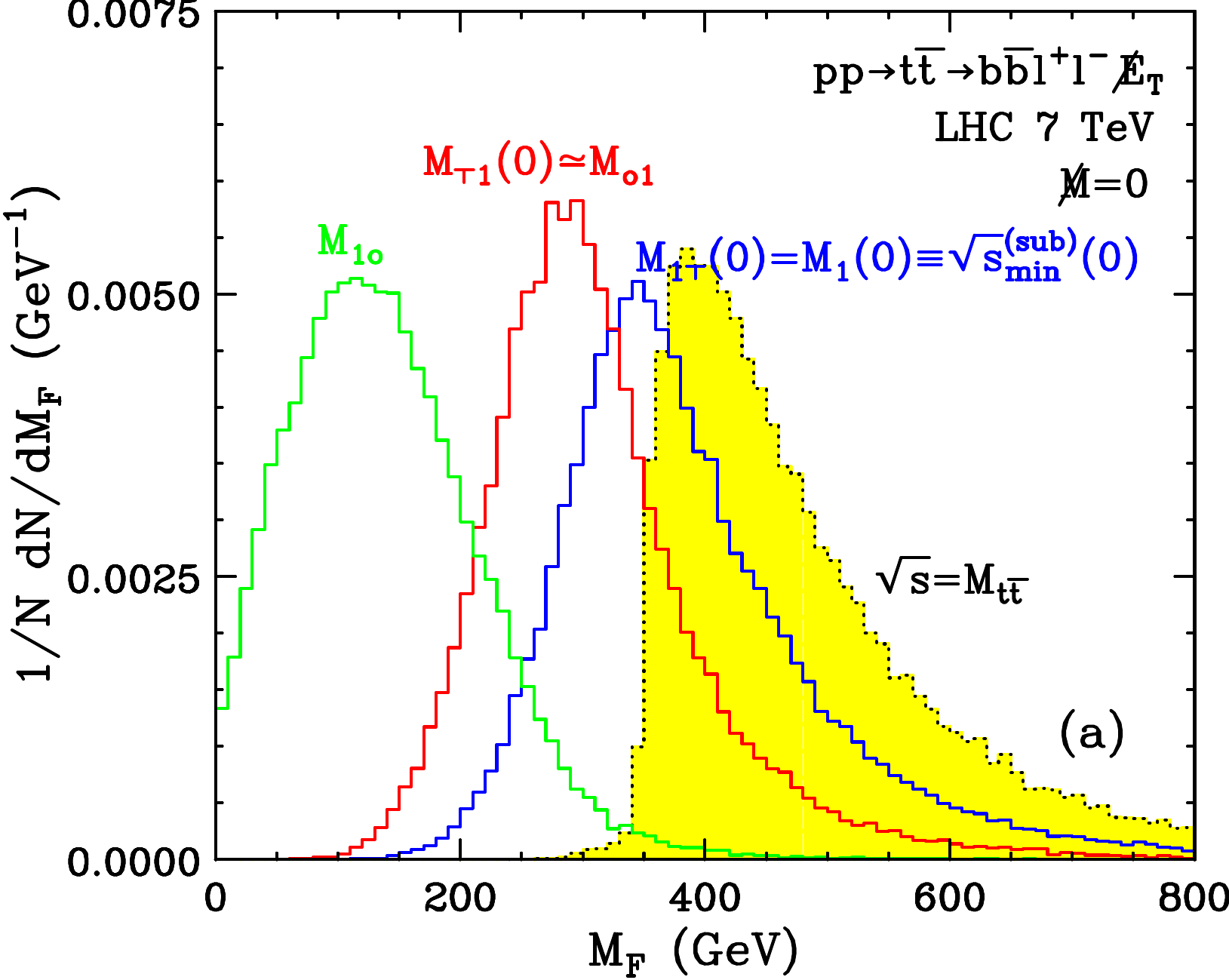}
\hspace{0.05\linewidth} 
\includegraphics[width=0.45\linewidth]{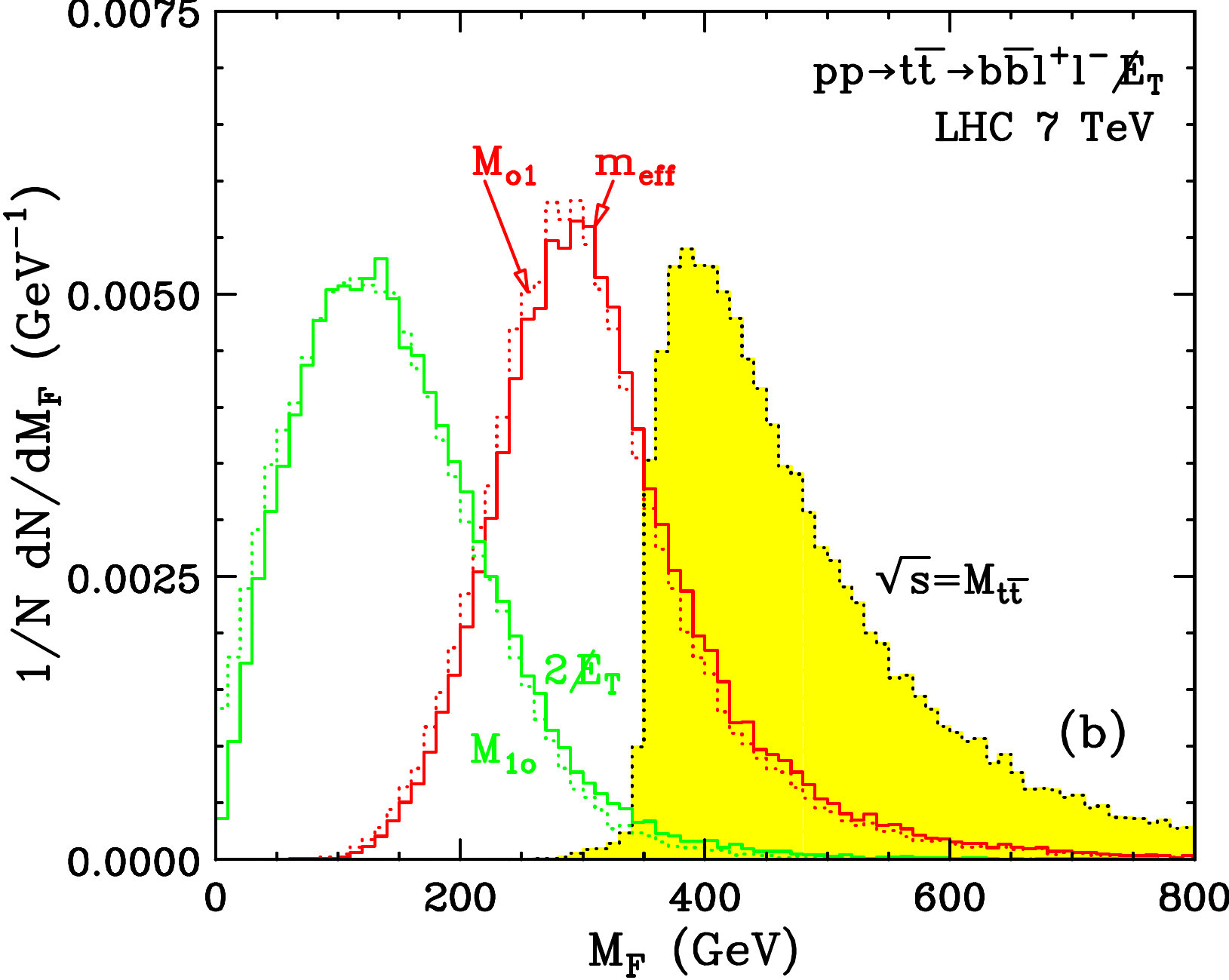} 
\end{center}
\caption[Distributions of N=1 mass variables for ttbar events]{\label{fig:ttb}
The same as Fig.~\ref{fig:hWW}, but for the 
$t\bar{t}$ example. The yellow shaded distribution 
now gives the true invariant mass of the $t\bar{t}$ pair.
}
\end{figure*}

We first reconsider the $N=1$ variables already 
studied in Sec.~\ref{sec:higgs},
and show that they can be useful even when 
there are multiple parents in the event.
For the purpose of constructing $N=1$ variables,
the event is partitioned simply as
\beas
{\cal V}_1 &=& \ourSet{b ,\bar{b}, \ell^+,\ell^- }, \\ [2mm]
{\cal I}_1 &=& \ourSet{\nu_\ell,\bar{\nu}_\ell }. 
\eeas
Fig.~\ref{fig:ttb}(a) displays the distributions of the 
resulting $N=1$ variables. Those distributions 
should be contrasted with the true $\sqrt{\hat{s}}$ 
distribution of the $t\bar{t}$ {\em pair},
which is shown in the figure with the yellow-shaded histogram.
Just like in Fig.~\ref{fig:hWW}(a), we find only three 
distinct distributions, since $M_1=M_{1\ourT}$ from (\ref{eq:MNeqMNTagain})
and $M_{\ourT 1}(\mmass_1=0)\simeq M_{\massless 1}$ from (\ref{eq:MTNeqM0N}).
The hierarchical ordering of the three distributions is
the same as in Fig.~\ref{fig:hWW}(a), the distribution
of $M_1=M_{1\ourT}$ being the hardest, and the distribution
of $M_{1\massless}$ being the softest. Since all of our
$N=1$ variables are defined through minimization, each 
variable provides a lower bound on the true
center-of-mass energy $\sqrt{\hat{s}}$ in the event. 
As one might have expected, it is the $M_1$ (or equivalently, 
the $M_{1\ourT}$) variable which offers the best (in the sense
of being most stringent and meaningful) bound.
Since $M_1$ and $M_{1\ourT}$ are identical to the 
$\sqrt{\hat{s}}_{\rm min}^{(\rm sub)}$ variable, their distribution
exhibits the interesting property first noted in
\cite{Konar:2008ei} in relation to $\sqrt{\hat{s}}_{\rm min}$:
that the peak of the distribution is located very near the 
mass threshold for producing the two heavy parents, 
in this case the two top quarks. Indeed, notice how the 
peak in the (blue) $M_1=M_{1\ourT}=\sqrt{\hat{s}}_{\rm min}^{(\rm sub)}$ 
histogram coincides with the onset of the (yellow-shaded)
true $\sqrt{\hat{s}}$ distribution. When applied to searches for 
new physics, one can then use the peak in the  
$M_1=M_{1\ourT}=\sqrt{\hat{s}}_{\rm min}^{(\rm sub)}$ distribution 
as a rough estimate of the new physics mass scale \cite{Konar:2008ei,Konar:2010ma}.

In analogy to Fig.~\ref{fig:hWW}(b) here we can also 
perform a comparison of the usual variables 
$\MEFF$ and $2\mpt$ to their mass-bound analogues 
$M_{\massless 1}$ and $M_{1\massless}$.
In Fig.~\ref{fig:ttb}(b) we compare
$\MEFF$ to $M_{\massless 1}$ (in red) and
$2\mpt$ to $M_{1\massless}$ (in green).
This time the differences are much less pronounced 
that the single resonance case shown in Fig.~\ref{fig:hWW}(b).
This suggests that for $\numparents=2$ processes, 
the variable $\MEFF$ ($2\mpt$) is on an equal footing with
$M_{\massless 1}$ ($M_{1\massless}$).

We remind the reader that the $\numparents=1$ variables 
shown in Fig.~\ref{fig:ttb}(a) do not exhibit 
any upper kinematic endpoints, since they are being 
applied to $\numparents=2$ events, i.e. they have the ``wrong'' 
value of $\numparents$ and so the bounding relations  
\eqref{eq:Mbound} do not apply to any individual parent. 
Thus let us now discuss the $\numparents=2$ variables, 
which have the correct value of $\numparents$ and
for which \eqref{eq:Mbound} holds.
In the case of $\numparents=2$, a $t\bar{t}$ dilepton 
event is partitioned as
\beas
{\cal V}_1 &=& \ourSet{     b , \ell^+ }, \\ [2mm]
{\cal V}_2 &=& \ourSet{\bar{b}, \ell^- }, \\ [2mm]
{\cal I}_1 &=& \ourSet{\nu_\ell },  \\ [2mm]
{\cal I}_2 &=& \ourSet{\bar{\nu}_\ell }.  
\eeas
This partitioning can be pictorially visualized 
in Fig.~\ref{fig:countingParticles2parents}.
Since we are primarily interested in the kinematical effects, 
for this illustrative example 
we make the simplifying assumption (unlikely to be realized in 
any real experiment) that each lepton can be associated
with its sibling $b$-jet.

\begin{figure*}[t]
\begin{center}
\includegraphics[width=0.45\linewidth]{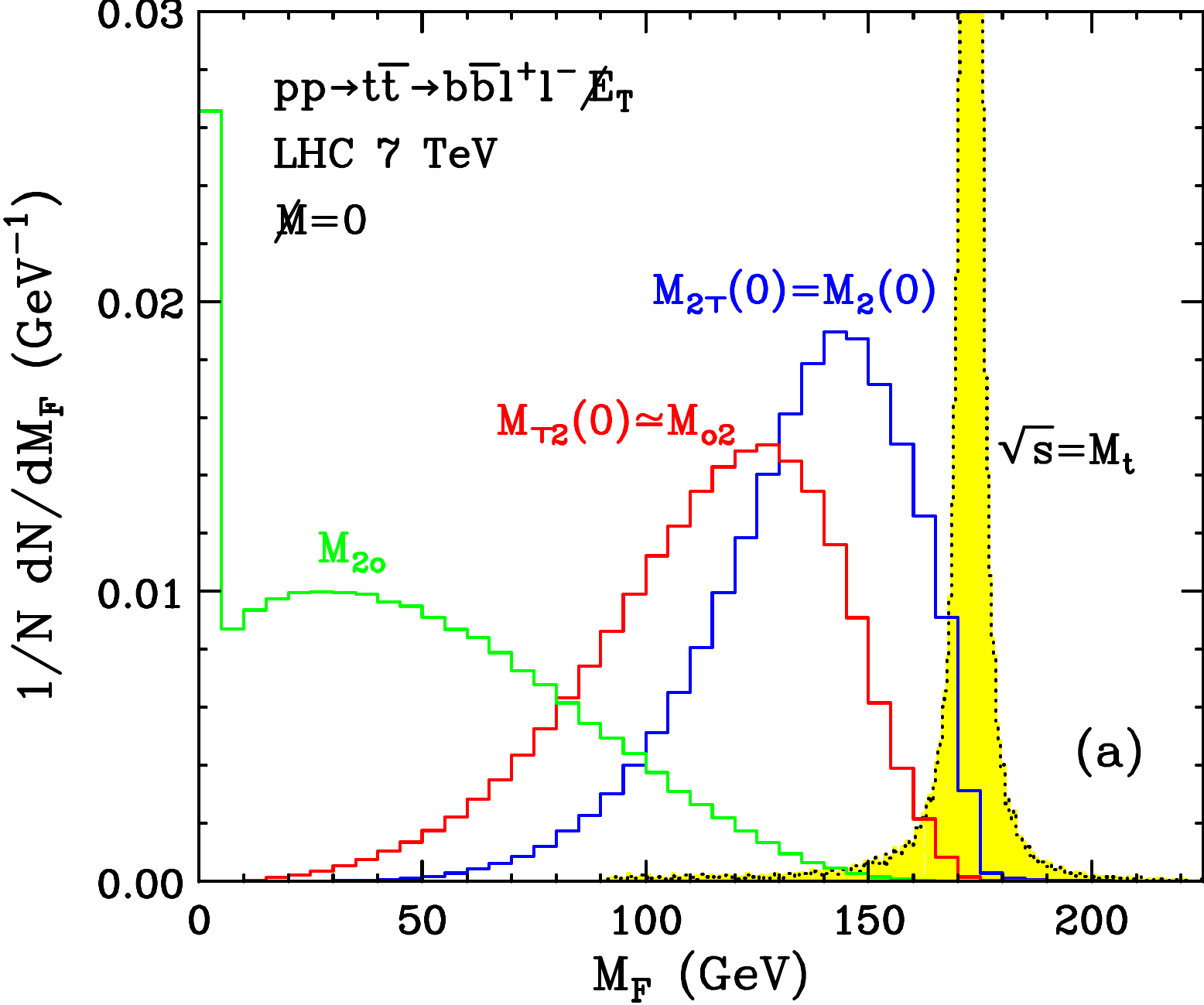} 
\hspace{0.05\linewidth}
\includegraphics[width=0.45\linewidth]{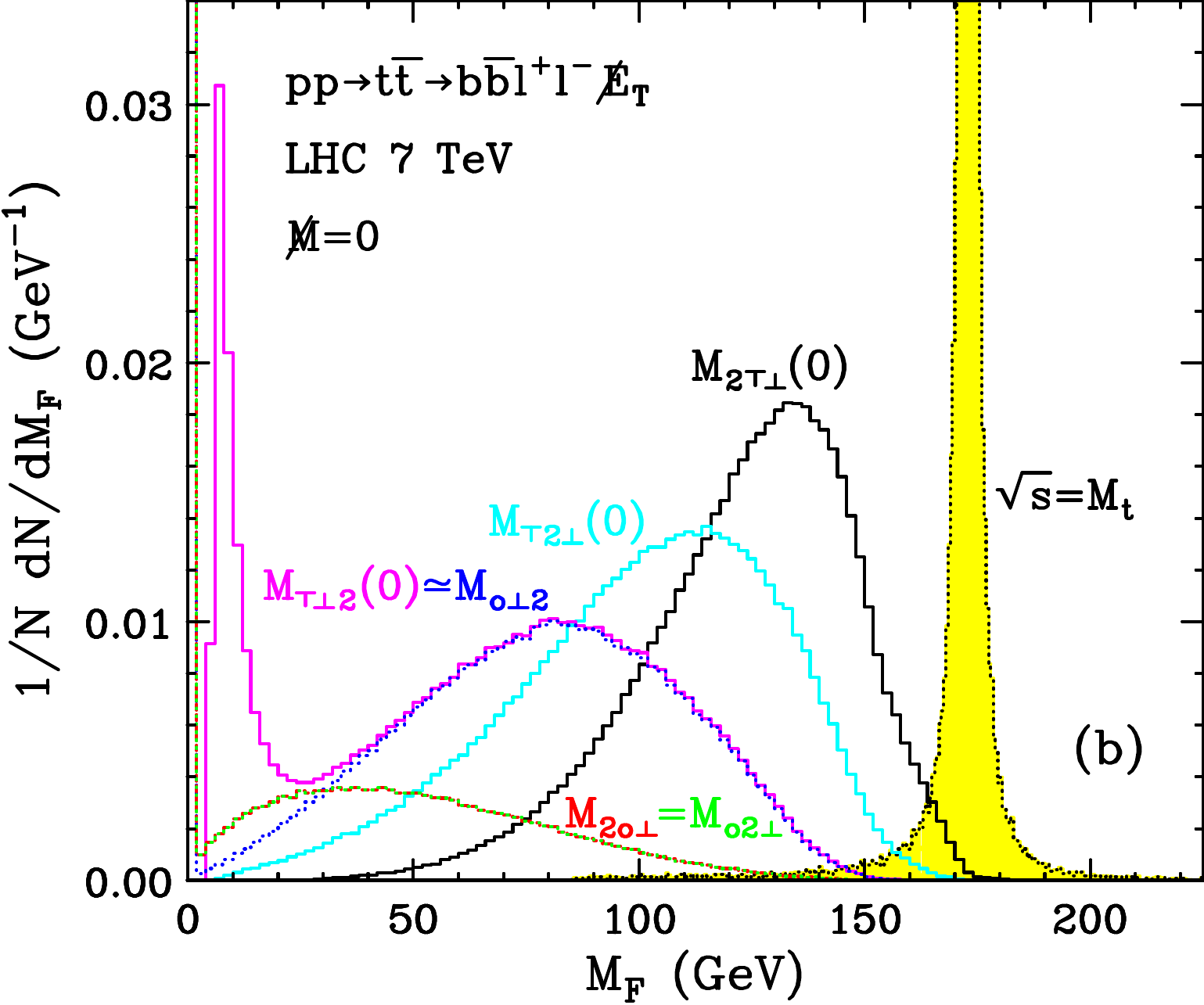} 
\end{center}
\caption[Distributions of N=2 mass variables for ttbar events]{\label{fig:ttb2}
The same as Fig.~\ref{fig:ttb}, but for 
$\numparents=2$ variables: (a) the unprojected $M_2$
and the singly projected variables
$M_{2\ourT}$, $M_{\ourT2}$, $M_{\massless2}$
and $M_{2\massless}$;
and (b) the doubly projected variables 
$M_{2 \ourT \myT}$ (black), 
$M_{\ourT 2  \myT}$ (cyan),
$M_{\ourT \myT 2 }$ (magenta),
$M_{2 \massless \myT}$ (red),
$M_{\massless 2  \myT}$ (green) 
and $M_{\massless \myT 2}$ (blue).
The yellow shaded distribution 
now gives the average top quark mass in the event.
In panel (b), ``$\ourT$''-projected quantities are 
denoted with solid lines 
and are evaluated with $\mmass_1=\mmass_2=0$, while 
``$\massless$''-projected quantities are denoted with
dotted lines.
}
\end{figure*}

The distributions of the corresponding five $N=2$ 
variables are shown in Fig.~\ref{fig:ttb2},
where for illustrative purposes we use Monte Carlo truth information to 
properly assign the correct $b$-jet to each lepton.
According to \eqref{eq:Mbound}, 
these distributions are bounded from above by the 
individual parent mass, which in this case is the mass
of the top quark. Correspondingly, in Fig.~\ref{fig:ttb2}
the reference yellow-shaded distribution 
now shows the (average) top quark mass in the event, which
follows the familiar Breit-Wigner shape
(compare to the Higgs resonance shape in Fig.~\ref{fig:hWW}).

As before, we observe three distinct distributions,
$M_2=M_{2\ourT}$ (in blue), $M_{\ourT 2}(\ourSet{\mmass_a=0})\simeq M_{\massless 2}$ (in red)
and $M_{2\massless}$ (in green).
All of them exhibit an upper endpoint 
less than or equal to the top quark mass $M_t$, in accordance with 
(\eqref{eq:Mbound}, but the three shapes are considerably different.
As before, and in agreement with the general hierarchy proven 
in the arguments leading up to \eqref{eq:thehierarchyisprovedforgeneraln} for any event,
the early partitioned versions $M_2$ and $M_{2\ourT}$
have the steepest endpoint, with the largest fraction of 
events near the endpoint. 

Again, we can show that we expect the $M_2$, $M_{2\ourT}$, $M_{\ourT2}$, $M_{\massless2}$
and $M_{2\massless}$ bounds to be saturated
by explicity constructing an extremal event $\event\in\allevents$
that satisfies the on-shell constraints of the $t$ and $\bar{t}$ quarks and $W^\pm$ bosons.
An example of such a configuration (see also Fig.~\ref{fig:configuration2}) is
\begin{gather*}
P_b        = P_{\bar{b}}   = (E_b,\,p_b,\,0,\,0)\\
P_{\ell^+} = P_{\ell^-}    = (E_\ell,\,p_\ell,\,0,\,0)\\
Q_\nu      = Q_{\bar{\nu}} = (E_\nu,\,-E_\nu,\,0,\,0)\,,
\end{gather*}
where, 
\begin{align*}
E_b^2&=p_b^2+M_b^2 &
E_\ell^2 &= p_\ell^2+M_\ell^2 \\
p_b &= \lambda(M_t,\,M_b,\,M_W) &
p_\ell&=M_\ell\sinh(\rho-\sigma)\\
E_\nu&=p^* e^\sigma &
p^*&= \lambda(M_W,\,M_\ell,\,0)\\
\sigma&=\sinh^{-1}\left(\frac{p_b}{M_W}\right) & 
\rho&=\sinh^{-1}\left( \frac{p^*}{M_\ell}\right),
\end{align*} and the two-body momentum function is given by
\beqs
\lambda(a,b,c) \equiv \frac{\sqrt{\left(a^2 - (b+c)^2 \right)  \left(a^2-(b-c)^2\right)}}{2a}.
\eeqs

\newcommand\configuration[5]{{
\begin{picture}(24,16)(-12,-8)
\linethickness{1pt}
\put(0,1){\circle{3}}            
\put(0,4.5){\makebox(0,0){#1}}   
\put(0,1){\vector(1,0){10}}      
\put(11,1){\makebox(1,0){#2}}    
\put(-4,-3){\circle{2}}          
\put(-2.5,-1){\vector(-1,0){3}}  
\put(-4,-7){\makebox(1,0){#3}}   
\put(-4,-3){\line(-1,0){2}}
\put(-7,-3){\line(-1,0){1}}
\put(-9,-3){\vector(-1,0){2}}    
\put(-12,-3){\makebox(-1,0){#4}} 
\put(-4,-3){\vector(+1,0){7}}    
\put(5,-3){\makebox(1,0){#5}}    
\end{picture}
}}

\begin{figure}[hbt]
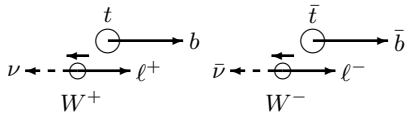

\begin{center}
\setlength{\unitlength}{1mm}
\configuration{$t$}{$b$}{$W^+$}{$\nu$}{$\ell^+$}
\configuration{$\bar{t}$}{$\bar{b}$}{$W^-$}{$\bar{\nu}$}{$\ell^-$}
\caption{\label{fig:configuration2}
Illustration of an example extremal configuration for the $N=2$ variables
when applied to the $t\bar{t}$ example.}
\end{center}
\end{figure}

Recall from eq.~(\ref{mt2eqM2})
that the early partitioned variables $M_2$ and $M_{2\ourT}$
are equivalent to the (1+3)-dimensional version $m_{T2}^{(1+3)}$ 
of the Cambridge $m_{T2}$ variable. It appears therefore that 
for studies like the one presented here, where the UVM 
contributions can be safely identified and accounted for, 
$m_{T2}^{(1+3)}$ is preferable over $m_{T2}^{(1+2)}$.

We use the $t\bar{t}$ example to also illustrate 
the doubly projected variables from Table~\ref{tab:floxbridge_double}.
Fig.~\ref{fig:ttb2}(b) shows the doubly projected 
``$\ourT$''-projections (solid lines)
and the ``$\massless$''-projections (dotted lines).
All ``$\ourT$''-projected quantities are 
evaluated with $\mmass_1=\mmass_2=0$.
Each type of projection can be done in three different ways:
early partitioning, $M_{2 \ourT \myT}$ (black) and 
$M_{2 \massless \myT}$ (red);
late partitioning, $M_{\ourT \myT 2 }$ (magenta) and
$M_{\massless \myT 2}$ (blue);
or in-between partitioning,
$M_{\ourT 2  \myT}$ (cyan) and
$M_{\massless 2  \myT}$ (green).

Similarly to the result from Fig.~\ref{fig:ttb2}(a),
Fig.~\ref{fig:ttb2}(b) also reveals that
the early partitioned, ``$\ourT$''-projected variable
$M_{2\ourT \myT}$ has the best defined endpoint structure, which
clearly indicates the value of the parent mass $M_t$.
As for the remaining variables, two are identically equal:
\beq
M_{2 \massless \myT} \equiv M_{\massless 2  \myT},
\eeq
which is a special case of the general identity (\ref{eq:N00eq0N0}),
while two others are approximately equal:
\beq
M_{\ourT \myT 2 }(\ourSet{\mmass_a=0}) \approx M_{\massless \myT 2},
\eeq
where a noticeable difference arises only at low values due to the
finite mass of the $b$-quark --- see eq.~(\ref{eq:MTNperpeqM0Nperp}).

\section{Conclusions}
\label{sec:conclusions}

\begin{table*}[tbh]
\begin{center}
\renewcommand\arraystretch{1.5}
\begin{tabular}{|c||c|c|c|c||c|c||}\cline{2-7}
\multicolumn{1}{c||}{}             & \multicolumn{6}{c||}{Mass-bound variable}  \\
\hline
Existing     & \multicolumn{4}{c||}{$N=1$ }        & \multicolumn{2}{c||}{$N=2$ }  \\
\cline{2-7}
variable     & $M_1(\mmass_1)=M_{1\ourT}(\mmass_1)$ & $M_{\ourT 1}(\mmass_1)$ & $M_{\massless 1}$ & $M_{1\massless}$ & $M_2(\chiM)=M_{2\ourT}(\chiM)$ & $M_{2\ourT\myT}(\chiM)$  \\ \hline\hline
$2\mpt=2\met$&                 &              &                  &    $u_T\to 0$   &                 &                  \\ \hline
$\MEFF$    &                 & $\mmass_1\to 0, u_T\to 0$  &   $u_T\to 0$     &                 &                 &                  \\ \hline
$\sqrt{\hat{s}}_{\rm min}^{(\rm sub)}(\mmass_1)$
             &   $\ourtick$       &              &                  &                 &                 &                  \\ \hline
$\sqrt{\hat{s}}_{\rm min}(\mmass_1)$
             & $u_T\to 0$      &              &                  &                 &                 &                  \\ \hline 
$m_{Te\nu}(M_e,M_\nu)$  
             &  $\ourtick$        &  $\ourtick$     &$M_e,M_\nu\to 0$  & $M_e,M_\nu\to 0$ &                 &                  \\ \hline 
$M_{T,ZZ}(M_Z)$& $\ourtick$       &  $\ourtick$      &                  &                 &                 &                  \\ \hline
$M_{C,WW}$   & $\mmass_1\to 0$ &              &                  &                 &                 &                  \\ \hline
$m_T^{\rm true}$ & $\mmass_1\to 0$ &              &                  &                 &                 &                  \\ \hline
$m_{TZ'}^{reco}(M_Z)$ &  $u_T\to 0$ &  $u_T\to 0$  &                  &                 &                 &                  \\ \hline\hline
$m_{T2}(\chiM)$&                 &              &                  &                 &  $\ourtick$        &                  \\ \hline
$m_{T2\perp}(\chiM)$&                 &              &                  &                 &                 &  $\ourtick$         \\ \hline
\hline
\end{tabular}
\caption[How existing variables fit the mass-bound classification]{\label{tab:summary}
Correspondence between some of the existing variables in the literature, which were
discussed in Section~\ref{sec:literature}, and the corresponding
mass-bound variables.
A checkmark ($\ourtick$) implies an exact equivalence, 
otherwise the relevant limiting condition is listed.
The last variable $m_{T2\perp}(\mmass_a)$ 
employs the doubly-projected $\myT$ construction described in Appendix~\ref{sec:doubleT}.
}
\end{center}
\end{table*}

The main ``result'' of this paper is the proposal made in 
Section~\ref{sec:imass} of a general scheme for 
constructing and categorizing the basic invariant mass variables
which are best suited for the study of missing energy events
at hadron colliders. As a demonstration of the utility
of this general scheme, in Section~\ref{sec:literature} 
we showed how a wide variety of widely used kinematic variables 
discussed in the literature can be properly accommodated in 
our framework. A short summary of this discussion is 
presented in Table~\ref{tab:summary}, which exhibits 
the connections between the variables discussed in
Section~\ref{sec:literature} and the corresponding
mass-bound variables from Tables~\ref{tab:floxbridge_basic} and \ref{tab:floxbridge_double}.
The table reveals that one can give a new meaning to well-known variables 
like $\mpt$ and $\MEFF$, which were originally introduced and defined 
in a way unrelated to any invariant mass considerations. Now we see that the
same variables allow an alternative interpretation in terms 
of bounds on Lorentz invariants of interest 
as long as one is using the ``massless'' ($\massless$) type 
of projection for the transversification.

Another lesson from Table~\ref{tab:summary} is that 
depending on the specific topology, the same bound may be constructed in different ways.
A perfect illustration is provided by the variable $M_{1}=M_{1\ourT}$. 
As discussed in detail in Sec.~\ref{sec:cluster},
even for the same final state (two leptons and missing energy),
the variable $M_{1}=M_{1\ourT}$ can emerge as differing bounds (either 
$M_{T,ZZ}$ or $M_{C,WW}$) depending on the choice of 
interpretation of the kinematical information.

But the value of the proposed scheme 
is not just in the accommodation of
existing techniques and variables. The primary benefit from our approach 
is that, having understood the main principles behind the construction
of a good invariant mass variable, the reader is now prepared to tackle 
almost any event topology, first by realizing what are the proper
invariant mass variables for the case at hand, and second,
knowing how to construct and calculate those variables.
As discussed in Sections~\ref{sec:projections}--\ref{sec:interpreting},
there are a number of choices to be made along the way, related to 
the method of transversification, 
the partitioning of the event, and the exact order in which one takes
all those operations. The main guiding principle through all this is
that at the end of the day, one is always going to construct
a bound on the mass of the heaviest parent.
In that sense we are extending the principles and methods of construction 
put forth in \cite{Lester:1999tx} for $m_{T2}$
and \cite{Konar:2008ei} for $\sqrt{\hat{s}}_{\rm min}$.

As we have seen, many of the generalized mass-bound variables are already 
in use at the LHC and elsewhere, but the majority have, for the moment, 
the status of solutions in search of problems.

\appendix

\section{Computer libraries offering ``transverse'' energy and mass variables}
\label{sec:libraries}

\begin{table*}[tbh]
\begin{center}
\renewcommand\arraystretch{1.5}
\begin{tabular}{|l|l|c|c|c|c|c|c|c|}\hline
\multirow{2}{*}{\bf Library } & \multirow{2}{*}{\bf Object } & \multicolumn{7}{c|}{\bf Method/function name} \\
\cline{3-9}
 &  & $e_\ourT$ & $e_\ourT^2$ & $m_\ourT$ & $m_\ourT^2$ & $m_{T2}$ & $e_\ourperp$ & $e_\ourperp^2$ \\
\hline\hline
{\tt CLHEP}\cite{Lonnblad:1994kt} & {\tt LorentzVector}& 
   {\tt mt()} &  {\tt mt2()} & -- & -- & -- &  
   {\tt et()} & {\tt et2()} \\
\hline
{\tt ROOT}~\cite{Antcheva:2009zz} & {\tt TLorentzVector} & 
   {\tt Mt()} & {\tt Mt2()}  & -- & -- & -- & 
   {\tt Et()} & {\tt Et2()} \\  
\hline
{\tt Fastjet}~\cite{Cacciari:2005hq} & {\tt Pseudojet} &  
 {\tt mperp()} & {\tt mperp2()} & -- & -- & -- & {\tt Et()} & {\tt Et2()}  \\ 
\hline
{\tt PGS}~\cite{PGS} & -- & -- & -- & -- & -- & -- & {\tt v4et(p)} & --  \\ 
\hline
Oxbridge & {\tt LorentzVector} & {\tt ET()} & {\tt ET2()} & {\tt LTV().mass() } & {\tt LTV().masssq() } & -- & -- & -- \\
\MTTWO~\cite{oxbridgeStransverseMassLibrary} & {\tt LorentzTransverseVector} & {\tt Et()} & {\tt Etsq()} & {\tt mass() } & {\tt masssq() } & -- & -- & -- \\
 & {\tt Mt2\_332\_Calculator}& -- & -- & -- & -- & {\tt mT2\_332()} & -- & -- \\
\hline
UCD \MTTWO~\cite{zenuhanStransverseMassLibrary} & {\tt mt2 }  & {\tt Ea}, {\tt Eb} &{\tt Easq}, {\tt Ebsq} & -- & -- & {\tt get\_mt2()} & -- & -- \\ \hline\hline
\multicolumn{2}{|l|}{Defining equation in this paper} & \multicolumn{2}{c|}{\eqref{eq:eTdef}} & \multicolumn{2}{c|}{(\ref{eq:MourTdef})} &
(\ref{mt2eqM2})  & \multicolumn{2}{c|}{\eqref{eq:eperpdef}}\\ \hline
\end{tabular}
\caption[How computer libraries define transverse methods]{\label{tab:libraries}
The versions of the transverse variables used in commonly used high-energy physics computer libraries and codes.
A brief survey of experimental collaborations' software suggests that most follow the conventions of {\tt CLHEP}. `LTV' is a shorthand for the method {\tt getLorentzTransverseVector()}.
}
\end{center}
\end{table*}

Though libraries should be a repository of human knowledge, 
any careful experimentalist will already have recognized 
that the computer libraries which support transverse projection
methods for Lorentz vectors do not always produce the expected behavior.
A selection of some of the most commonly used libraries 
and some of their methods for calculating transverse variables 
can be found in Table~\ref{tab:libraries}. In many cases the 
method of projection used (i.e. ``$\ourT$'' or ``$\ourperp$'') 
is undocumented and can only be determined by excavating the 
implementation. What is more, the names of the methods and 
functions in some cases produce output very different from 
what the user might expect. The result is that use of a 
plausible-sounding method can land the unwary user 
with a totally unexpected result -- for example the 
{\tt CLHEP} method called {\tt mt()} returns the $\ourT$-projected 
transverse {\em energy} $(e_\ourT=\sqrt{M^2+p_T^2})$, 
not the transverse mass they might 
have anticipated. Of course, because of the right-hand expression
in eq.~(\ref{eq:eTdef}), one might fittingly call 
this quantity a ``mass'', but in that case the proper
nomenclature should probably be a ``longitudinal'' mass and 
not a ``transverse'' mass.

To the extent that there is agreement on the conventions, 
one can see that the most commonly-used libraries ({\tt ROOT} 
and {\tt CLHEP}) use the $\ourperp$ convention when calculating 
``transverse energy'' quantities. The Tevatron and LHC experimental 
collaborations tend to follow the ``$\ourperp$'' conventions when 
talking about ``transverse energy'' in calorimeters. For analyses 
where the transverse mass really matters, e.g. for $W\to \ell\nu$, 
the ({\tt ROOT} and {\tt CLHEP}) libraries have no function to 
return the `usual' transverse mass of  
Refs~\cite{vanNeerven:1982mz,Arnison:1983rp,Banner:1983jy,Smith:1983aa,Barger:1987du}: 
$m_T$ must instead be calculated explicitly by the user.

\section{Mass bounds on collections of momenta}
\label{sec:massbounds}






In this section we present derivations of mass bounds on collections of arbitrary momenta, 
which may be represented by unprojected vectors and/or vectors transversified by any of the 
projections \ourT{}, \ourperp{} and \massless{}. These cover the cases mentioned in 
\ref{sec:imass}, and justify the representation of multibody decays to visible and invisible 
particles in the form of a pair of composite momenta, where all visibles are projected 
identically (if at all) and all invisibles are likewise projected identically, though not necessarily 
by the same method as the visibles.

The question of {\em what} goes into the set of momenta from which we
wish to generate the parental mass bound is not a mathematical
question at all.  However, once that set of momenta is formed,
the question of {\em how to calculate the best bound making maximum
use of the information contained in that set} is entirely
mathematical.  It is this mathematical question that we solve in the
this section.

In essence, we try to answer the following question:
\begin{quote}
Given a particular set of vectors, what is the greatest possible lower bound that we can place on the mass of any parent particle which could have have decayed to daughters characterized by that set? In particular, how does that bound depend on the dimensionalities and projection-types of the vectors characterizing the information about the daughters?
\end{quote}

We shall denote the answer to that question as $\ourmm{\ldots} $,
where $\ourSet{\ldots}$ is the set of vectors.  We do not wish to restrict
the set to contain only momenta of the same type (e.g.~only
four-momenta).  Instead, we permit the set, if so desired, to be a
heterogeneous mixture containing any number of four-momenta,
$\ourT$-momenta, $\ourperp$-momenta, $\massless$-momenta or 2-momenta.
For example, $\ourmm{ A^\mu, B^\mu, c^\alpha_\ourT,
d^\alpha_\ourT,e^\alpha_\ourT,
f^\alpha_\ourperp,g^\alpha_\massless, \vec{h}_T}$ would denote be the greatest possible
lower bound on the mass of a particle assumed to have decayed to (at least)
eight daughters, under the assumption that the only information from which we would wish that bound to be 
constructed were to comprise: the four-momenta of two daughters $a$ and $b$; the 
masses and transverse two-momenta of three daughters $c$, $d$ and $e$; the three-speed 
and transverse two-momentum of daughter $f$; and the transverse two-momenta of particles 
$g$ and $h$.\footnote{Note that it makes no difference whether we use $\vec{h}_T$ instead 
of $h^\alpha_\massless$ as an input, as the information content of each is identical.}

\subsection{Parental mass bounds from sets containing any two objects}\label{sec:boundsonpairs}
Before considering parental bounds from arbitrary sets of momenta, we shall first consider the bound one obtains for each of the ten pair-wise combinations of the various types of vectors, i.e.:$$
\begin{array}{|cccc|}\hline
\ourmm{A^\mu,B^\mu} &
\ourmm{A^\mu,b^\alpha_\ourT} &
\ourmm{A^\mu,b^\alpha_\ourperp} &
\ourmm{A^\mu,\vec{b}_\genericT} \\
- &
\ourmm{a^\alpha_\ourT,b^\alpha_\ourT} & 
\ourmm{a^\alpha_\ourT,b^\alpha_\ourperp} & 
\ourmm{a^\alpha_\ourT,\vec{b}_\genericT} \\
- &- &
\ourmm{a^\alpha_\ourperp,b^\alpha_\ourperp} &
\ourmm{a^\alpha_\ourperp,\vec{b}_\genericT} \\
- &- &- &
\ourmm{\vec{a}_\genericT,\vec{b}_\genericT} \\
\hline
\end{array}.
$$
To avoid imposing a physical interpretation on the vectors (other than that they are momenta), we generally work with $A$'s and $B$'s, as opposed to the $P$'s and $Q$'s used in the main text. The latter carry implications of visibility/invisibility that are irrelevant to the considerations of this section.

The list above appears to leave out the massless \massless{}-projection, but 
this is simply a special case of the \ourT{} and \ourperp{} projections, so the results for 
\massless{} can be derived from the other two cases. In its place, we allow for combinations 
of vectors including transverse two-momenta $\vec{a}_\genericT, \vec{b}_\genericT$, in 
which the timelike component is simply unspecified. It will be seen that the bounds from 
combinations involving $\vec{a}_\genericT, \vec{b}_\genericT$ simply emerge to be the 
massless case.

\subsubsection{\pairtitle{A^\mu}{B^\mu}} \label{sec:ABbound}
We start with a straightforward case, taking care to be explicit about the sequence of operations that will also be required for the construction of the bound in the less trivial cases.
The best parental mass bound\footnote{Note that it is simplest to calculate the bound for the squared of the parental mass $\ourmmsq{\ldots}$ rather than for the parental mass itself $\ourmm{\ldots}$.  This difference is of no consequence, and so for brevity we will talk only of ``mass bounds'' in the text, ignoring the square.} given a pair of daughter 1+3 momenta $A^\mu$ and $B^\mu$ is given by 
\bea
\ourmmsq{A^\mu, B^\mu} &=& \min{\ourMaxMinBracs{M^2}} \nonumber\\
              &=&  \min{\ourMaxMinBracs{ P^\mu P_\mu }} \nonumber\\
              &=& \min{\ourMaxMinBracs{(A^\mu+B^\mu)(A_\mu+B_\mu)}} \nonumber\\
              &=& (A^\mu+B^\mu)(A_\mu+B_\mu),\nonumber \\
              &\equiv& (A^\mu+B^\mu)^2  \label{eq:mmlorlor}
\eea
where the first equality is simply a rephrasing of the meaning of $\ourmm{}$ as the minimum mass consistent with the constraints. The second equality is from the definition of the inner product (or physically the definition of the mass), and the third equality is from the definition of a vector space  (physically representing energy-momentum conservation). The fourth equality is a statement that the vectors $A^\mu$ and $B^\mu$ are fully specified, so the minimization is trivial (no parameters need be changed).  The hopefully unsurprising outcome, then, is that the best lower bound on the parental mass is given by the invariant mass of the two daughter momenta $A^\mu$ and $B^\mu$.

\subsubsection{\pairtitle{A^\mu}{b^\alpha_\ourT}} \label{sec:AbTbound}

To calculate the bound
\beq
\ourmm{A^\mu,b^\alpha_\ourT} \label{eq:mmlorTintro}
\eeq
we note that $b^\alpha_\ourT$ contains partial information about some 1+3 vector $B^\mu$
which projects to $b^\alpha_\ourT$ under the $\ourT$-projetion, about which the $x$ and $y$ components are known, but
the $z$ component, $b_z$ is completely unspecified. 
The bound \eqref{eq:mmlorTintro} can therefore be rephrased,
\beq
\ourmmsq{A^\mu,b^\alpha_\ourT} = \min_{b_z} \ourMaxMinBracs{(A^\mu+B^\mu)^2}.
\eeq
For the minimization we recognise that provided either $M_A \ne 0$ or $|\vec{a}_\genericT|\ne 0$, then $M$ is unbounded above as $b_z\to\pm\infty$. Provided that we are dealing with particles produced with non-zero transverse momentum (which we shall assume hereafter), the solution must then be given by the local minimum
\beas
0 &=& \frac{\partial}{\partial b_z}(A^\mu+B^\mu)^2  \\
  &=& \frac{\partial}{\partial b_z} \left( M_A^2 + M_B^2
                           + 2\left(E_A E_B  - \vec{a}_\genericT \cdot \vec{b}_\genericT - a_z b_z \right) \right). 
\eeas
The minimization selects $b_z/E_B = a_z/E_A$ such that $B^\mu$ has equal rapidity to $A^\mu$,
\beq \label{eq:lorTrapidity}
y_B  = y_A.
\eeq
To calculate the value of the mass bound we recognise that, by the definition of the Lorentz transformation, the inner product of $A^\mu$ and $B^\mu$, which we might denote by $g(A,B) \equiv A^\mu g_{\mu\nu} B^\nu$, is invariant under identical Lorentz transforms $\Lambda$ of both vectors
\beq
g(A,B) = g(\Lambda A, \Lambda B) .
\eeq
By letting $\Lambda$ be a boost along the $z$-axis corresponding to rapidity change $-y_A$, which will then set both rapidities to zero, one finds that 
\beq
g(A,B) = e_\ourT^{(A)} e_\ourT^{(B)} - \vec{a}_\ourT \cdot \vec{b}_\ourT.\label{eq:ohnonotagain}
\eeq
The best lower bound on the parent mass for daughters specified by a 1+3 momentum $A^\mu$ and a $\ourT$-projected 1+2 momentum $b^\alpha_\ourT$ is then given by
\bea \label{eq:mmlorT}
\ourmmsq{A^\mu,b^\alpha_\ourT} &=& M_A^2 + M_B^2 + 2\left( e_\ourT^{(A)} e_\ourT^{(B)} - \vec{a}_\ourT \cdot \vec{b}_\ourT\right) \nonumber \\
& = & (a_\ourT + b_\ourT)^\alpha (a_\ourT + b_\ourT)_\alpha \nonumber \\
& \equiv & (a_\ourT^\alpha + b_\ourT^\alpha)^2. \label{eq:aReducedDimMassSq}
\eea

\subsubsection{\pairtitle{A^\mu}{\vec{b}_\genericT}}
The bound on a 1+3 Lorentz vector with a transverse two-vector can be found in a similar manner, but the 1+3 vector which projects to $\vec{b}_\genericT$ is now given by some $B$ which has both unknown $z$ component {\em and} unknown mass. The bound is given by 
\beq
\ourmm{A, \vec{b}_\genericT} = \ourmm{A, B} = \min_{b_z, M_B} \ourMaxMinBracs{A+B}.
\eeq
A similar argument to that which led to \eqref{eq:lorTrapidity} shows that the $b_z$ component must be such that $y_B=y_A$.
The $M_B$ minimization selects $M_B=0$, so that 
\bea \label{eq:mmlorvec}
\ourmmsq{A, \vec{b}_\genericT} &=& M_A^2 + 2\left( e_\genericT^{(A)} |\vec{b}_\genericT| - \vec{a}_\genericT \cdot \vec{b}_\genericT\right) \nonumber \\
                          & \equiv & (a_\ourT^\alpha + b_\massless^\alpha)^2,
 \label{eq:mtlorvecdef}
\eea
so the bound is formed by turning the transverse two-momentum into a \massless{}-projected 1+2 momentum. Comparing with \eqref{eq:aReducedDimMassSq}, we see that if $b^\alpha_\ourT$ is made massless $M_B = 0 $, then $e^{(B)}_\ourT = |\vec{b}_\genericT|$ and \eqref{eq:mtlorvecdef} is reproduced.

\subsubsection{\pairtitle{a^\alpha_\ourT}{b^\alpha_\ourT}}
For each of the 1+2 $\ourT$-projected vectors, the corresponding set of 1+3 dimensional objects
shares the same transverse components and inner product (mass) as their $\ourT$-projected counterpart, 
but has arbitrary $z$ momentum.
The bound is then given by
\beq
\ourmmsq{a_\genericT, b_\genericT} = \min_{a_z,b_z}\ourMaxMinBracs{(A+B)^2} 
\eeq
This time the minimizations force the rapidities of $A$ and $B$ to be equal, but leave the value of that rapidity $y_A=y_B$ free. Similarly to the previous cases,
\bea\label{eq:mmTT}
\ourmmsq{a_\genericT, b_\genericT} &=& M_A^2 + M_B^2 + 2\left( e_\genericT^{(A)} e_\genericT^{(B)} - \vec{a}_\genericT \cdot \vec{b}_\genericT\right) \nonumber \\
                                &=&  (a_\ourT^\alpha + b_\ourT^\alpha)^2. \label{eq:mtTTdef}
\eea

\subsubsection{\pairtitle{a^\alpha_\ourT}{\vec{b}_\genericT}}

The limit on $M$ is given by 
\beq 
\ourmmsq{a_\genericT, \vec{b}_\genericT} = \min_{a_z,b_z,M_B}\ourMaxMinBracs{ (A+B)^2}.
\eeq
The minimizations set the rapidities to be equal $y_A=y_B$ (but undefined) and $M_B=0$.
The limit again appears in the form,
\bea \label{eq:mmTvec}
\ourmmsq{a_\genericT, \vec{b}_\genericT} &=& M_A^2 + 2\left( e_\genericT^{(A)} |\vec{b}_\genericT| - \vec{a}_\genericT \cdot \vec{b}_\genericT\right) \nonumber \\
                                &\equiv&  (a_\ourT^\alpha + b_\massless^\alpha)^2. \label{eq:mtTvecdef}
\eea

\subsubsection{\pairtitle{\vec{a}_\genericT}{\vec{b}_\genericT}}
The limit on $M$ for a pair of transverse two-momenta is given by 
\beq
\ourmmsq{\vec{a}_\genericT, \vec{b}_\genericT} = \min_{a_z,b_z,M_A,M_B} \ourMaxMinBracs{(A+B)^2}.
\eeq
The $z$ minimizations again set the relative rapidities equal but arbitrary $y_A=y_B$, 
and the mass minimizations set $M_A=M_B=0$.
\bea \label{eq:mmvecvec}
\ourmmsq{\vec{a}_\genericT, \vec{b}_\genericT} &=& 2\left( |\vec{a}_\genericT|\,|\vec{b}_\genericT| - \vec{a}_\genericT \cdot \vec{b}_\genericT\right) \nonumber \\
                                &\equiv&  (a_\massless^\alpha + b_\massless^\alpha)^2. \label{eq:mtvecvecdef}
\eea

\subsubsection{\pairtitle{A^\mu}{b^\alpha_\ourperp}}

The $\ourperp$ projection described in section \ref{sec:perpmalism} maps all 1+3 vectors 
$B^\mu$ with the same transverse momentum $\vec{b}_{\genericT}$ and velocity 
$V_B = |\vec{b}|/E_B$ to the same 1+2 vector $b^\alpha_\ourperp$. Therefore the 
longitudinal momentum component $b_z$ is unspecified and the parental mass bound 
$\ourmm{A^\mu, b_\ourperp^\alpha }$ is given by 
\bea
\ourmmsq{A^\mu, b_\ourperp^\alpha } &=& \min_{b_z}\ourMaxMinBracs{(A + B)^2}\, ,
\eea
From equations \ref{eq:perpEsq} and \ref{eq:perpMsq}, we see that we can 
decompose the full (1+3)-dimensional energy and mass
\bea
E_B^2 &=& (e^B_\ourperp)^2 + (e^B_{z})^2, \label{eq:EBdecomp} \\ [2mm]
M_B^2 &=& (m^B_{\ourperp})^2 + (m^B_{z})^2\ . \label{eq:MBdecomp}
\eea
each in terms of a transverse quantity ($e^B_\ourperp$, $m^B_\ourperp$) 
and a longitudinal quantity ($e^B_z = |\vec{b}_z| / V_B$, $m^B_z = |\vec{b}_
z| / (V_B \gamma_B)$) with $\gamma_B$ denoting the Lorentz factor $1/\sqrt{1-V_B^2}$.

Using these relations, we can write the Lorentz-invariant quantity $(A+B)^2$ as
\bea
(A^\mu+B^\mu)^2 &=&  M_A^2 + (m_\ourperp^B)^2 + b_z^2 / (V_B \gamma_B)^2  \label{eq:Abperp}\\
  &&  + \;  2 \left( \frac{E_A}{V_B} \sqrt{b_\genericT^2+b_z^2} - \vec{a}_\genericT\cdot\vec{b}_\genericT - a_z b_z \right) \, .\nonumber
\eea

Leaving aside the trivial case of $b_\genericT = 0$, we now attempt the minimization over $b_z$, requiring
\bea
0 &=& \frac{\partial}{\partial b_z} (A^\mu+B^\mu)^2 \\
   &=& 2\left(\frac{E_A}{V_B} \frac{ b_z}{\sqrt{b_\genericT^2 + b_z^2}} - a_z
   + \; \frac{b_z}{V_B^2 \gamma_B^2}  \right) \ . \nonumber
\eea
This gives rise to a quartic in $b_z$,
\beq
(b_\genericT^2 + b_z^2)(b_z - \alpha)^2 - \epsilon^2 \, b_z^2 = 0 \ , \label{eq:bzquartic}
\eeq
where the constants
\bea
\alpha &=& a_z\,V_B^2 \,\gamma_B^2, \\
\epsilon &=& E_A\, V_B \, \gamma_B^2\, .
\eea

The need to solve this quartic makes the $\ourmmsq{A^\mu, b_\ourperp^\alpha}$ bound 
intractable in comparison with the similar $\ourmmsq{A^\mu, b_\ourT^\alpha}$ bound. 
Similar difficulties are encountered in the following $\ourmmsq{a_\ourperp^\mu, b_\ourperp^\alpha}$ case.

One might guess that the $y_B = y_A$ condition resulting from $\ourmmsq{A^\mu, b_\ourT^\alpha}$ bound could represent the correct solution, since we are again working with a
fully (1+3)-dimensional vector combined with a (1+2)-dimensional vector. But the solution 
this condition gives for $b_z$ is not a root of the quartic in \eqref{eq:bzquartic}. One can
show that $b'_z = 0$ implies 
\beq
b_z = \frac{b_\genericT \alpha}{\sqrt{\epsilon^2 V_B^2 - \alpha^2}} \ ,
\eeq
which when substituted into the LHS of \eqref{eq:bzquartic} gives
\bea
\frac{b_\genericT^2 \alpha^2 \epsilon^2} {(\alpha^2 - \epsilon^2 V_B^2)^2}
\bigg( b_\genericT^2 V_B^4 &+& (V_B^4 - 1) (\epsilon^2 V_B^2 - \alpha^2)
\nonumber \\
	&-&  2 \, b_\genericT \, V_B^4 \, \sqrt{\epsilon^2 V_B^2 - \alpha^2} \, \bigg), \nonumber
\eea
which is in general non-zero, i.e. the equal rapidities condition only minimizes $b_z$ under certain special conditions.

\subsubsection{\pairtitle{a^\alpha_\ourperp}{b^\alpha_\ourperp}}

In the case of two $\ourperp$-projections, the mass bound is given by
\beq
\ourmmsq{ a^\alpha_\ourperp, b^\alpha_\ourperp } =  \min_{a_z, b_z}\ourMaxMinBracs{(A + B)^2}  \, ,
\eeq
the minimization of which involves finding $a_z$ and $b_z$, with both $V_A$ and $V_B$ 
being held fixed, such that each of $a_z$ and $b_z$ is a root of a quartic like that in
\eqref{eq:bzquartic}.

In this situation, we can apply the same mass/energy decompositions \eqref{eq:EBdecomp} and \eqref{eq:MBdecomp} to $a_z$, to get a variant of \eqref{eq:Abperp},
\begin{multline}
(A^\mu+B^\mu)^2 \label{eq:aperpbperp} = \\
\shoveleft	=  (m_\ourperp^A)^2 + \frac{a_z^2}{V_A^2 \gamma_A^2}
	+ (m_\ourperp^B)^2 + \frac{b_z^2}{V_B^2 \gamma_B^2}  
		\\
	+ 2 \left( \frac{\sqrt{a_\genericT^2+a_z^2}}{V_A}
		\frac{\sqrt{b_\genericT^2+b_z^2}}{V_B}
		- \vec{a}_\genericT\cdot\vec{b}_\genericT - a_z b_z \right)  .
\end{multline}

Differentiating by $a_z$ and by $b_z$ separately, the minimization imposes
\bea
0 &=& \frac{\partial}{\partial a_z} (A^\mu+B^\mu)^2 \label{eq:diffaz} \\
   &=& 2\left(\frac{\sqrt{b_\genericT^2+b_z^2}}{\sqrt{a_\genericT^2 + a_z^2}} 
   	\frac{ a_z}{V_A V_B} - b_z + \; \frac{a_z}{V_A^2 \gamma_A^2}  \right) \ , \nonumber
\eea
and simultaneously
\bea
0 &=& \frac{\partial}{\partial b_z} (A^\mu+B^\mu)^2 \label{eq:diffbz} \\
   &=& 2\left(\frac{\sqrt{a_\genericT^2+a_z^2}}{\sqrt{b_\genericT^2 + b_z^2}}
   	\frac{ b_z}{V_A V_B} - a_z
   + \; \frac{b_z}{V_B^2 \gamma_B^2}  \right) \ . \nonumber
\eea

Note that the 
fraction $|\vec{a}| / |\vec{b}|$ appears in both \eqref{eq:diffaz} and \eqref{eq:diffbz},
albeit as a reciprocal in the latter. So, we can combine the two minimization constraints in 
the form of a quadratic in $a_z$ and $b_z$:
\beq
\frac{a_z^2}{V_A^2 \gamma_A^2} + \frac{b_z^2}{V_B^2 \gamma_B^2}
+ a_z b_z \left( \frac{1}{V_A^2 \gamma_A^2} + \frac{1}{V_B^2 \gamma_B^2} \right) = 0 \ .
\eeq
This can be solved to give $a_z = -c \, b_z$, with
\beq
c = 1\quad \text{or} \quad \frac{V_A^2 \gamma_A^2}{V_B^2 \gamma_B^2} \ .
\eeq

Substituting this solution back into \eqref{eq:aperpbperp} gives a pleasingly simple result
\begin{multline}
A^\mu+B^\mu)^2 \label{eq:2perpmin} = \\
	=  (m_\ourperp^A)^2 + (m_\ourperp^B)^2
	 + b_z^2\left (2\,c + \frac{c^2}{V_A^2 \gamma_A^2} + \frac{1}{V_B^2 \gamma_B^2} \right) \\
	+ 2 \left( \frac{\sqrt{a_\genericT^2+c^2 b_z^2}}{V_A}
		\frac{\sqrt{b_\genericT^2+b_z^2}}{V_B}
		- \vec{a}_\genericT\cdot\vec{b}_\genericT \right)  .
\end{multline}
Since $c$ was chosen to be positive, this expression is clearly minimised for $b_z = 0$, 
implying that $a_z = 0$ as well. If we then make the replacements $a_\genericT / V_A = e^A_\ourperp$ and $b_\genericT / V_B = e^B_\ourperp$, we find that the choice $a_z = b_z = 0$ gives, quite simply and in tune with our intuition and inductive 
sense,
\beq
\ourmmsq{a^\alpha_\ourperp, b^\alpha_\ourperp } = (a^\alpha_\ourperp+b^\alpha_\ourperp)^2 \, .
\eeq

\subsubsection{\pairtitle{a^\alpha_\ourT}{b^\alpha_\ourperp}}

The bound $\ourmmsq{a^\alpha_\ourT, b^\alpha_\ourperp }$ requires minimization over 
both $a_z$ and $b_z$.
\beq
\ourmmsq{a^\alpha_\ourT, b^\alpha_\ourperp } = \min_{a_z,b_z}\ourMaxMinBracs{(A^\mu+B^\mu)^2} \, ,
\eeq
with $(A^\mu+B^\mu)^2$ defined as before in \eqref{eq:Abperp}. First the minimization over $a_z$ forces the rapidities of $A$ and $B$ to be equal, $a_z/E_A = b_z /E_B$.
Plugging this into \eqref{eq:Abperp}, we obtain
\bea
(A^\mu+B^\mu)^2
  &=&  M_A^2 + (m_\ourperp^B)^2 + b_z^2 / (V_B \gamma_B)^2 \\
  &&  + \; 2 \left(E_A E_B - \vec{a}_\genericT\cdot\vec{b}_\genericT - b_z^2 \frac{E_A}{E_B} \right) \, . \nonumber \\
  &=&  M_A^2 + (m_\ourperp^B)^2 + b_z^2 / (V_B \gamma_B)^2 \\
  &&  + \; 2 \left(\frac{E_A}{E_B} (E_B^2 - b_z^2) - \vec{a}_\genericT\cdot\vec{b}_\genericT \right) \, . \nonumber
\eea
The expression $(E_B^2 - b_z^2)$ can be written one of two ways -- either as
$(e^B_\ourT)^2$ or as $(e^B_\ourperp)^2 - b_z^2/(V_B \gamma_B)^2$. We choose the
 latter, since we have fixed $e^B_\ourperp$, but if we were to fix instead $e^B_\ourT$, we
 would rederive \eqref{eq:mtlorvecdef}.

A further simplification is implied by the equal rapidities condition, since $E_A = e^A_\ourT \, E_B / \sqrt{E_B^2 - b_z^2}$. Using this and \eqref{eq:EBdecomp}, we find
\bea
(A^\mu+B^\mu)^2
  &=&  M_A^2 + (m_\ourperp^B)^2 + b_z^2 / (V_B \gamma_B)^2 \\
  &&  + \; 2 \left(e^A_\ourT \, e^B_\ourperp \sqrt{1+\frac{b_z^2}{b_T^2 \gamma_B^2}} - \vec{a}_\genericT\cdot\vec{b}_\genericT \right) \, . \nonumber
\eea
Recognising that $b_\genericT^2 \gamma_B^2$ is positive, we see that the $b_z$ minimization simply gives $a_z = b_z = 0$, and thus
\beq
\ourmmsq{a^\alpha_\ourT, b^\alpha_\ourperp } = (a^\alpha_\ourT+b^\alpha_\ourperp)^2 \, .
\eeq

\subsubsection{\pairtitle{a^\alpha_\ourperp}{\vec{b}_\genericT}}

This mass bound is similar to $\ourmm{ a^\alpha_\ourperp, b^\alpha_\ourT }$ with 
the additional minimization of $M_B \mapsto 0$.
\bea
\ourmmsq{ a^\alpha_\ourperp, \vec{b}_\genericT } &=&  \min_{a_z, b_z, M_B}\ourMaxMinBracs{(A + B)^2}  \, \\ 
&=& (m_\ourperp^A)^2 
   +  2 \left(e_\ourperp^A \, b_\genericT  - \vec{a}_\genericT\cdot\vec{b}_\genericT  \right) \,  \\
&=& (a_\ourperp^\alpha + b_\massless^\alpha)^2 \, .
\eea

\subsection{Arbitrarily large sets of (1+3)-, (1+2)$_\ourT$- and 2-vectors}\label{sec:boundsonsets}

The generalization of \eqref{eq:mmlorlor} to an arbitrarily large set of fully specified 1+3 vectors $\mathcal{A}=\ourSet{A^\mu_i \mid 1 \le i \le |\mathcal{A}| }$ is
\bea
\ourmmsq{\mathcal{A}} &=& \min\ourMaxMinBracs{M^2} = \min\ourMaxMinBracs{ P^2 }  \nonumber\\
             &=& \min \ourMaxMinBracs{(\Sigma_i A^\mu_i)^2} = (\Sigma_i A_i)^2.  \label{eq:mmsetlor}
\eea
Note that in the special case of fully specified 1+3 vectors, the mass bound for the set is the same as the mass bound of the single object formed of the sum of those vectors
\beq \label{eq:mmsetlorsum}
\ourmm{\mathcal{A}} = \ourmm{\Sigma_i A_i}.
\eeq

Let us further generalize our results to an arbitrary set of (1+3)-vectors 
$\mathcal{A}$ and $(1+2)_\ourT$-projected vectors 
$\mathcal{B}_\ourT=\ourSet{b^\alpha_{j\ourT} \mid 1 \le j \le |\mathcal{B}_\ourT| }$. 
Each of the $b^\alpha_{j\ourT}$ has a (1+3)-vector equivalence class $B^\mu_j$ for which the $z$ components can take any value. Writing $\mathcal{B}=\ourSet{B^\mu_j \mid 1 \le j \le |\mathcal{B}_\ourT| }$ and $\mathcal{B}_z=\ourSet{b_{jz} \mid 1 \le j \le |\mathcal{B}_z| }$, we can therefore write the mass bound as 
\beq
\ourmm{\mathcal{A}, \mathcal{B_\ourT} } = \ourmm{ \mathcal{A}, \mathcal{B} } = \min_{\mathcal{B}_z}\ourMaxMinBracs{\left(\Sigma_i A^\mu_i + \Sigma_j B^\mu_j\right)^2}.
\eeq
where each of the $B^\mu_j$ has a free $z$ component.
The result can be found by induction. 
The bound $\ourmm{K^\mu_1,\,\ldots,\,K^\mu_m}$ for some set of fully specified (1+3)-vectors is given by the sum $\ourmm{\Sigma_{i=1,m} K^\mu_i}$ by \eqref{eq:mmsetlorsum}.
Adding a further 1+3 vector $K^\mu_{m+1}$ which has free $z$ momentum to that set gives a bound $\ourmm{K^\mu_1,\,\ldots,\,K^\mu_{m+1}}$. A similar argument to that which led to \eqref{eq:lorTrapidity} shows that the rapidity of $K^\mu_{m+1}$ must be equal to that of $\Sigma_{i=1,m} K^\mu_i$. With this constraint applied $K^\mu_{m+1}$ becomes a fully specified 1+3 vector, so we can treat it as one of the known 1+3 vectors and proceed with the next 1+2 ($\ourT$-projected) vector in the set.

Applying this argument sequentially to the $B_j$ we find that 
\beq
\ourmm{\mathcal{A}, b^\alpha_{j\ourT}} = \ourmm{(\Sigma_i{A^\mu_i}), B^\mu_j}, \label{eq:mmlorsetTset}
\eeq
where each of the $B_j$ has the same rapidity as $\Sigma_i{A^\mu_i}$. 

Since the set of 1+2 $\ourT$-projected vectors is isomorphic
to the set of 1+3 vectors with fixed (but arbitrary) rapidity under the operations of addition and inner product, we can rewrite this bound as
\bea
\ourmm{\mathcal{A}, \mathcal{B}_\ourT} = \ourmm{(\Sigma_i A^\mu_i), (\Sigma_j b^\alpha_{j\ourT})}, \label{eq:sum4vbound}
\eea
the bound for the summed 1+3 vector $(\Sigma_i{A^\mu_i})$ and the 1+2 $\ourT$-projected vector $(\Sigma_j b^\alpha_{j\ourT})$, the explicit formula for which is given in \eqref{eq:mmlorT}.

We can further extend the argument by allowing some other daughters parameterized only by their two-momentum to be added to the set,
\beq
\ourmm{\mathcal{A}, \mathcal{B}_\ourT, \mathcal{C}_\genericT },
\eeq
with $\mathcal{C}_\genericT=\ourSet{\vec{c}_{k\genericT} \mid 1 \le k \le |\mathcal{C}_\genericT| }$. Each 2-vector $\vec{c}_{k\genericT}$ has a corresponding equivalence class which can be represented by a 1+2 $\ourT$-projected vector $c^\alpha_{k\ourT}$ with unknown mass. The arguments which led to  \eqref{eq:mmlorsetTset} apply equally to the $c_{k\ourT}$, so the corresponding $C^\mu_k$ 1+3 vector rapidities are set equal to $\Sigma_i A\mu_i$, but now we have the extra minimization over the masses which fixes $m_C=0$ for each $C^\mu_k$ (or indeed $c^\alpha_{k\ourT}$) .

Therefore
\beq
\ourmm{\mathcal{A}, \mathcal{B}_\ourT, \mathcal{C}_\genericT } 
= \ourmm{(\Sigma_i A^\mu_i), (\Sigma_j b^\alpha_{j\ourT} + \Sigma_k c^\alpha_{k\massless} ) }\ .
\label{eq:mmlorsetTsetvecset}
\eeq
Now \eqref{eq:mmlorsetTsetvecset} has the same form as all the previous bounds, 
but in obtaining the result we have found out something non-trivial:
one would {\em not} get the best bound on $M$ if 
one were simply to replace the 
set of 2-vectors $\mathcal{C}_\genericT$ by their sum $(\Sigma_k \vec{c}_{k\genericT})$:
One must instead add the corresponding massless 1+2 vectors $c_{k\massless}^\alpha$.

In principle we could now try to extend our bounds to include (arbitrarily large numbers of) $\ourperp$-projected 1+2 vectors.
However we shall {\em not} do so for two reasons. The first reason is that in collider experiments such as the LHC, situations for which $\ourperp$ projection is appropriate are rare. It is only in very unusual cases where we might find ourselves knowing just the transverse momentum components and the size of the three-velocity, but not the azimuthal angle $\theta$, the $z$-momentum, or the mass. 

The second reason we do not pursue the $\ourperp$ vectors further is that one ends up with
a real mess, as we have seen. The most basic pair-wise combination $\ourmm{A,b_\ourperp}$ requires solution of a quartic equation in $b_z$. Only if one is solely interested in combining (\ourT{}, \ourperp{}, \massless)-projected vectors might the expressions be tractable, but the utility of such a combination is unclear.

\subsection{Mass bound hierarchies}
\label{sec:hierproofs}

The similarity in the expressions for the mass bounds derived in the preceding sections 
allows for a further observation -- that as progressively more information is neglected or 
unknown, the mass bound is lowered. Intuitively one would expect this, since the absence of
hard information causes one to have to be progressively more conservative, but we can, with  
little additional work, show this explicitly to be the case.

We set out, therefore, to prove the hierarchy that was seen earlier in \eqref{eq:sneakpreviewofhierarchy}.
Our proof proceeds in two stages.  In the first stage we demonstrate the result for the case $\numparents=1$, in which the hierarchy becomes:
\beq
M_{1} =  M_{1\ourT} \ge M_{\ourT1}\ge M_{\massless1} \ge M_{1\massless}. \label{eq:thenequals1hierarchy}
\eeq  In the second stage we extend this to general $\numparents$.

Using the results of the previous section, we can treat each of the mass bound variables in
terms of the composite visible and composite invisible objects described in section 
\ref{sec:composite}. The equality in \eqref{eq:thenequals1hierarchy} then results from the definition of $M_{1}$ as a concrete case of \eqref{eq:mmlorT}, where $A^\mu$ represents the visible $P^\mu$, and $b^\alpha_\ourT$ the invisible $q^\alpha_\ourT$. Similarly, $M_{1\ourT}$ is just \eqref{eq:mtTTdef}, where $a^\alpha_\ourT$ and $b^\alpha_\ourT$ stand in for $p^\alpha_\ourT$ and $q^\alpha_\ourT$. On comparing \eqref{eq:mmlorT} with \eqref{eq:mtTTdef}, we see that they are identical, and hence $M_1 = M_{1\ourT}$.

For the next statement, $M_{1\ourT} \ge M_{\ourT1}$, we have to consider the difference between ``early'' and ``late'' partition, i.e. whether we retain information about the relative longitudinal momenta of the visibles. Let our visible composite $\mathbf{P}^\mu_a$ of parent  $\parent a$ be composed of constituents $P^\mu_i$, i.e.
\beas
\comp{P}^\mu_a &=& \sum_{i \in \visassign{a} } P^\mu_i \\
&=& \left( \comp{E}_a, \vec{\comp{p}}_{a \genericT}, \comp{p}_{a z} \right) \, ,
\eeas
with
\beas
\comp{E}_a &=& \sum_{i \in \visassign{a}} E_i \ \, , \\
\vec{\comp{p}}_{a \genericT} &=& \sum_{i \in \visassign{a}} \vec{p}_{i\genericT} \, , \\
\comp{p}_{a z} &=& \sum_{i \in \visassign{a}} p_{iz} \, . 
\eeas
We form the early-partitioned composite 
\beqs
\comp{p}^\alpha_{a \ourT} = \left( \sum_{i \in \visassign{a}} P^\mu_i \right)_\ourT \, \\
    = \left( \comp{e}_{a \ourT},  \vec{\comp{p}}_{a \genericT} \right) \, ,
\eeqs
and the late-partitioned composite
\beqs
\comp{p}^\alpha_{\ourT a} = \sum_{i \in \visassign{a}} p^\alpha_{i\ourT} \, , \\
    = \left( \comp{e}_{\ourT a},  \vec{\comp{p}}_{\genericT a}\right) \, , 
\eeqs
differing only in their energy components
\beas
\comp{e}_{a \ourT} &=& \sqrt{\comp{E}^2_a - \comp{p}_{a z}^2} \\
	&=& \sqrt{ \comp{M}^2_a + \comp{p}_{a \genericT}^2 } \, ,  \\[2mm]
\comp{e}_{\ourT a} &=& \sum_{i \in \visassign{a}} e_{i\ourT} \nonumber \\
	&=& \sqrt{ \comp{m}^2_{\ourT a} + \comp{p}_{\genericT a}^2 } \, , 
\eeas
where
\beas
\comp{M}^2_a = \left( \sum_{i \in \visassign{a}} P_i^\mu \right)^2 \, , \\
\comp{m}_{\ourT a}^2 = \left( \sum_{i \in \visassign{a}} p_{i\ourT}^\alpha \right)^2 \, . 
\eeas
Of course, it is established in the preceding sections \ref{sec:ABbound} and 
\ref{sec:AbTbound} that $\comp{M}^2_a \ge \comp{m}^2_{\ourT a}$, since $\comp{m}_{\ourT a}$ 
could be constructed by repeated minimizations of $\comp{M}_a$ over the longitudinal 
momentum components $(p_z)_i$. Hence, $\comp{e}_{a \ourT} \ge \comp{e}_{\ourT a}$.

If we now define analogous quantities $\tilde{\comp{e}}_{a\ourT}, \tilde{\comp{e}}_{\ourT a}$ 
and $\vec{\comp{q}}_{a\genericT}$ for the composite invisible particle, 
then all the same arguments apply. Forming the two mass variables as 
in \eqref{eq:MaourTdef} and \eqref{eq:MaourTdef2},
\bea
M_{1\ourT}^2 = (\comp{e}_{1 \ourT}+\tilde{ \comp{e}}_{1 \ourT })^2 
    &-& (\vec{\comp{p}}_{1\genericT}+\vec{\comp{q}}_{1\genericT})^2 \nonumber \\
    &\ge& \nonumber \\
    (\comp{e}_{\ourT 1}+\tilde{\comp{e}}_{\ourT 1})^2 
    &-& (\vec{\comp{p}}_{\genericT 1}+\vec{\comp{q}}_{\genericT 1})^2
     = M_{\ourT1}^2 \, . \nonumber
\eea

Moving next to $M_{\massless 1}$, we note that this is simply the previous case, with an 
additional minimization over the masses $M_i$ of the constituent particles, which must 
reduce the size of the bound, forcing $M_{\massless 1} \le M_{\ourT 1}$.

For the final inequality, we recall the statement due to \eqref{eq:mmlorsetTsetvecset},
that says the bound is weakened (i.e. made smaller) if we base the bound on the sum of the 
transverse two-vectors, rather than promoting them to \massless{}-projected (1+2)-vectors 
before summing. The difference is solely in the energy component -- the late-partitioned 
$\comp{p}^\alpha_{\massless a}$ has energy component
\beqs
\comp{e}_{\massless a} = \sum_{i \in \visassign{a} } (p_\genericT)_i \, ,  \\
\eeqs
whereas the early-partitioned $\comp{p}^\alpha_{a \massless}$ has energy component
\beqs
\comp{e}_{a \massless} = \comp{p}_{a\genericT} \, . \\
\eeqs
By the triangle inequality, $\comp{e}_{\massless a} \ge \comp{e}_{a \massless}$, yielding
the final required result, that
\beas
M_{\massless1}^2 = (\comp{e}_{\massless 1}+\tilde{\comp{e}}_{\massless 1})^2 
    &-& (\vec{\comp{p}}_{\genericT 1}+\vec{\comp{q}}_{\genericT 1})^2  \\
    &\ge& \\
    (\comp{e}_{1\massless}+\tilde{\comp{e}}_{1\massless})^2 
    &-& (\vec{\comp{p}}_{1\genericT}+\vec{\comp{q}}_{1\genericT})^2
     = M_{1\massless}^2 \, . 
\eeas

Armed with this knowledge, we tackle the hierarchy when $N>1$. We revisit the definitions of 
$M_\numparents, M_{\numparents\ourT}, M_{\ourT\numparents}, 
M_{\numparents\massless}$, and $M_{\massless\numparents}$, from section 
\ref{sec:TMNT}, as
\beas
M_\numparents(\chiM)&\equiv& \min_{\substack{\sum \vec{q}_{iT} = \mptvec}}
    \ourMaxMinBracs{ \max_a
    \ourMaxMinBracs{{\cal M}_{a}(\comp{P}_{a},\comp{Q}_{a},\invismassseti a)} },  \\ 
M_{\numparents\ourT}(\chiM)&\equiv& \min_{\substack{\sum \vec{q}_{iT} = \mptvec}}
    \ourMaxMinBracs{ \max_a
    \ourMaxMinBracs{{\cal M}_{a\ourT}(\comp{p}_{a\ourT},\comp{q}_{a\ourT},\invismassseti a)} }, \\ 
M_{\ourT\numparents}(\chiM)&\equiv& \min_{\substack{\sum \vec{q}_{iT} = \mptvec}}
    \ourMaxMinBracs{ \max_a
    \ourMaxMinBracs{{\cal M}_{\ourT a}(\comp{p}_{\ourT a},\comp{q}_{a\ourT a},\invismassseti a)} }, \\
M_{\numparents\massless}(\chiM)&\equiv& \min_{\substack{\sum \vec{q}_{iT} = \mptvec}}
    \ourMaxMinBracs{ \max_a
    \ourMaxMinBracs{{\cal M}_{a\massless}(\comp{p}_{a\massless},\comp{q}_{a\massless},\invismassseti a)} },  \\ 
M_{\massless\numparents}(\chiM)&\equiv& \min_{\substack{\sum \vec{q}_{iT} = \mptvec}}
    \ourMaxMinBracs{ \max_a
    \ourMaxMinBracs{{\cal M}_{\massless a}(\comp{p}_{\massless a},\comp{q}_{\massless a},\invismassseti a)} }. 
\eeas

At first glance, it might seem alarming that we assert $M_1 = M_{1\ourT}$, when $M_1$ 
seems to be built of a fully (1+3)-dimensional object ${\cal M}_1(\comp{P}_1,\comp{Q}_1,\invismassseti 1)$.
But in fact, with the components $(q_z)_i$ left free, the minimization will (for reasons identical
to those in the discussion of early and late partitioning) be achieved when all the constituents
of $\comp{Q}_1$ have equal rapidity to $\comp{P}_1$, meaning
\beas
{\cal M}_1(\comp{P}_1,\comp{Q}_1,\invismassseti 1)
    &=& {\cal M}_{1\ourT}(\comp{p}_{1\ourT},\comp{q}_{1\ourT},\invismassseti 1).
\eeas

But this should apply to all $N$, since the only constraint on the invisibles of each parent 
$\comp{Q}_a$ is on their transverse momentum components. That is, for each of the $N$ 
parents, given our inputs we will get 
\beqs
{\cal M}_a(\comp{P}_a,\comp{Q}_a,\invismassseti a) 
= {\cal M}_{1\ourT}(\comp{p}_{1\ourT},\comp{q}_{1\ourT},\invismassseti 1), 
\eeqs
and therefore we immediately see that
\bea
M_\numparents(\chiM) &\equiv& \min_{\substack{\sum \vec{q}_{iT} = \mptvec}}
    \ourMaxMinBracs{ \max_a
    \ourMaxMinBracs{{\cal M}_{a}(\comp{P}_{a},\comp{Q}_{a},\invismassseti a)} }
\nonumber \\ 
&=& \min_{\substack{\sum \vec{q}_{iT} = \mptvec}}
    \ourMaxMinBracs{ \max_a
    \ourMaxMinBracs{{\cal M}_{a\ourT}(\comp{p}_{a\ourT},\comp{q}_{a\ourT},\invismassseti a)} }\nonumber \\
&\equiv& M_{\numparents\ourT}(\chiM) \ . \label{eq:proofofMNtisMNT}
\eea

Next one might ask whether the successive inequalities still hold. The very first one follows
straightforwardly. Only in the input vectors to each of the $N$ parental mass bounds
$\mathcal{M}_{a\ourT}$ do $M_{\numparents\ourT}$ and $M_{\ourT\numparents}$ differ. 
Furthermore, since the late-partitioned input vectors $\comp{p}_{\ourT a}$ and 
$\comp{q}_{\ourT a}$ will have smaller energy components than their early-partitioned
counterparts $\comp{p}_{a\ourT}$ and $\comp{q}_{a\ourT}$, each of the individual parental
bounds follows the relation
\beq
\mathcal{M}_{a\ourT} (\comp{p}_{a\ourT}, \comp{q}_{a\ourT},\invismassseti a) \ge
\mathcal{M}_{\ourT a}(\comp{p}_{\ourT a},\comp{q}_{\ourT a},\invismassseti a) \label{eq:nobigger} \, ,
\eeq
for every possible choice of unprojected inputs $\comp{P}_{a},\comp{Q}_a$.

To complete the argument, we need to establish that the global minimum considering all 
trial $\vec{q}_{i\genericT}$ cannot increase if any or all of the parental bounds decrease. 

The minimization probes the full space of $\ourSet{\vec{q}_{i\genericT}}$, subject to the 
constraint that their sum is the missing transverse momentum vector, with all other 
parameters having been specified. For the minimization to pick out a larger value for
$M_{\ourT\numparents}$ than for $M_{\numparents\ourT}$, we must have
\beq
\max_a
    \ourMaxMinBracs{{\cal M}_{\ourT a}(\comp{p}_{\ourT a},\comp{q}_{\ourT a},\invismassseti a)} >
\max_a
    \ourMaxMinBracs{{\cal M}_{a\ourT}(\comp{p}_{a\ourT},\comp{q}_{a\ourT},\invismassseti a)}
\eeq
for the same values of $\ourSet{\vec{q}_{i\genericT}}$ that give the value of 
$M_{\numparents\ourT}$, if nowhere else. But we have already established 
\eqref{eq:nobigger} for all $a$ and all inputs. So we are led to the conclusion 
\beq
M_{\numparents\ourT}(\chiM) \ge M_{\ourT\numparents}(\chiM)\ .
\eeq

Actually, we have achieved more than that. The same argument holds for the remaining 
levels of the hierarchy involving the \massless{}-projection. So we can boldly claim our 
final result and can retire to a well-deserved cuppa
\bea
M_{\numparents} &=&  M_{\numparents\ourT} \nonumber \\
&\ge& \nonumber \\
& M_{\ourT\numparents}& \nonumber \\
&\ge& \nonumber \\
& M_{\massless\numparents} & \nonumber \\
&\ge& \nonumber \\
& M_{\numparents\massless} . & \label{eq:thehierarchyisprovedforgeneraln}  
\eea

\vfill

\begin{acknowledgments}
This work is supported in part by a US Department of Energy grant DE-FG02-97ER41029, 
and by the Science and Technology Research Council of the United Kingdom.
TJK is supported by a Dr. Herchel Smith Fellowship from Williams College.
KCK is partially supported by the National Science Foundation under Award No. EPS-0903806 and 
matching funds from the State of Kansas through Kansas Technology Enterprise Corporation.
We would like to thank W. Buttinger and B. Gripaios for useful discussions.
We are grateful to Joe and Mary Ann McDonald for interrupting their Indian tiger 
safari to give permission for the use of the \beastie{} photograph.
\end{acknowledgments}

\begin{widetext}
\bibliography{ProjectionMethods}
\end{widetext}



\end{document}